\documentclass[sort&compress]{elsarticle}
\usepackage[utf8]{inputenc}

\usepackage[capposition=top]{floatrow}
\usepackage{setspace}
\usepackage[margin=0.6in]{geometry}
\usepackage[colorlinks=true]{hyperref}
\hypersetup{linkcolor=blue}
\hypersetup{citecolor=blue}
\usepackage[small, justification=justified, singlelinecheck=false]{caption}
\usepackage{url}
\usepackage{booktabs}
\usepackage{multirow}
\usepackage{graphicx}
\usepackage{adjustbox}
\usepackage{amssymb}
\usepackage{amsmath}
\usepackage{dutchcal}
\usepackage{tabularx}
\usepackage[table,xcdraw]{xcolor}
\usepackage{algorithm}
\usepackage{algpseudocode}
\usepackage{comment}
\usepackage{subcaption}
\usepackage{numprint}
\npthousandsep{\,}
\usepackage{tocbibind}
\usepackage{enumitem}
\usepackage{bigfoot}
\usepackage[toc,page]{appendix}

\makeatletter
 \def\ps@pprintTitle{%
  \let\@oddhead\@empty
  \let\@evenhead\@empty
  \def\@oddfoot{}%
  \let\@evenfoot\@oddfoot
 }
\makeatother

\begin{document}
	
\begin{frontmatter}
\title{Deep Limit Order Book Forecasting \\[1ex] \large A microstructural guide}
		
\author[myaddress2,myaddress4]{Antonio Briola\corref{mycorrespondingauthor}}
\ead{antonio.briola.20@ucl.ac.uk}

\author[myaddress2,myaddress4]{Silvia Bartolucci}

\author[myaddress2,myaddress3,myaddress4]{Tomaso Aste}

\cortext[mycorrespondingauthor]{Corresponding author(s)}

\address[myaddress2]{Department of Computer Science, University College London, Gower Street, WC1E 6EA - London, United Kingdom.}
\address[myaddress3]{Systemic Risk Centre, London School of Economics, London, United Kingdom.}
\address[myaddress4]{UCL Centre for Blockchain Technologies, London, United Kingdom.}
	
\begin{abstract}
We exploit cutting-edge deep learning methodologies to explore the predictability of high-frequency Limit Order Book mid-price changes for a heterogeneous set of stocks traded on the NASDAQ exchange. In so doing, we release `LOBFrame', an open-source code base to efficiently process large-scale Limit Order Book data and quantitatively assess state-of-the-art deep learning models' forecasting capabilities. Our results are twofold. We demonstrate that the stocks' microstructural characteristics influence the efficacy of deep learning methods and that their high forecasting power does not necessarily correspond to actionable trading signals. We argue that traditional machine learning metrics fail to adequately assess the quality of forecasts in the Limit Order Book context. As an alternative, we propose an innovative operational framework that evaluates predictions' practicality by focusing on the probability of accurately forecasting complete transactions. This work offers academics and practitioners an avenue to make informed and robust decisions on the application of deep learning techniques, their scope and limitations, effectively exploiting emergent statistical properties of the Limit Order Book.

\hfill
	
\end{abstract}
	
\begin{keyword}
Limit Order Book $\cdot$ Econophysics $\cdot$ High Frequency Trading $\cdot$ Market Microstructure $\cdot$ Deep Learning
\end{keyword}	

\end{frontmatter}

\newpage

\section{Introduction}\label{sec:Introduction}
Financial markets are highly stochastic environments with a generally low signal-to-noise ratio, characterized by the interplay of a heterogeneous group of actors competing at different time scales, possessing asymmetric levels of information and diverse technical abilities to trade financial securities. To efficiently manage the resulting emerging complexity, modern exchanges rely on computerized systems that collect and organize the continuous flux of incoming and exiting orders to provide eased matching mechanics between demand and supply while guaranteeing transaction fairness. These systems usually exploit a \textit{FIFO} (first-in, first-out) mechanism \cite{bouchaud2018trades} to establish execution's priority, guaranteeing, at each point in time, access to the market's status to all the participants through the `Limit Order Book' (LOB) \cite{abergel2016limit}. The way such information is used and the consequent market's response to the exploitation of inefficiencies has been studied from an ecological perspective in many research works \cite{farmer2013ecological, bouchaud2009markets, scholl2021market}. In this paper, we are particularly interested in the role of high-frequency trading (HFT). HFT refers to strategies that gain an edge through speed, allowing certain traders to act on information not yet accessible to others \cite{lehalle2018market}. These actors exploit the market's microstructure imperfections to the detriment of other traders, suggesting a predator–prey relationship with other strategies \cite{farmer2013ecological}. HFT integrates and exploits market data at different scales instead of directly handling external information, ultimately producing noise and thereby maintaining unpredictability in price trends \cite{bouchaud2009markets}. The impact of this practice on markets' stability has been widely discussed in the literature, yet a unique consensus regarding its beneficial or detrimental effects has not been reached \cite{markets2009impact, zhang2010high, zhang2011impact, jarrow2012dysfunctional, cartea2012value}.

Recent achievements in deep learning for uni- and multi-variate time-series forecasting have opened new and exciting scenarios in high-frequency LOB forecasting \cite{dixon2018sequence, sirignano2019deep, briola2020deep, zhang2019deeplob}. The diversity of approaches to the problem has fostered a vibrant yet evolving research environment. A major driver of such divergent approaches is the difference between the focus of academic efforts and practitioners' interests in the field. On one side, this distance plunges its roots and is raised by technical problems that academics face in this application domain, including, but not limited to, the costs of accessing and storing live and historical high-frequency data as well as the costs of accessing considerably large-scale computational resources to process them. On the other side, the approach of the newborn community of scholars interested in applying deep learning techniques to LOB forecasting often overlooked the complex nature of the system under analysis while exploiting it as another playground for challenging the continuously evolving state-of-the-art in time series forecasting.

Statistical properties of Limit Order Book data emerging from the mechanics of trading at the micro-scale (i.e., at the level of order submissions), including, for instance, the analysis of liquidity and spread patterns in different types of stocks, have been extensively studied in the context of the microstructural analysis and modelling of markets \cite{bouchaud2018trades, o1998market}. Market microstructure analysis is aimed at tackling problems related to how information and trading intentions are aggregated into prices, the drivers of transaction costs in markets, and how to optimally design markets, to name a few. Such information on the typical emergent behaviour of stocks' dynamics has been so far largely unexploited in problems related to LOB forecasting. In this work, we aim to bridge this gap, providing a straightforward way to link stocks' `predictability rate’ (i.e., the probability of correctly forecasting the direction of mid-price movements over a certain time horizon) directly to their microstructural properties. Starting from a set of $15$ stocks traded on the NASDAQ exchange, we first define and compute their main microstructural properties, and then we analyze under which conditions a well-known state-of-the-art deep learning model (i.e., DeepLOB \cite{zhang2019deeplob}) succeeds or fails in predicting the direction of mid-price changes at non-homogeneous time horizons. We finally discuss the usability of such forecasts in real-world scenarios, providing a clear-cut and general way to quantify the simulation-to-reality gap \cite{prata2023lob}. In so doing, we provide the academic and practitioners community with `LOBFrame'\,\footnote{\url{https://github.com/FinancialComputingUCL/LOBFrame}.}, an open-source code base with high usability and versatility, which eases the integration with new forecasting models. This step is needed since, despite the optimistic environment surrounding deep learning for LOB forecasting, the field continues to grapple with the lack of standardized protocols. Even state-of-the-art methodologies fall short in providing easily accessible, complete and operable open-source software as well as data-processing pipelines, hindering their potential to set definitive standards for the community. The current paper and the associated code aim to establish a benchmark resource in the field.

The rest of the paper is organised as follows. Section \ref{sec:Limit_Order_Book} introduces the technical priors to understand LOB's mechanics. Section \ref{sec:Related_Work} presents an overview of essential scientific works describing statistical and automated modelling of LOB's dynamics. In Section \ref{sec:Data}, we describe in details the dataset used for our experiments, while in Section \ref{sec:Methods}, we provide technical insights into the novel framework used to train and validate the DeepLOB model. In Section \ref{sec:Microstructural_Priors}, we describe a set of microstructural properties of the stocks used in our analysis, while in Section \ref{sec:Results}, we present the results of the forecasting experiments. Lastly, in Section \ref{sec:Conclusion}, we wrap up our findings, providing a unified view of market-microstructure-informed deep learning methods for LOB forecasting, with an overview on open challenges in the field.

\section{Limit Order Book}\label{sec:Limit_Order_Book}

The majority of modern exchanges utilize an electronic system that stores and matches agents' trading intentions. This system operates on a data structure known as the `Limit Order Book' (LOB). Each security has its own LOB, which gives traders simultaneous access to the currently visible market's supply and demand. In this context, the price formation of an arbitrary security is a self-organized process driven by the submission and cancellation of orders \cite{briola2021deep}.

An order can be considered a visible declaration of a market participant’s intention to buy or sell a fixed amount of an asset's shares at a specified price. Its execution is subordinated to finding a counterpart willing to trade at the same conditions. An order $\mathbcal{o}$ is formally defined as a tuple $(\epsilon_{\mathbcal{o}}, p_{\mathbcal{o}}, v_{\mathbcal{o}}, \tau_{\mathbcal{o}})$ \cite{bouchaud2018trades}. $\epsilon_{\mathbcal{o}}$ indicates the sign or direction at which a given asset is traded. Conventionally, buy (or bid) orders have a positive sign $\epsilon = +1$, while sell (or ask) orders have a negative sign $\epsilon = -1$. $p_{\mathbcal{o}}$ indicates the price a trader wants to trade a given asset. Orders can be submitted at prices belonging to a discrete set, constituting the LOB's price levels (or quotes). The smallest price increment is known as `tick size' ($\theta$), which, on the NASDAQ exchange, is fixed and equal to $\$ 0.01$ for all the securities\,\footnote{The value of $\theta$ varies across exchanges and, within the same exchange, for a single stock, it can vary across time as a function of the price attained by the asset.}. $v_{\mathbcal{o}}$ indicates the number of asset shares a trader wants to exchange. Orders can be submitted on a discrete set of volumes, constituting LOB's volume levels. The smallest volume increment, which determines the minimum distance between two consecutive volume levels, is known as `lot size', and, on the NASDAQ exchange, it is fixed and equal to $1$ for all the securities. $\tau_{\mathbcal{o}}$ indicates the time an order is submitted, and it is a continuous variable (typically known with a precision of up to the nanoseconds).

There are three main families of orders that can be submitted: (i) limit orders; (ii) market orders; (iii) cancellation orders. A limit order represents an intention to buy or sell a fixed amount of an asset at a price different from the current best available matching price on the opposite side of the LOB. There is no guarantee of execution for this type of orders. A limit order is typically subject to lower transaction costs (i.e., the costs of transferring ownership rights \cite{niehans1989transaction}) since it actively provides liquidity to the LOB. A market order represents an intention to buy or sell a fixed amount of shares at the current best available matching price. If its volume is higher than the one supporting the best quote on the opposite side of the LOB, the remaining amount is executed against active orders at deeper price levels (sitting further away from the best quotes). A market order is typically subject to higher transaction costs since it reduces the liquidity available in the LOB. A cancellation order represents an intention to fully or partially delete an active limit order. It is typically not subject to any transaction cost. 

\begin{figure}[H]\label{fig:lob_schema}
    \centering
    \includegraphics[scale=0.6]{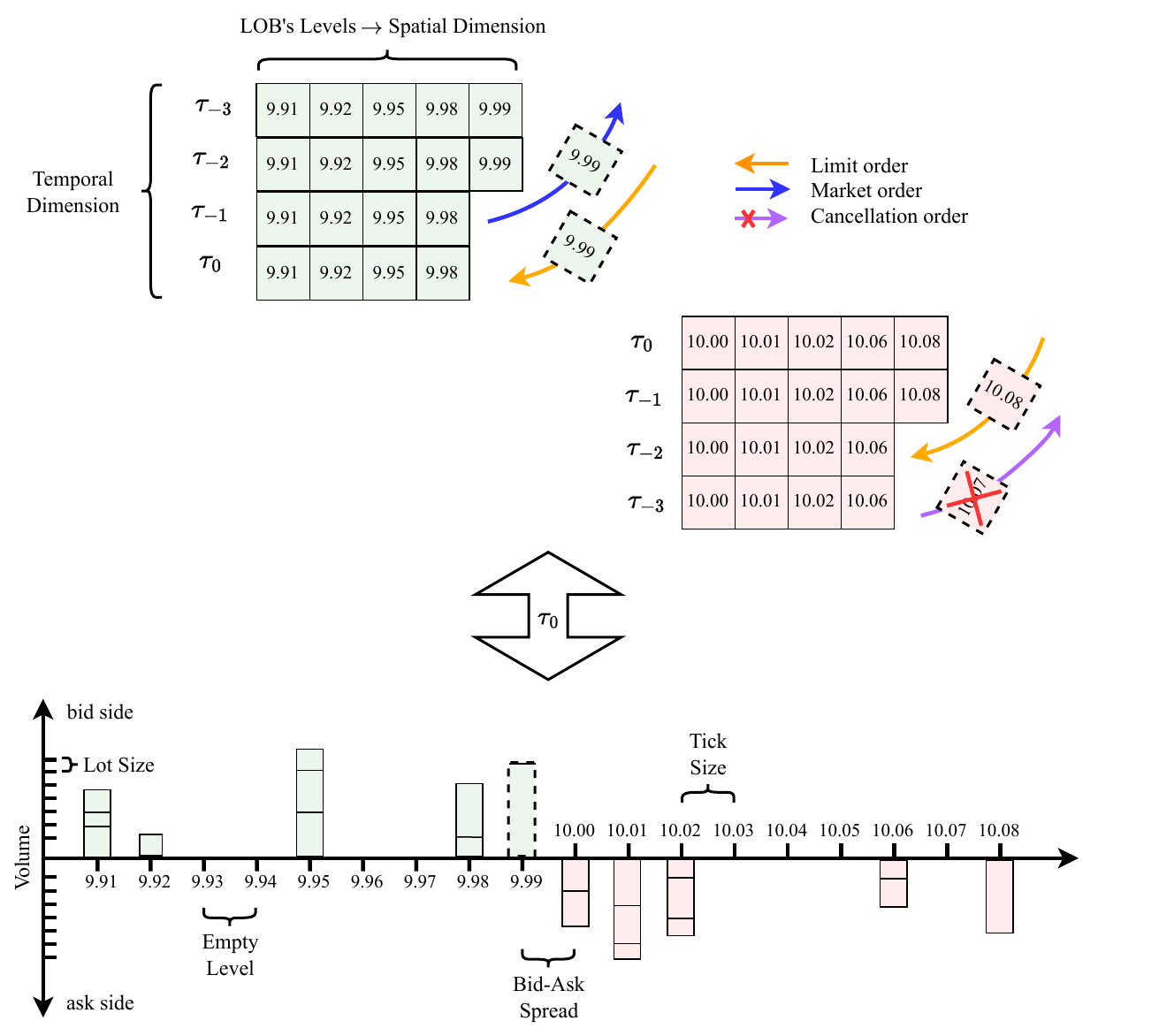}
    \caption{Pictorial representation of the LOB. In the upper part of the Figure, we show the dynamical evolution of the LOB price levels as a consequence of the submission of limit, market and cancellation orders, while, in the lower part, we show a static view of a LOB snapshot (i.e., $\mathbb{L}(\tau_0)$) including also the volumes.}
    \label{fig:lob_schema}
\end{figure}

All the elements introduced above allow us to define the LOB as a collection of unmatched (active) limit orders for a given asset, on a given platform, at time $\tau$ \cite{briola2021deep} (see Figure \ref{fig:lob_schema}). We represent it as a multivariate time series $\mathbb{L}$, where each $\mathbb{L}(\tau) \in \mathbb{R}^{4L}$ is a LOB record characterised by $L$ price/volume levels \cite{prata2023lob}. More specifically, $\mathbb{L}(\tau)$ can be written in the form of $\{\mathbf{P}^s(\tau), \mathbf{V}^s(\tau)\}_{s \in \{\text{ask}, \text{bid}\}}$, where $\textbf{P}^{\text{ask}}(\tau), \textbf{P}^{\text{bid}}(\tau) \in \mathbb{R}^L$ are the sets of prices on the ask and bid side, and $\textbf{V}^{\text{ask}}(\tau), \textbf{V}^{\text{bid}}(\tau) \in \mathbb{R}^L$ are the sets of volumes on the ask and bid side, respectively:

\begin{equation}\label{eq:Compressed_Representation_LOB}
    \mathbb{L}(\tau) = \{p_\ell^{\text{ask}}(\tau), v_\ell^{\text{ask}}(\tau), p_\ell^{\text{bid}}(\tau), v_\ell^{\text{bid}}(\tau)\}_{\ell=1}^L \ .
\end{equation}

This means that, $\forall \, \tau \in \{1,\dots,N\}$ and $\forall \, \ell \in \{1,\dots, L\}$ on the $s$ side, $v_\ell^{\text{s}}(\tau)$ shares can be sold or bought at price $p_\ell^{\text{s}}(\tau)$. 
The mid-price $m_\tau$ of the stock at time $\tau$, is defined as the average between the best ask price (i.e., $p_1^{\text{ask}}(\tau)$) and the best bid price (i.e., $p_1^{\text{bid}}(\tau)$),  $m_\tau=\frac{p_1^{\text{ask}}(\tau) + p_1^{\text{bid}}(\tau)}{2}$. The bid-ask spread $\sigma_\tau$ of the stock at time $\tau$, is defined as the difference between the best ask price and the best bid price, $\sigma_\tau=p_1^{\text{ask}}(\tau) - p_1^{\text{bid}}(\tau)$.

\section{Related Work}\label{sec:Related_Work}
Market microstructure analysis and automated learning modelling of LOB dynamics are two continuously evolving research areas. In this Section, we provide a subset of core references to works that allow the reader to navigate the broader universe of related literature. 

Market microstructure entails the analysis of how  traders’ intentions are translated into prices and volumes \cite{madhavan2000market, biais2005market}. The aim is to understand fundamental issues and phenomena, such as characterising price formation mechanisms \cite{lillo2021order, bonart2018continuous, mike2005empirical} and quantifying the impact of asymmetric information \cite{glosten1985bid, kyle1985continuous}. In terms of markets' dynamics, price jumps \cite{marcaccioli2022exogenous, toth2009studies, zheng2012price, cont2013price} and flash crashes events \cite{turiel2021self, turiel2022heterogeneous, brewer2013market, kirilenko2017flash, paddrik2017effects} have been extensively studied. Modelling and analysing transaction costs (e.g. price impact \cite{eisler2012price, cont2014price, avellaneda2008high, mastromatteo2014agent}), and optimal order execution is also among core areas of investigation, especially in the context of HFT \cite{cartea2018enhancing, cartea2015optimal, lehalle2017limit, hollifield2004empirical}. In this paper, we analyse a set of microstructural properties that can be used to characterise and classify stocks. For a complete review and deep discussion of the emergent statistical properties of stocks, we refer the reader to the comprehensive book by Bouchaud et al. \cite{bouchaud2018trades}. Specifically, we focus on spread and liquidity (e.g., depth at best), whose behaviours have been extensively studied for different types of stocks in various markets, evidencing typical intra-day behaviours and statistical properties \cite{abergel2016limit, chakraborti2011econophysics1, lehalle2017limit}. In our analysis, we will also reference a more recently introduced measure, namely the `information richness' (IR) \cite{kolm2023deep}, which characterises the stocks' activity in terms of the number of events occurring at the best levels of the LOB. 

In this paper, we directly link the microstructural assessment of the LOB properties with mid-price changes forecasting performance of a state-of-the-art deep learning model, namely the DeepLOB \cite{zhang2019deeplob}, specifically crafted to handle such data. Regarding the automated learning modelling of the LOB, it is useful to organize the related literature into three main areas of interest: (i) the study of linear models and regression analysis tools for LOB features extraction \cite{zheng2012price, alvim2010daily, detollenaere2017identifying, cenesizoglu2014effects, panayi2018designating}; (ii) the study of non-linear deep learning models for short-term price forecasting \cite{tsantekidis2017using, zhang2019deeplob, passalis2017time, nousi2019machine, tsantekidis2020using, briola2020deep, tran2018temporal, tran2021data, passalis2020temporal, zhang2021multi, guo2023forecasting, kearns2013machine}; (iii) the study of reinforcement learning methods for automated trading \cite{nevmyvaka2006reinforcement, gavsperov2021market, kumar2020deep, briola2021deep, kumar2023deep, gavsperov2021reinforcement, frey2023jax, nagy2023asynchronous, tsantekidis2023modelling, tsantekidis2021diversity, zarkias2019deep}. In this paper, we are interested in the second macro-area. Assessing non-linear deep learning models for short-term price forecasting, we underline $3$ main issues that are common to the majority of referenced research works: (i) the usage of only one simplistic dataset, namely the FI-2010 dataset \cite{ntakaris2018benchmark} as benchmark dataset; (ii) the lack of data analysis for proprietary LOB data; (iii) the difficulty in experiments' reproducibility. FI-2010 consists of $10$ trading days LOB data from $5$ Finnish companies traded on the NASDAQ Nordic stock market. It records $4$M events sampled at intervals of $10$, resulting in $\approx 395$K events. This dataset represents the first and unique experiment to provide a standard benchmark for research in LOB forecasting. Even if it is remarkable, the outcome of this attempt presents some significant limitations. The dataset comes in an already pre-processed (filtered, normalized, and labelled) format so that the original LOB cannot be backtracked, thus hampering thorough experimentation. In addition to this, the dataset is too simplistic, leaving ample space for models' overfitting \cite{prata2023lob}, consequently undermining their robustness when tested in real-world scenarios. Using this dataset as a benchmark for deep learning models represents the first cause of the so-called `simulation-to-reality' gap \cite{liu2022finrl, zaznov2022predicting}. The singular characteristics of this benchmark dataset lead us to discuss the second type of criticality. Proprietary LOB data are considered sensible data, owned and managed by private financial institutions \cite{briola2022dependency} with few third-party vendors, who only distribute exchange-specific historical samples. This makes academic research in the field highly dependent on data sources and the generalization capabilities of developed models questionable. Moreover, an accurate description and quantitative analysis of the dataset are often lacking, making comparisons of models' performances on stocks traded on different exchanges even more unreliable, thus representing a barrier towards experiments' reproducibility \cite{prata2023lob}. Similarly, the code used to conduct the analysis is also rarely shared, directly hampering a meaningful comparison between different approaches.

In the broader context of the questions addressed in this paper, closely related works are those by \cite{lucchese2022short, prata2023lob, kolm2023deep, ait2022and}. The common aspect that links all these research papers is a significant effort in investigating the reasons why deep learning models are effective only in specific scenarios. In the work by \cite{lucchese2022short}, the authors highlight the most important factors that guarantee a successful forecast, including working with what they define `high-frequency stocks', L2 data (i.e., all available bid and ask prices and corresponding volumes) and an order-flow representation of the LOB. In their narrative, the authors are particularly interested in statistically assessing the performance's degradation at longer prediction horizons. The work by \cite{prata2023lob} anticipates some of the technical drawbacks discussed in this Section and highlights the influence of volatility clusters on forecasting models' performances. In the work by \cite{kolm2023deep}, the authors introduce the concept of `information-rich stocks' and show how automated learning models can handle them more easily. Lastly, in the work by \cite{ait2022and}, the authors succeed in isolating some of the variables that are thought to be among the more responsible for driving stocks' predictability.

\section{Data}\label{sec:Data}
In this work we consider $15$ stocks from different sectors and industries, all traded on the NASDAQ exchange \cite{nasdaq_website}. For each of them, we use high quality, tick-by-tick, LOB data from the LOBSTER provider \cite{lobsterdata_what_is_lobster}. To determine stocks' sector and industry affiliation, we follow the taxonomy proposed by the NASDAQ exchange itself \cite{nasdaq_stock_screener}; in this context, the strong heterogeneity of our choices ensures robustness to results. As one can see from Table \ref{tab:stocks_introduction}, we consider $5$ stocks belonging to the `Technology' sector (i.e., AAPL, GOOG, IBM, NVDA, ORCL), $3$ stocks belonging to the `Health Care' sector (i.e., ABBV, PFE, PM), $3$ stocks belonging to the `Telecommunications' sector (i.e., CHTR, CSCO, VZ), $2$ stocks belonging to the `Finance' sector (i.e., BAC, GS), $1$ stock belonging to the `Consumer Staples' sector (i.e., KO) and $1$ stock belonging to the `Consumer Discretionary' sector (i.e., MCD). We consider the entire trading period from $2017$ to $2019$, ensuring to treat only stocks maintaining a large- (i.e., $10$B-$200$B) to mega- (i.e., $\geq 200$B) capitalization. During the $2017$, the average capitalization is $\$253.71$B, the median capitalization is $\$189.34$B, while the standard deviation is $\$220.29$B. During the $2018$, the average capitalization is $\$228.90$B, the median capitalization is $\$162.03$B, while the standard deviation is $\$207.36$B. During the $2019$, the average capitalization is $\$297.05$B, the median capitalization is $\$169.94$B, while the standard deviation is $\$329.04$B. As one can notice, the average capitalization is always larger than the median, indicating the presence of right skewed distributions with `positive' outliers pulling up the means. Across the $3$-year analysis period, the stock with the highest capitalization is AAPL, which breaks the $\$1$T barrier in $2019$. In $2017$, the stock with the lowest capitalization is CHTR ($\$83.94$B), while, in $2018$ and $2019$, the stock with the lowest capitalization is GS ($\$61.43$B and $\$79.86$B, respectively).

{
\renewcommand{\arraystretch}{1.3}
\begin{table}[H]
    \centering
    \caption{Overview of the stocks used in the papers. For each asset we report the ticker, the extended name, the sector, the industry and the capitalization during $2017$, $2018$ and $2019$. To determine stocks' sector and industry affiliation, we follow the taxonomy proposed by the NASDAQ exchange itself \cite{nasdaq_stock_screener}. To determine stock's capitalization we rely on the data provided by CompaniesMarketCap.com \cite{capitalization_provider}. }
    \label{tab:stocks_introduction}
    \resizebox{\columnwidth}{!}{%
    \begin{tabular}{@{}ccccccc@{}}
    \toprule
    \textbf{Ticker} &
      \textbf{Stock Name} &
      \textbf{Sector} &
      \textbf{Industry} &
      \textbf{Capitalization (2017)} &
      \textbf{Capitalization (2018)} &
      \textbf{Capitalization (2019)} \\ \midrule
    AAPL & Apple, Inc.                                 & Technology             & Computer Manufacturing                          & \$860.88 B & \$746.07 B & \$1.287 T  \\
    ABBV & AbbVie, Inc.                                & Health Care            & Biotechnology: Pharmaceutical Preparations      & \$154.39 B & \$136.33 B & \$130.94 B \\
    BAC  & Bank of America Corporation                 & Finance                & Major Banks                                     & \$307.91 B & \$238.25 B & \$311.20 B \\
    CHTR & Charter Communications, Inc.                & Telecommunications     & Cable \& Other Pay Television Services          & \$83.94 B  & \$64.21 B  & \$101.85 B \\
    CSCO & Cisco Systems, Inc.                         & Telecommunications     & Computer Communications Equipment               & \$189.34 B & \$194.81 B & \$203.45 B \\
    GOOG & Alphabet, Inc.                              & Technology             & Computer Software: Programming, Data Processing & \$729.45 B & \$723.55 B & \$921.13 B \\
    GS   & Goldman Sachs Group, Inc.                   & Finance                & Investment Bankers/Brokers/Service              & \$96.09 B  & \$61.43 B  & \$79.86 B  \\
    IBM  & International Business Machines Corporation & Technology             & Computer Manufacturing                          & \$142.03 B & \$101.44 B & \$118.90 B \\
    KO   & Coca-Cola Company                           & Consumer Staples       & Beverages (Production/Distribution)             & \$195.47 B & \$202.08 B & \$236.89 B \\
    MCD  & McDonald's Corporation                      & Consumer Discretionary & Restaurants                                     & \$137.21 B & \$136.21 B & \$147.47 B \\
    NVDA & NVIDIA Corporation                          & Technology             & Semiconductors                                  & \$117.26 B & \$81.43 B  & \$144.00 B \\
    ORCL & Oracle Corporation                          & Technology             & Computer Software: Prepackaged Software         & \$195.72 B & \$162.03 B & \$169.94 B \\
    PFE  & Pfizer, Inc.                                & Health Care            & Biotechnology: Pharmaceutical Preparations      & \$215.89 B & \$249.54 B & \$216.82 B \\
    PM   & Philip Morris International, Inc.           & Health Care            & Medicinal Chemicals and Botanical Products      & \$164.09 B & \$103.78 B & \$132.39 B \\
    VZ   & Verizon Communications, Inc.                & Telecommunications     & Telecommunications Equipment                    & \$215.92 B & \$232.30 B & \$253.93 B \\ \bottomrule
    \end{tabular}%
    }
\end{table}
}

To train our model, we use only a portion of the entire dataset. For each year, we choose $45$ consecutive days of training, $5$ days of validation and $10$ consecutive days of testing (see Table \ref{tab:training_validation_test_split}). It is worth noting that the $5$ days of the validation set are not consecutive and are randomly chosen from the same period of the training set. This choice guarantees greater robustness in the validation step, and it is made possible by the adopted standardization procedure, which prevents any data leakage. In line with what is suggested by Lucchese et al. \cite{lucchese2022short}, a $5$-days feature-wise rolling window $z$-score normalization is applied to the data. This procedure differs from the others used in most of the related literature \cite{zhang2019deeplob} (which usually standardizes the entire training dataset at once based on the overall statistics) and guarantees greater effectiveness in an evolving and strongly non-stationary environment like the LOB. All the experiments presented in the current work are conducted on $L = 10$ LOB's levels (see Equation \eqref{eq:Compressed_Representation_LOB}). Data \cite{lobsterdata_what_is_lobster} are originally made of two separate files: (i) the `message file' lists every market-, limit- and cancellation order, reporting the arrival time, event type, id, size, price and direction; (ii) the `orderbook file' describes the market state (i.e., the total volume of buy or sell orders at each price level) immediately after an event occurs. These files are jointly processed as described by Lucchese et al. \cite{lucchese2022short} by (i) removing crossed quotes; (ii) collapsing states occurring at the same timestamp (to nanoseconds precision) to the last state; and (iii) removing the effects of potential auction calls by considering only events happening between 9:40 am (Eastern Time) and 03:50 pm (Eastern Time). This last choice is made following the suggestion by Briola et al. \cite{briola2021deep} to exclude from experiments the first and the last $10$ minutes of each trading day due to the widely different dynamics and higher volatility that usually affect the market's opening and closing periods. The reader should be aware that trading does not occur on weekends or public holidays, so these days are excluded from all the analyses. 

{
\renewcommand{\arraystretch}{1.5}
\begin{table}[h!]
\centering
\caption{Basic structure of the datasets used during the training, validation and test stages. For each year, for the training and test set, we report the starting and the ending day (which are included), while, for the validation set, we report all the days in an extended way. It is worth noting that weekends and public holidays are not trading days and, consequently, do not belong to any of the datasets.}
\scriptsize
\label{tab:training_validation_test_split}
\begin{tabular}{c|cc|c|cc}
\hline
\textbf{year} & \multicolumn{2}{c|}{\textbf{training}} & \textbf{validation}                                                           & \multicolumn{2}{c}{\textbf{test}} \\ \cline{2-6} 
\textbf{}     & \textbf{from}       & \textbf{to}      & \textbf{days}                                                                 & \textbf{from}    & \textbf{to}    \\ \hline
2017          & 03-13               & 05-22            & \begin{tabular}[c]{@{}c@{}}03-23, 04-05,\\ 04-13, 04-18,\\ 05-02\end{tabular} & 05-23            & 06-06          \\ \hline
2018          & 08-09               & 10-18            & \begin{tabular}[c]{@{}c@{}}08-15, 08-16,\\ 09-19, 09-26\\ 10-03\end{tabular}  & 10-19            & 11-01          \\ \hline
2019          & 06-04               & 08-13            & \begin{tabular}[c]{@{}c@{}}06-14, 06-27,\\ 07-08, 07-10,\\ 07-24\end{tabular} & 08-14            & 08-27          \\ \hline
\end{tabular}
\end{table}
}

In this work, based on the microstructural properties of a given stock, we are interested in studying the predictability of the direction of mid-price changes at different time horizons when such a movement is larger than or equal to a threshold $\theta$. For the sake of readability, we will refer to these mid-price differences as `returns', stressing that we are neither referencing relative returns nor logarithmic ones. We decide to use the simple difference in mid-prices to gain higher control over the amplitude of the change at different time horizons, preserving, at the same time, the stationarity property of the resulting time series. Many alternatives have been proposed as target variables in the literature (see \cite{ntakaris2018benchmark, tsantekidis2017forecasting, zhang2019deeplob, lucchese2022short}). All of them are based on the usage of the log-return as a fundamental quantity and apply different smoothing methods to prevent a strong fit between labels and actual prices. Even if acceptable from an academic perspective, the practicability of these choices is unclear since they are designed to characterise mid-price trends (not immediate changes), leaving a reduced control over tick-by-tick changes, which are of higher interest in the development of high-frequency trading strategies. 

In this paper, we consider $3$ different horizons $\text{H}{\Delta \tau} \in \{10, 50, 100\}$ and, for each of them, the labelling step can be described as follows:

\vspace{0.1cm}
\begin{equation}\label{eq:labelling_step}
\left\{
    \begin{array}{ll}
        (m_{\tau+\Delta \tau} - m_\tau) \leq - \, \theta \; \rightarrow \; -1 \; \rightarrow \; \text{Down} \ , \\
        - \, \theta < (m_{\tau+\Delta \tau} - m_\tau) < + \, \theta \; \rightarrow \; 0 \; \rightarrow \; \text{Stable}\ , \\
        (m_{\tau+\Delta \tau} - m_\tau) \geq + \, \theta \; \rightarrow \; 1 \; \rightarrow \; \text{Up} \ ,
    \end{array}
\right.
\end{equation}
\vspace{0.1cm}

where $\theta$ is the tick size and $m_\tau$ is the mid-price at time $\tau$\,\footnote{The reader should be aware that, in the Python code related to the current research paper, the labelling schema is: $0 \rightarrow \text{Down}$, $1 \rightarrow \text{Stable}$, $2 \rightarrow \text{Up}$. In Equation \eqref{eq:labelling_step} we use labels $-1$, $0$ and $1$, respectively, to enhance the paper's readability and remain coherent with the standard used in the related literature \cite{zhang2019deeplob}.}. It is worth noting that horizons are always defined in terms of LOB updates (which are unevenly spaced), while physical time is never used.

Tables \ref{tab:average_training_classes}, \ref{tab:average_validation_classes} and \ref{tab:average_test_classes} report the stocks' average daily class distribution for the training, validation and test set, computed across the $3$-year analysis period, for $\text{H}{\Delta \tau} \in \{10, 50, 100\}$. Generally speaking, it is always possible to detect imbalances. Their evolution across horizons is, however, different for different groups (or sets) of stocks (notice that, in Tables \ref{tab:average_training_classes}, \ref{tab:average_validation_classes} and \ref{tab:average_test_classes}, groups are separated by horizontal lines). The groups' separation into so-called small-tick stocks (group $1$), medium-tick stocks (group $2$) and large-tick stocks (group $3$) will be formally described in Section \ref{sec:Microstructural_Priors} in relation to the microstructural properties displayed by the financial assets.

{
\renewcommand{\arraystretch}{1.35}
\begin{table}[H]
\centering
\caption{Stocks' average daily class distribution for the training set, computed across the $3$-year analysis period, for $\text{H}{\Delta \tau} \in \{10, 50, 100\}$.}
\label{tab:average_training_classes}
\scriptsize
\begin{tabular}{c|ccc|ccc|ccc}
\hline
\textbf{Ticker} & \multicolumn{3}{c|}{\textbf{H10}}              & \multicolumn{3}{c|}{\textbf{H50}}              & \multicolumn{3}{c}{\textbf{H100}}             \\ \cline{2-10} 
\textbf{}       & \textbf{Down} & \textbf{Stable} & \textbf{Up} & \textbf{Down} & \textbf{Stable} & \textbf{Up} & \textbf{Down} & \textbf{Stable} & \textbf{Up} \\ \hline
CHTR & 2.19e+04 & 1.93e+04 & 2.11e+04 & 2.95e+04 & 4.67e+03 & 2.81e+04 & 3.05e+04 & 2.24e+03 & 2.96e+04 \\
GOOG & 8.82e+04 & 1.92e+05 & 8.66e+04 & 1.47e+05 & 7.31e+04 & 1.47e+05 & 1.64e+05 & 3.71e+04 & 1.66e+05 \\
GS   & 3.96e+04 & 4.18e+04 & 3.96e+04 & 5.50e+04 & 1.12e+04 & 5.49e+04 & 5.77e+04 & 6.12e+03 & 5.72e+04 \\
IBM  & 4.10e+04 & 7.29e+04 & 4.13e+04 & 6.56e+04 & 2.40e+04 & 6.56e+04 & 7.04e+04 & 1.46e+04 & 7.03e+04 \\
MCD  & 3.46e+04 & 5.60e+04 & 3.50e+04 & 5.32e+04 & 1.84e+04 & 5.39e+04 & 5.69e+04 & 1.10e+04 & 5.77e+04 \\
NVDA & 1.18e+05 & 1.27e+05 & 1.18e+05 & 1.62e+05 & 3.80e+04 & 1.63e+05 & 1.69e+05 & 2.42e+04 & 1.70e+05 \\ \hline
AAPL & 2.06e+05 & 4.59e+05 & 2.06e+05 & 3.36e+05 & 1.98e+05 & 3.37e+05 & 3.67e+05 & 1.33e+05 & 3.70e+05 \\
ABBV & 4.00e+04 & 1.07e+05 & 3.98e+04 & 6.95e+04 & 4.82e+04 & 6.90e+04 & 7.77e+04 & 3.17e+04 & 7.73e+04 \\
PM   & 3.68e+04 & 9.05e+04 & 3.69e+04 & 6.37e+04 & 3.63e+04 & 6.42e+04 & 7.02e+04 & 2.30e+04 & 7.09e+04 \\ \hline
BAC  & 1.24e+04 & 4.59e+05 & 1.23e+04 & 4.32e+04 & 3.98e+05 & 4.30e+04 & 6.91e+04 & 3.46e+05 & 6.87e+04 \\
CSCO & 2.36e+04 & 4.51e+05 & 2.39e+04 & 7.32e+04 & 3.52e+05 & 7.33e+04 & 1.12e+05 & 2.75e+05 & 1.12e+05 \\
KO   & 1.44e+04 & 2.14e+05 & 1.44e+04 & 4.17e+04 & 1.59e+05 & 4.15e+04 & 6.03e+04 & 1.22e+05 & 6.00e+04 \\
ORCL & 2.63e+04 & 3.15e+05 & 2.62e+04 & 6.93e+04 & 2.29e+05 & 6.93e+04 & 9.75e+04 & 1.73e+05 & 9.75e+04 \\
PFE  & 1.85e+04 & 2.97e+05 & 1.85e+04 & 5.25e+04 & 2.29e+05 & 5.25e+04 & 7.65e+04 & 1.82e+05 & 7.62e+04 \\
VZ   & 2.45e+04 & 2.62e+05 & 2.44e+04 & 6.52e+04 & 1.81e+05 & 6.49e+04 & 8.97e+04 & 1.33e+05 & 8.91e+04 \\ \hline
\end{tabular}
\end{table}
}

{
\renewcommand{\arraystretch}{1.35}
\begin{table}[H]
\centering
\caption{Stocks' average daily class distribution for the validation set, computed across the $3$-year analysis period, for $\text{H}{\Delta \tau} \in \{10, 50, 100\}$.}
\label{tab:average_validation_classes}
\scriptsize
\begin{tabular}{c|ccc|ccc|ccc}
\hline
\textbf{Ticker} & \multicolumn{3}{c|}{\textbf{H10}}              & \multicolumn{3}{c|}{\textbf{H50}}              & \multicolumn{3}{c}{\textbf{H100}}              \\ \cline{2-10} 
\textbf{}       & \textbf{Down} & \textbf{Stable} & \textbf{Up} & \textbf{Down} & \textbf{Stable} & \textbf{Up} & \textbf{Down} & \textbf{Stable} & \textbf{Up} \\ \hline
CHTR & 1.90e+04 & 1.82e+04 & 1.87e+04 & 2.66e+04 & 3.95e+03 & 2.54e+04 & 2.76e+04 & 1.71e+03 & 2.66e+04 \\
GOOG & 6.73e+04 & 1.56e+05 & 6.70e+04 & 1.15e+05 & 6.13e+04 & 1.15e+05 & 1.30e+05 & 3.16e+04 & 1.29e+05 \\
GS   & 4.10e+04 & 4.77e+04 & 4.10e+04 & 5.83e+04 & 1.38e+04 & 5.76e+04 & 6.12e+04 & 7.58e+03 & 6.10e+04 \\
IBM  & 3.59e+04 & 6.92e+04 & 3.58e+04 & 5.86e+04 & 2.34e+04 & 5.90e+04 & 6.31e+04 & 1.43e+04 & 6.35e+04 \\
MCD  & 3.19e+04 & 5.73e+04 & 3.24e+04 & 5.03e+04 & 1.92e+04 & 5.21e+04 & 5.41e+04 & 1.12e+04 & 5.63e+04 \\
NVDA & 1.33e+05 & 1.37e+05 & 1.32e+05 & 1.80e+05 & 4.29e+04 & 1.79e+05 & 1.87e+05 & 2.77e+04 & 1.87e+05 \\ \hline
AAPL & 1.79e+05 & 4.63e+05 & 1.79e+05 & 3.11e+05 & 2.00e+05 & 3.11e+05 & 3.42e+05 & 1.36e+05 & 3.43e+05 \\
ABBV & 4.80e+04 & 1.31e+05 & 4.77e+04 & 8.31e+04 & 6.01e+04 & 8.34e+04 & 9.29e+04 & 3.99e+04 & 9.38e+04 \\
PM   & 3.60e+04 & 9.34e+04 & 3.59e+04 & 6.38e+04 & 3.80e+04 & 6.36e+04 & 7.06e+04 & 2.41e+04 & 7.06e+04 \\ \hline
BAC  & 1.17e+04 & 4.38e+05 & 1.17e+04 & 4.06e+04 & 3.80e+05 & 4.07e+04 & 6.50e+04 & 3.31e+05 & 6.50e+04 \\
CSCO & 1.90e+04 & 3.98e+05 & 1.85e+04 & 5.88e+04 & 3.21e+05 & 5.57e+04 & 9.09e+04 & 2.57e+05 & 8.72e+04 \\
KO   & 1.16e+04 & 1.97e+05 & 1.14e+04 & 3.45e+04 & 1.52e+05 & 3.35e+04 & 5.10e+04 & 1.19e+05 & 4.98e+04 \\
ORCL & 1.95e+04 & 2.69e+05 & 1.92e+04 & 5.25e+04 & 2.03e+05 & 5.26e+04 & 7.44e+04 & 1.58e+05 & 7.56e+04 \\
PFE  & 1.49e+04 & 2.68e+05 & 1.50e+04 & 4.31e+04 & 2.12e+05 & 4.30e+04 & 6.40e+04 & 1.70e+05 & 6.39e+04 \\
VZ   & 2.04e+04 & 2.42e+05 & 2.04e+04 & 5.62e+04 & 1.69e+05 & 5.70e+04 & 7.95e+04 & 1.22e+05 & 8.09e+04 \\ \hline
\end{tabular}
\end{table}
}

{
\renewcommand{\arraystretch}{1.35}
\begin{table}[h!]
\scriptsize
\centering
\caption{Stocks' average daily class distribution for the test set, computed across the $3$-year analysis period, for $\text{H}{\Delta \tau} \in \{10, 50, 100\}$.}
\label{tab:average_test_classes}
\begin{tabular}{c|ccc|ccc|ccc}
\hline
\textbf{Ticker} & \multicolumn{3}{c|}{\textbf{H10}}              & \multicolumn{3}{c|}{\textbf{H50}}              & \multicolumn{3}{c}{\textbf{H100}}              \\ \cline{2-10} 
\textbf{}       & \textbf{Down} & \textbf{Stable} & \textbf{Up} & \textbf{Down} & \textbf{Stable} & \textbf{Up} & \textbf{Down} & \textbf{Stable} & \textbf{Up} \\ \hline
CHTR & 3.47e+04 & 4.49e+04 & 3.20e+04 & 4.87e+04 & 1.73e+04 & 4.57e+04 & 5.22e+04 & 9.96e+03 & 4.94e+04 \\
GOOG & 1.76e+05 & 3.02e+05 & 1.63e+05 & 2.72e+05 & 1.05e+05 & 2.63e+05 & 2.95e+05 & 5.42e+04 & 2.91e+05 \\
GS   & 4.54e+04 & 5.19e+04 & 4.53e+04 & 6.45e+04 & 1.33e+04 & 6.48e+04 & 6.79e+04 & 6.62e+03 & 6.81e+04 \\
IBM  & 5.82e+04 & 7.68e+04 & 5.90e+04 & 8.51e+04 & 2.42e+04 & 8.47e+04 & 9.01e+04 & 1.57e+04 & 8.82e+04 \\
MCD  & 5.12e+04 & 5.91e+04 & 5.14e+04 & 7.20e+04 & 1.79e+04 & 7.18e+04 & 7.57e+04 & 1.08e+04 & 7.52e+04 \\
NVDA & 1.09e+05 & 7.34e+04 & 1.10e+05 & 1.37e+05 & 1.78e+04 & 1.37e+05 & 1.41e+05 & 1.12e+04 & 1.40e+05 \\ \hline
AAPL & 2.71e+05 & 3.50e+05 & 2.70e+05 & 3.75e+05 & 1.39e+05 & 3.76e+05 & 3.98e+05 & 9.35e+04 & 4.00e+05 \\
ABBV & 4.45e+04 & 8.43e+04 & 4.45e+04 & 7.09e+04 & 3.16e+04 & 7.08e+04 & 7.71e+04 & 1.99e+04 & 7.63e+04 \\
PM   & 4.57e+04 & 8.92e+04 & 4.66e+04 & 7.62e+04 & 3.01e+04 & 7.53e+04 & 8.25e+04 & 1.85e+04 & 8.05e+04 \\ \hline
BAC  & 1.88e+04 & 5.93e+05 & 1.88e+04 & 6.37e+04 & 5.03e+05 & 6.40e+04 & 1.02e+05 & 4.26e+05 & 1.02e+05 \\
CSCO & 4.82e+04 & 6.26e+05 & 4.84e+04 & 1.47e+05 & 4.29e+05 & 1.47e+05 & 2.10e+05 & 3.06e+05 & 2.07e+05 \\
KO   & 2.54e+04 & 2.87e+05 & 2.55e+04 & 7.18e+04 & 1.94e+05 & 7.27e+04 & 9.93e+04 & 1.39e+05 & 1.00e+05 \\
ORCL & 4.22e+04 & 4.43e+05 & 4.18e+04 & 1.18e+05 & 2.92e+05 & 1.18e+05 & 1.65e+05 & 1.97e+05 & 1.65e+05 \\
PFE  & 2.78e+04 & 3.51e+05 & 2.78e+04 & 7.64e+04 & 2.53e+05 & 7.64e+04 & 1.07e+05 & 1.92e+05 & 1.07e+05 \\
VZ   & 4.59e+04 & 3.42e+05 & 4.63e+04 & 1.13e+05 & 2.09e+05 & 1.13e+05 & 1.43e+05 & 1.47e+05 & 1.43e+05 \\ \hline
\end{tabular}
\end{table}
}

The first set (group $1$, small-tick stocks) has cardinality equal to $6$ and is made of CHTR, GOOG, GS, IBM, MCD and NVDA. At $\text{H}10$, the order of magnitude for the daily average number of samples for each label remains constant for the training, validation and test set, with only minor oscillations. At $\text{H}50$ and $\text{H}100$, the order of magnitude of representatives for classes `Up' and `Down' gradually increases, highlighting a more pronounced imbalance towards the two `active' classes. This pattern can be detected in the training, validation and test set. The second group of stocks (group $2$, medium-tick stocks) has cardinality equal to $3$ and is made of AAPL, ABBV and PM. In this case, the order of magnitude of labels' representatives remains stable across horizons and for the training, validation and test set. Lastly, the third group of stocks (group $3$, large-tick stocks) has a cardinality equal to $6$ and is made of BAC, CSCO, KO, ORCL, PFE and VZ. The order of magnitude of representatives for the `Stable' class is higher than the one for classes `Down' and `Up' at $\text{H}10$, while a stability is gradually matured moving to horizons $\text{H}{\Delta \tau} \in \{50, 100\}$. We highlight that this behaviour is symmetrical with respect to the one detected for group 1. As already underlined for the other two groups of assets, also in this case, the described pattern remains constant for the training, validation and test set. 

\section{Methods}\label{sec:Methods}
From a practical perspective, this paper aims to provide a straightforward way to estimate a given stock's predictability based on its LOB microstructural properties. This aim can be achieved by splitting the research process into two steps: (i) extract and classify the microstructural properties of a heterogeneous set of stocks (see Section \ref{sec:Microstructural_Priors}); (ii) accomplish the forecasting task on each of them and review the obtained results in relation to the outcomes of the previous step. 

\begin{figure}[H]
    \centering
    \includegraphics[scale=0.59]{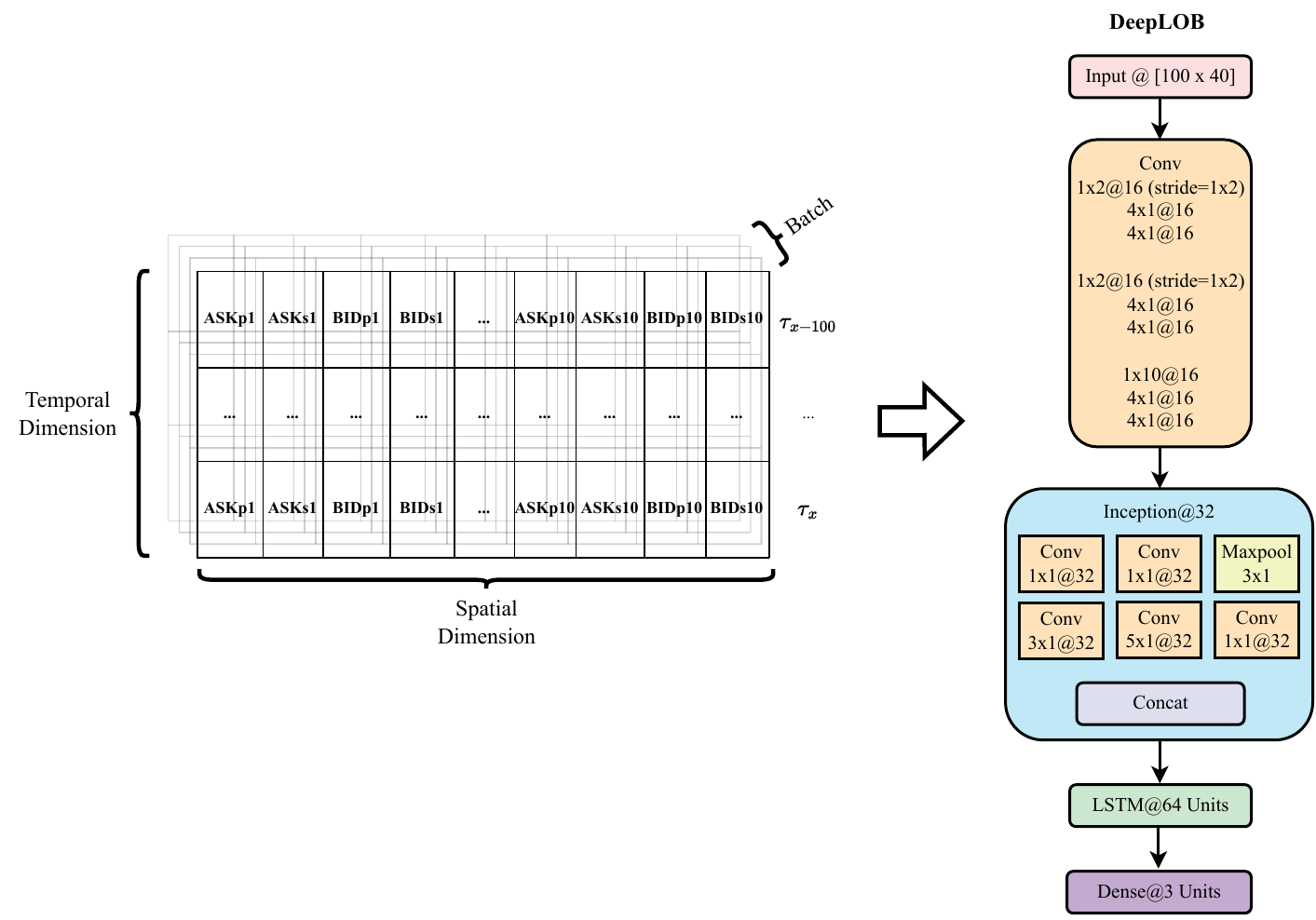}
    \caption{Pictorial representation of an input batch for the DeepLOB model (left-hand side), and of the architecture itself (right-hand side).}
    \label{fig:deepLOB_schema}
\end{figure}

To perform the forecasting task, we release `LOBFrame', a novel, open-source code base which presents a renewed way to process large-scale LOB data. This framework integrates all the latest cutting-edge insights from related scientific research into a cohesive system. Its strength lies in the comprehensive nature of the implemented pipeline, which includes the data transformation and processing stage, an ultra-fast implementation of the training, validation, and testing steps, as well as the evaluation of the quality of a model's outputs through trading simulations\,\footnote{We remark that results of trading simulations are not part of this work because of the reasons reported in Section \ref{sec:Practicability_Forecasts}.}. Moreover, it offers flexibility by accommodating the integration of new models, ensuring adaptability to future advancements in the field. Results discussed in the current paper come from the usage of a state-of-the-art model in literature: DeepLOB \cite{zhang2019deeplob}\,\footnote{We stress that theoretical findings discussed in Section \ref{sec:Results} are independent of the choice of the model. Indeed, in this work, we decide to use DeepLOB only because it fully respects the following three criteria: (i) availability of the original code used in the experiments; (ii) data-driven design of the model's architecture; (iii) community's recognition as a state-of-the-art model in the field.}. This architecture mainly relies on two well-known deep learning modules: it exploits (i) the power of convolutional neural networks (CNNs) to model inter-levels, spatial LOB's dynamics \cite{lecun1998gradient, o2015introduction, albawi2017understanding}; and (ii) the memory of the LSTM  module to handle the temporal dimension of the input \cite{hochreiter1997long, van2020review}. For a detailed overview of the architecture, the reader is referred to the original work Zhang et al. \cite{zhang2019deeplob}, while, in this Section, our efforts are towards providing the intuition behind the model. The main idea of using CNNs is to automate the feature extraction process in a notoriously noisy and with a low signal-to-noise ratio context \cite{briola2021deep} such as the one provided in LOB, without any strong initial assumption. Indeed, weights are learned during inference, and derived features -- learned from the training set -- are data-adaptive. The LSTM layer, on the other side, is used to capture residual time dependencies among the resulting features. It is worth underlining that very short-time dependencies are already captured in the convolutional layer, which takes LOB's snapshots as inputs (see Figure \ref{fig:deepLOB_schema}). To train, validate and test the DeepLOB model, we design a high-performance data loader, which samples mini-batches of size $32$ (as per in the original model's implementation), each made of inputs with size $100 \times 40$. Dimension $100$ (i.e., the temporal dimension) represents the history length and corresponds to the number of historically consecutive LOB updates constituting each sample. Dimension $40$, instead, is the number of spatial constituents for each LOB's snapshot (see Equation \eqref{eq:Compressed_Representation_LOB}). The sampling process differs for the training, validation, and test sets. During training, the (sub)-sampling is random and balanced. From each trading day, we detect the number of samples for the less represented class and (i) if this value is $\geq 5000$, then we sample $5000$ random representatives (a representative is a $100 \times 40$ input) for each of the three classes (see Equation \ref{eq:labelling_step}), otherwise, (ii) if this value is $< 5000$, then we sample a number of random representatives for each class which is equal to the number of samples for the less represented class. 

During validation and test stages, we still sample batches with a size of $32$, but they are always sequential and cover the totality of data in the two sets. In line with the related literature \cite{zhang2019deeplob}, the model is always trained for a maximum number of epochs equal to $100$, with patience equal to $15$ epochs. We use a modified version of the Adam optimizer \cite{kingma2014adam} with decoupled weight decay \cite{loshchilov2017decoupled}, commonly known as `AdamW'. Following the latest applied research findings \cite{brown2020language, nanoGPT}, we use a learning rate equal to $6 \times 10^{-5}$, a $\beta_1$ decay rate equal to $0.9$ and a $\beta_2$ decay rate equal to $0.95$. The choices of values for these parameters are determined by the training pipeline described above, which is different from the one proposed in the original work \cite{zhang2019deeplob} and relies on a smaller number of training samples to reduce the model's exposition to the noise characterizing the LOB. The entire framework described in this paper is coded in Python using the PyTorch deep learning library \cite{paszke2019pytorch}. A total number of $135$ experiments have been run on the University College London Computer Science Department's High-Performance Computing Cluster \cite{UCL_HPC_Cluster} for a cumulative GPU runtime of $959$ hours, $16$ minutes and $27$ seconds. Six different types of GPUs have been used: (i) NVIDIA GeForce GTX 1080 Ti; (ii) NVIDIA GeForce RTX 2080 Ti; (iii) NVIDIA TITAN X (Pascal); (iv) NVIDIA TITAN Xp; (v) Tesla V100-PCIE-16GB; and (vi) Tesla V100-PCIE-32GB.

\section{Microstructural Priors}\label{sec:Microstructural_Priors}
As a first microstructural property, we study the relationship between the stocks' average spread $\langle\sigma\rangle$ and the tick size $\theta$ across the $3$-year analysis period. In literature, a stock is differently classified based on a general definition which establishes that if $\langle \sigma \rangle \gg \theta$, than the asset is a small-tick stock, if $\langle \sigma \rangle \simeq \theta$, than it is a large-tick stock \cite{bouchaud2018trades}. Even if widespread, this definition is not quantitative and possibly too restrictive to characterise the more nuanced behaviour of stocks traded in the NASDAQ exchange.

{
\renewcommand{\arraystretch}{1.35}
\begin{table}[h!]
\centering
\caption{The $15$ small-, medium- and large-tick stocks that we include in our analysis, along with their mean price and mean bid-ask spread during $2017$, $2018$ and $2019$.}
\label{tab:stock_size_general}
\scriptsize
\begin{tabular}{c|cc|cc|cc|c}
\hline
\textbf{Ticker} &
  \multicolumn{2}{c|}{\textbf{2017}} &
  \multicolumn{2}{c|}{\textbf{2018}} &
  \multicolumn{2}{c|}{\textbf{2019}} &
  \textbf{Size} \\ \cline{2-7}
 &
  \textbf{\begin{tabular}[c]{@{}c@{}}mean \\ price \\ {[}\${]}\end{tabular}} &
  \textbf{\begin{tabular}[c]{@{}c@{}}mean \\ spread \\ {[}\${]}\end{tabular}} &
  \textbf{\begin{tabular}[c]{@{}c@{}}mean \\ price\\ {[}\${]}\end{tabular}} &
  \textbf{\begin{tabular}[c]{@{}c@{}}mean \\ spread\\ {[}\${]}\end{tabular}} &
  \textbf{\begin{tabular}[c]{@{}c@{}}mean \\ price\\ {[}\${]}\end{tabular}} &
  \textbf{\begin{tabular}[c]{@{}c@{}}mean \\ spread\\ {[}\${]}\end{tabular}} &
   \\ \hline
CHTR & 343.65 & 0.2869 & 312.61  & 0.3475 & 394.76  & 0.2206 & small  \\
GOOG & 934.44 & 0.4362 & 1099.33 & 0.7898 & 1186.57 & 0.5511 & small  \\
GS   & 232.82 & 0.0965 & 223.35  & 0.1111 & 204.19  & 0.0759 & small  \\
IBM  & 157.90 & 0.0362 & 140.23  & 0.0444 & 137.94  & 0.0316 & small  \\
MCD  & 146.71 & 0.0321 & 166.39  & 0.0542 & 198.29  & 0.0531 & small  \\
NVDA & 144.12 & 0.0437 & 233.82  & 0.0844 & 172.13  & 0.0500 & small  \\ \hline
AAPL & 151.97 & 0.0145 & 190.11  & 0.0223 & 208.62  & 0.0190 & medium \\
ABBV & 71.59  & 0.0211 & 94.98   & 0.0422 & 76.86   & 0.0212 & medium \\
PM   & 110.78 & 0.0231 & 86.96   & 0.0293 & 81.89   & 0.0240 & medium \\ \hline
BAC  & 24.69  & 0.0109 & 29.31   & 0.0109 & 29.40   & 0.0105 & large  \\
CSCO & 33.20  & 0.0106 & 44.25   & 0.0110 & 51.31   & 0.0107 & large  \\
KO   & 44.11  & 0.0112 & 45.84   & 0.0116 & 51.00   & 0.0111 & large  \\
ORCL & 46.51  & 0.0115 & 47.89   & 0.0117 & 54.05   & 0.0111 & large  \\
PFE  & 33.93  & 0.0111 & 39.85   & 0.0114 & 39.99   & 0.0109 & large  \\
VZ   & 48.30  & 0.0119 & 52.80   & 0.0121 & 57.92   & 0.0112 & large  \\ \hline
\end{tabular}
\end{table}
}

In this paper, we provide a practical classification which establishes that if (i) $\langle \sigma \rangle \gtrsim 3\theta$, we are dealing with a small-tick stock; if (ii) $\langle \sigma \rangle \lesssim 1.5\theta$, we are dealing with a large-tick stock; if (iii) $1.5\theta \lesssim \langle \sigma \rangle \lesssim 3\theta$, we are dealing with a medium-tick stock. In this way, we impose quantitative boundaries for stock classification that allow the introduction of an extra family (i.e. medium-tick stocks) that groups `borderline` assets. This category was previously identified in \cite{bonart2017optimal,bouchaud2018trades}. Considering that we analyse $3$ years of data, one of the previously mentioned conditions should remain valid for at least $2$ of the $3$ considered years. Looking at Table \ref{tab:stock_size_general}, we have $6$ representatives of small-tick stocks (i.e., CHTR, GOOG, GS, IBM, MCD, NVDA), $3$ representatives of medium-tick stocks (i.e., AAPL, ABBV, PM) and $6$ representatives of large-tick stocks (i.e., BAC, CSCO, KO, ORCL, PFE, VZ). It is evident that, for small-tick stocks, the yearly average spread is subject to non-negligible fluctuations, while medium- and large-tick stocks are more stable across years. As we will point out several times in this paper, specific properties of small-, medium- and large-tick stocks highly impact their predictability.

\begin{figure}[H]
    \centering
    \includegraphics[scale=0.3]{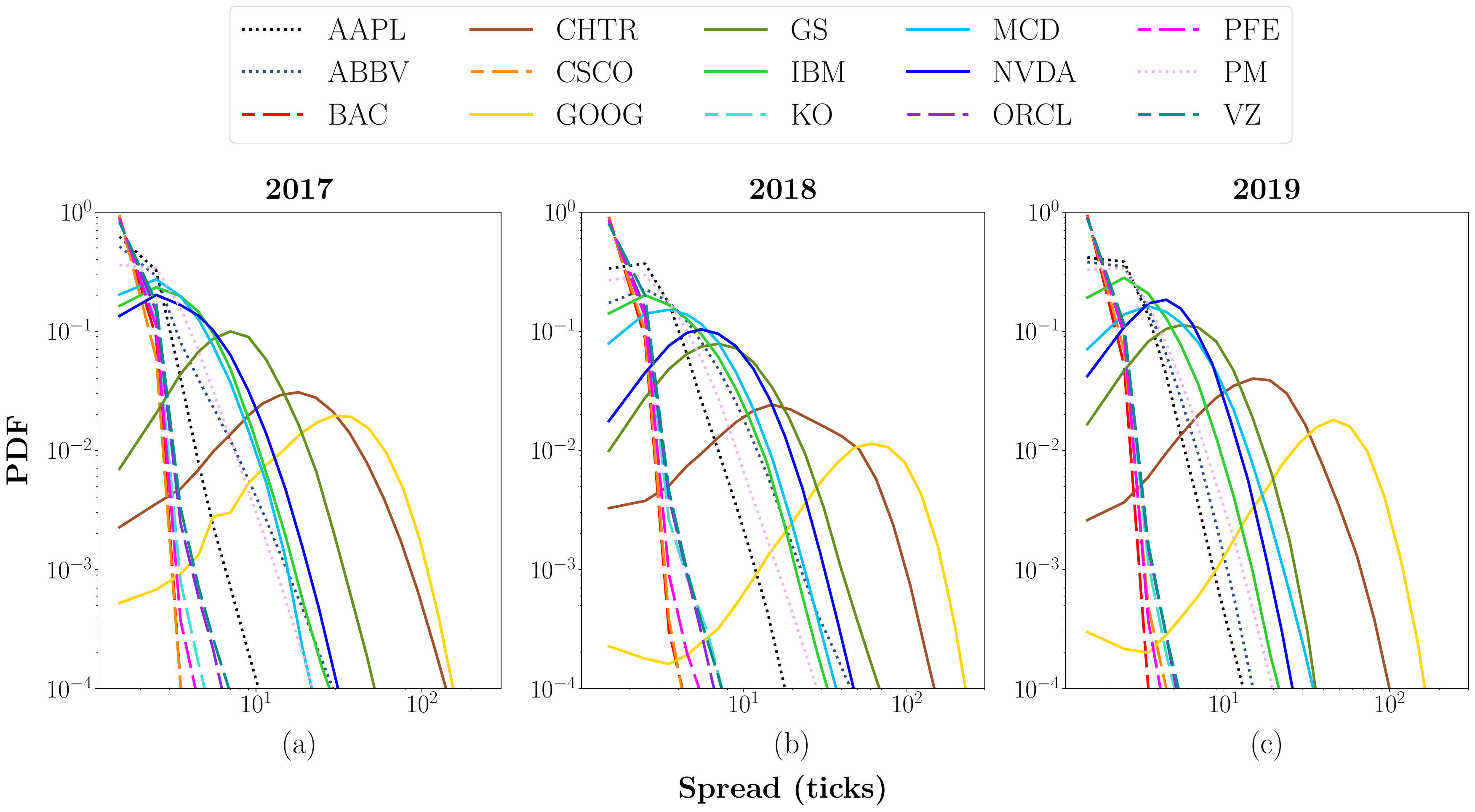}
    \caption{PDF of the spread (expressed in number of ticks) for the $15$ stocks of interest, in the $3$-year analysis period.}
    \label{fig:pdf_spreads_in_ticks_partial}
\end{figure}

Extending the previous analysis, in Figure \ref{fig:pdf_spreads_in_ticks_partial}, we report the PDF of the spread (expressed in number of ticks) for each considered stock. As one can notice, distributions are different for small-, medium- and large-tick stocks, defining evident behavioural clusters. For large-tick stocks, distributions are peaked at an average value of $1.5$ ticks (extremely close to the minimum spread allowed of $1$ tick), with rare openings to larger realizations. This finding is consistent across the $3$ years. It is worth noting that, from a practical perspective, tighter spreads are beneficial for traders looking for stocks allowing to enter and exit positions quickly and with minimal impact on the transaction costs. For medium-tick stocks, distributions are peaked slightly over the minimum spread: during $2017$, the average peak value is equal to $1.50$ ticks; during $2018$, the average peak value is equal to $2.50$ ticks; while, during $2019$, the average peak value is equal to $1.83$ ticks. Notably, these distributions express more significant variations than those describing large-tick stocks. Among these assets, AAPL is characterized by a distribution with a shape more similar to that of large-tick stocks, while ABBV and PM show a behaviour more similar to that of small-tick stocks. This result is expected since, by definition, medium-tick stocks are `borderline' assets characterized by behavioural patterns that do not clearly belong to the class of small- nor large-tick stocks. Lastly, small-tick stocks show consistently broader distributions. In this family, we distinguish two different subsets of assets: the first one is made of CHTR, GOOG, and GS, while the second one is made of IBM, MCD, and NVDA. Distributions characterizing the first subset have an average peak of $18.16$ ticks in $2017$, $27.50$ ticks in $2018$ and $22.60$ ticks in $2019$. Distributions characterizing the second subset, instead, have an average peak equal to $2.50$ ticks in $2017$, $3.83$ ticks in $2018$ and $3.50$ ticks in $2019$. In both cases, small-tick stocks express more significant variances than large-tick stocks, suggesting less frequent trading activity or larger orders that could move the market \cite{bouchaud2018trades} and, consequently, an higher exposition to market impact for actors placing trades. It is worth noting that, for each class of stocks, the shape of the spread's distribution remains consistent over the three years, suggesting an overall macro-stability.

\begin{figure}[h!]
    \centering
    \includegraphics[scale=0.3]{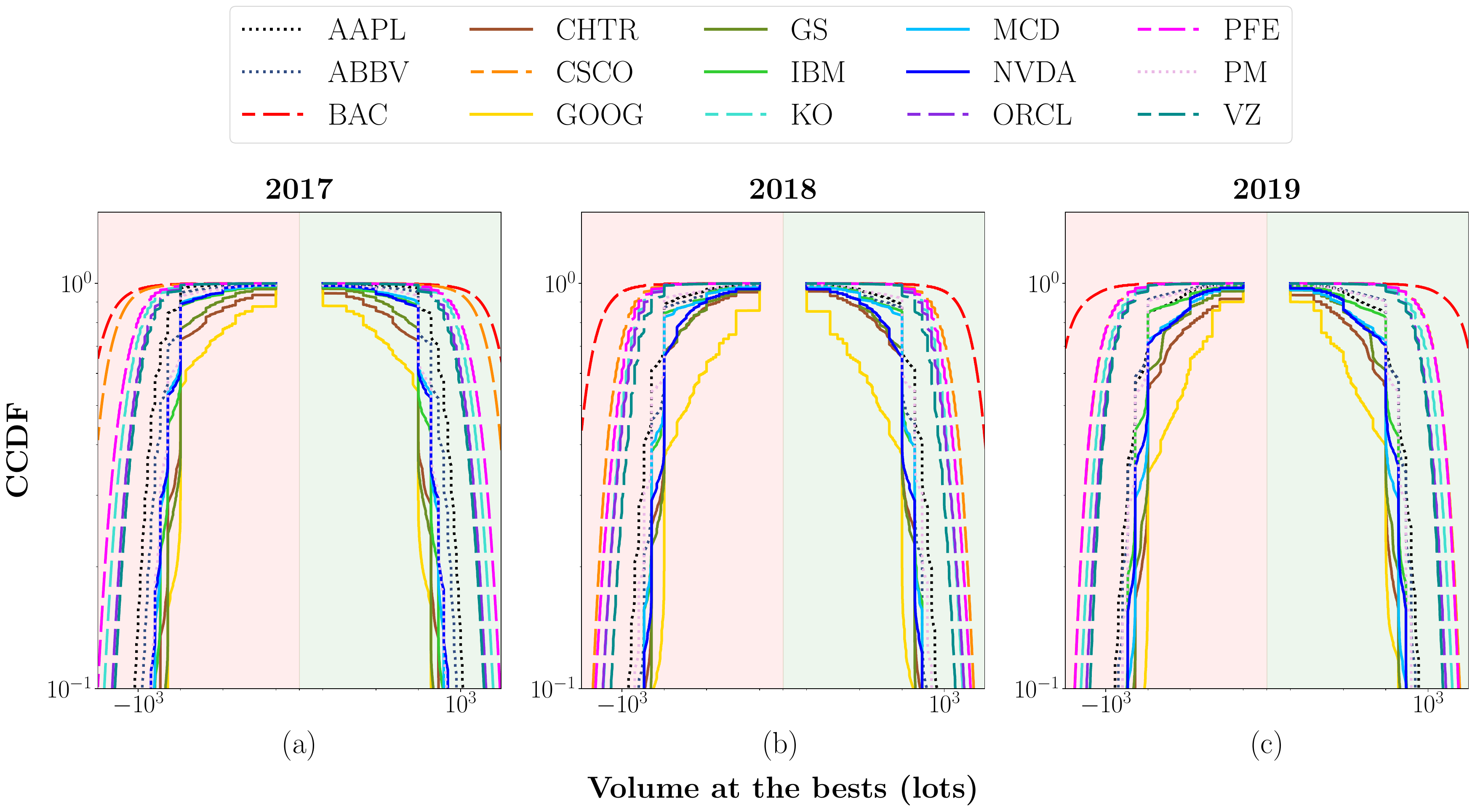}
    \caption{CCDF of the volumes available at the best quotes for the $15$ stocks of interest, in the $3$-year analysis period.}
    \label{fig:ccdf_best_volume_in_ticks_partial}
\end{figure}

A second microstructural aspect to investigate concerns the liquidity at the best levels. Indeed, once the impact of stocks' tick size on the potential costs related to fast trading is clarified, it is relevant to study the CCDF of volumes available at the best quotes to understand if there is the necessary liquidity to perform such an activity. Figure \ref{fig:ccdf_best_volume_in_ticks_partial} reports the results of this analysis. The $x$-axis utilizes a symmetric log-scale to study both the ask side (negative part, red area) and the bid side (positive part, green area) of the LOB, while underlying the broadness of distributions. As one can notice, distributions are roughly symmetric for the two sides; also, in this case, a behavioural clustering directly dependent on the tick size of the stocks is evident. Distributions characterizing large-tick stocks are significantly wider, highlighting a condition of higher liquidity at best quotes. Even if not visible in Figure \ref{fig:ccdf_best_volume_in_ticks_partial}, as explained by Bouchaud et al. \cite{bouchaud2018trades}, it is relevant to underline that in the case of large-tick stocks, the volume of the queues decreases before transactions since liquidity takers rush to take the remaining volumes before it disappears. This phenomenon provides more information on the direction of future price changes, potentially contributing to an improved forecast accuracy of deep learning models. One more time, it is possible to highlight the `borderline' behaviour of medium-tick stocks. They exhibit distributions that fall in the middle between the ones characterizing large- and small-tick stocks. Lastly, the curves characterizing small-tick stocks are the steepest, highlighting an overall condition of lower liquidity and potentially higher volatility. Also in this case, even if not immediately visible from Figure \ref{fig:ccdf_best_volume_in_ticks_partial}, as explained by Bouchaud et al. \cite{bouchaud2018trades}, it is relevant to underline that the volume at the best quote increases immediately before being hit by a market order, indicating that liquidity takers choose to submit their orders when the opposite volume is relatively high. 

\begin{figure}[h!]
    \centering
    \includegraphics[scale=0.3]{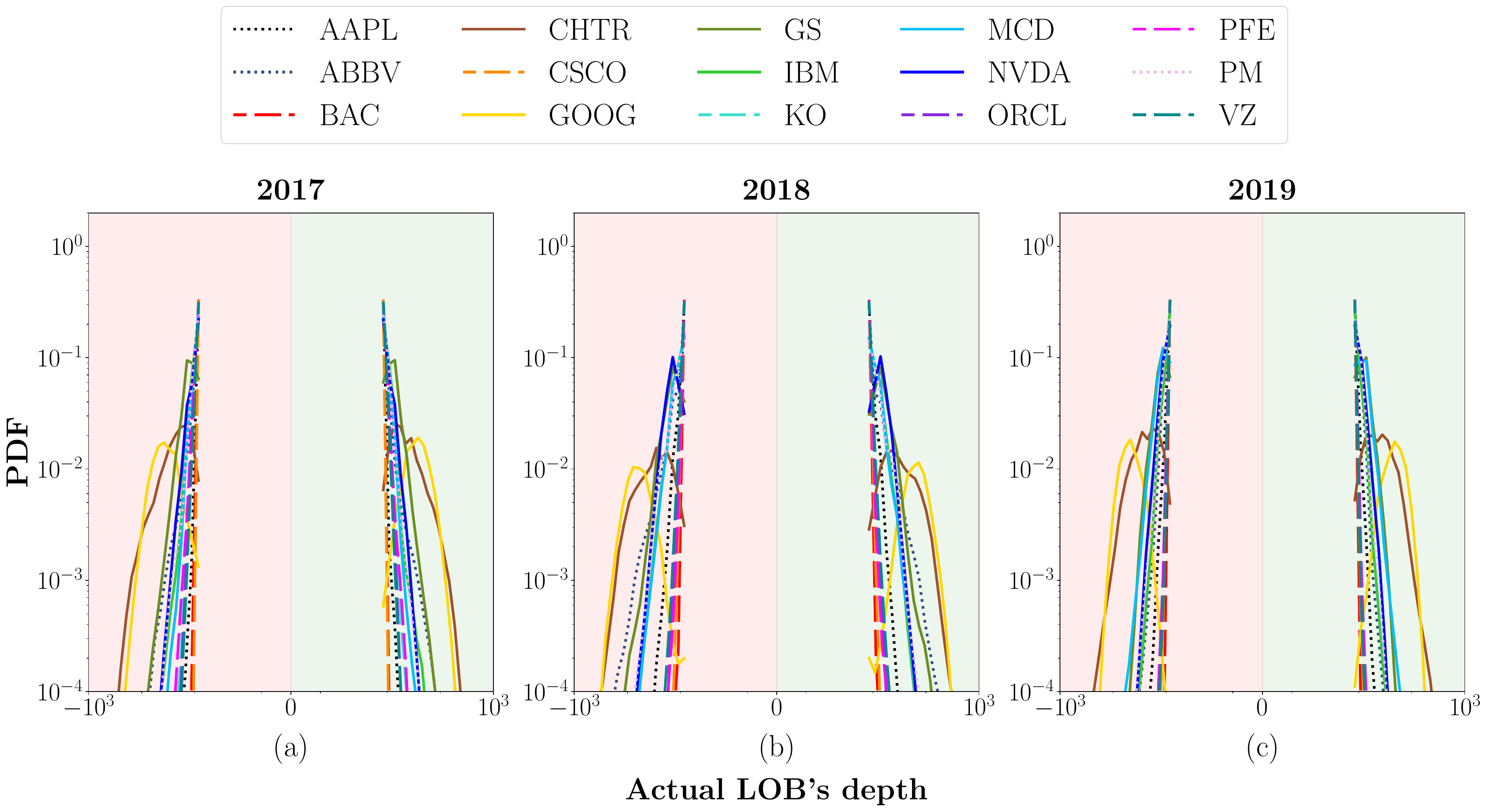}
    \caption{PDF of the `actual LOB depth' ($ \Xi$) for the $15$ stocks of interest, in the $3$-year analysis period.}
    \label{fig:pdf_relative_number_levels_partial}
\end{figure}

As pointed out in the work by Wu et al. \cite{wu2021towards}, the LOB representation adopted in the current paper (see Equation \ref{eq:Compressed_Representation_LOB}), which is commonly referred to as `compressed representation' \cite{wu2021towards}, presents a major drawback: its spatial structure is not homogeneous (see Figure \ref{fig:lob_schema}) since there is no assumption for adjacent price levels to have fixed intervals, while only a monotonic order is guaranteed \cite{wu2021towards}. This representation is prone to dramatic changes due to occasional price-level shifts, significantly impacting predictability when treated as input for deep learning models. Indeed, in \cite{wu2021towards}, the authors underline that one of the main assumptions in deep learning is that signals from the same channel (or input dimension) are from the same source. In our case, a `level' is an artefact strictly related to a single snapshot of the LOB and it is not associated with a constant source, especially when its information shifts due to aggressive orders. To measure stocks' exposure to this issue, we compute a metric defined as the `actual LOB depth' ($\Xi$). Given a snapshot $\mathbb{L}(\tau)$, this measure is computed for the two sides of the market as follows:

\begin{align*}
 \Xi^{\rm Ask}_\tau = \frac{p^{\text{ask}}_{10}(\tau) - p^{\text{ask}}_{1}(\tau)}{0.01} \quad  {\rm ask ~ side}\ , \\  \Xi^{\rm Bid}_\tau =\frac{p^{\text{bid}}_{1}(\tau) - p^{\text{bid}}_{10}(\tau)}{0.01} \quad  {\rm bid ~  side}\ .
\end{align*}

In Figure \ref{fig:pdf_relative_number_levels_partial}, we report the PDF for $\Xi^{\rm Bid}$ and $\Xi^{\rm Ask}$ for each stock, across the $3$-year period of analysis.  Even if less evident than in previous analyses, detecting separate clusters of stocks displaying a different behaviour is still possible. One more time, distributions are roughly symmetric for the two sides of the LOB. The likelihood of having a homogeneous spatial structure across different levels is higher for large-tick stocks. In this case, distributions have an average peak equal to $9.50$ price levels (slightly more than the minimum allowed distance between the two extreme levels of the LOB) for both the ask and bid side across all the $3$-year analysis period. The same behaviour is detected for medium-tick stocks even if distributions are slightly wider, especially for PM and ABBV, suggesting a higher likelihood of extreme events. Lastly, when analyzing small-tick stocks, it is useful to divide them into two separate subsets as we did above: the first set is made of CHTR, GOOG and GS, while the second is made of IBM, MCD and NVDA. Distributions characterizing the first subset have an average peak equal to $23.10$ and $23.00$ price levels for the bid and ask side, respectively, in $2017$, $37.50$ and $40.83$ in $2018$, and $31.00$ and $26.66$ in $2019$. Distributions characterizing the second subset have an average peak equal to $9.50$ for both bid and ask sides in $2017$, $11.33$ for both sides in $2018$, and $11.33$ and $10.50$ in $2019$. In this case, distributions are much wider than those characterizing large-tick stocks, underlying a higher likelihood of heterogeneous spatial structure across different levels of the LOB. This means that the corresponding stocks are characterized by a sparse LOB structure with empty levels, potentially inflating the inner representation of deep learning models.

In addition to these foundational microstructural properties, many derived ones have been recently introduced. Despite the goal of all of them is digging into a specific (sub)-aspect of the LOB microstructural structure, it is easily demonstrable that most of them can be directly mapped to one or more of the fundamental quantities introduced earlier in this Section. An example is the `information richness' (IR) score \cite{kolm2023deep}. In the original paper, the authors claim it is a measure of stocks' predictability; this is only partially true. As we empirically show in \ref{sec:Appendix_A}, there is a direct mapping between the IR score of a stock and its tick size; consequently, the tick size itself could be used as a proxy measure of a stock's predictability. 

{
\renewcommand{\arraystretch}{1.35}
\begin{table}[h!]
\centering
\caption{Average probability (computed across the $3$-year analysis period) that the number of updates characterising the three horizons $\text{H}{\Delta \tau} \in \{10, 50, 100\}$, happens in a physical time (i) $< 1$ second; (ii) $\geq 1$ second and $< 10$ seconds; or (iii) $\geq 10$ seconds.}
\scriptsize
\label{tab:probability_number_seconds_horizons}
\begin{tabular}{c|ccc|ccc|ccc}
\hline
\textbf{Ticker} &
  \multicolumn{3}{c|}{\textbf{\textless 1s}} &
  \multicolumn{3}{c|}{\textbf{\textgreater{}=1s  \& \textless{}10s}} &
  \multicolumn{3}{c}{\textbf{\textgreater{}= 10s}} \\ \cline{2-10} 
 &
  \textbf{H10} &
  \textbf{H50} &
  \textbf{H100} &
  \textbf{H10} &
  \textbf{H50} &
  \textbf{H100} &
  \textbf{H10} &
  \textbf{H50} &
  \textbf{H100} \\ \hline
CHTR &
  {\color[HTML]{32CB00} 0.46} &
  0.06 &
  0.01 &
  0.41 &
  0.36 &
  0.18 &
  0.13 &
  {\color[HTML]{FE0000} 0.58} &
  {\color[HTML]{3166FF} 0.81} \\
GOOG &
  {\color[HTML]{32CB00} 0.76} &
  0.32 &
  0.14 &
  0.18 &
  {\color[HTML]{FE0000} 0.53} &
  {\color[HTML]{3166FF} 0.56} &
  0.06 &
  0.15 &
  0.30 \\
GS &
  {\color[HTML]{32CB00} 0.51} &
  0.05 &
  0.01 &
  0.39 &
  {\color[HTML]{FE0000} 0.56} &
  0.29 &
  0.10 &
  0.39 &
  {\color[HTML]{3166FF} 0.70} \\
IBM &
  {\color[HTML]{32CB00} 0.57} &
  0.07 &
  0.01 &
  0.35 &
  {\color[HTML]{FE0000} 0.65} &
  0.40 &
  0.08 &
  0.28 &
  {\color[HTML]{3166FF} 0.59} \\
MCD &
  0.00 &
  0.00 &
  0.00 &
  0.00 &
  0.00 &
  0.00 &
  {\color[HTML]{32CB00} 1.00} &
  {\color[HTML]{FE0000} 1.00} &
  {\color[HTML]{3166FF} 1.00} \\
NVDA &
  {\color[HTML]{32CB00} 0.72} &
  0.24 &
  0.06 &
  0.18 &
  {\color[HTML]{FE0000} 0.68} &
  {\color[HTML]{3166FF} 0.73} &
  0.10 &
  0.08 &
  0.21 \\ \hline
AAPL &
  {\color[HTML]{32CB00} 0.92} &
  {\color[HTML]{FE0000} 0.55} &
  0.23 &
  0.04 &
  0.41 &
  {\color[HTML]{3166FF} 0.73} &
  0.04 &
  0.04 &
  0.04 \\
ABBV &
  0.00 &
  0.00 &
  0.00 &
  0.31 &
  {\color[HTML]{FE0000} 0.63} &
  0.40 &
  {\color[HTML]{32CB00} 0.69} &
  0.37 &
  {\color[HTML]{3166FF} 0.60} \\
PM &
  {\color[HTML]{32CB00} 0.63} &
  0.07 &
  0.01 &
  0.32 &
  {\color[HTML]{FE0000} 0.69} &
  0.42 &
  0.05 &
  0.24 &
  {\color[HTML]{3166FF} 0.57} \\ \hline
BAC &
  {\color[HTML]{32CB00} 0.78} &
  0.43 &
  0.30 &
  0.12 &
  {\color[HTML]{FE0000} 0.47} &
  {\color[HTML]{3166FF} 0.55} &
  0.10 &
  0.10 &
  0.15 \\
CSCO &
  {\color[HTML]{32CB00} 0.80} &
  0.45 &
  0.24 &
  0.13 &
  {\color[HTML]{FE0000} 0.46} &
  {\color[HTML]{3166FF} 0.58} &
  0.07 &
  0.09 &
  0.18 \\
KO &
  {\color[HTML]{32CB00} 0.68} &
  0.33 &
  0.14 &
  0.23 &
  {\color[HTML]{FE0000} 0.49} &
  {\color[HTML]{3166FF} 0.51} &
  0.09 &
  0.18 &
  0.35 \\
ORCL &
  {\color[HTML]{32CB00} 0.75} &
  0.39 &
  0.18 &
  0.17 &
  {\color[HTML]{FE0000} 0.45} &
  {\color[HTML]{3166FF} 0.56} &
  0.08 &
  0.16 &
  0.26 \\
PFE &
  {\color[HTML]{32CB00} 0.72} &
  0.38 &
  0.17 &
  0.19 &
  {\color[HTML]{FE0000} 0.48} &
  {\color[HTML]{3166FF} 0.54} &
  0.09 &
  0.14 &
  0.29 \\
VZ &
  {\color[HTML]{32CB00} 0.73} &
  0.38 &
  0.16 &
  0.19 &
  {\color[HTML]{FE0000} 0.48} &
  {\color[HTML]{3166FF} 0.57} &
  0.08 &
  0.14 &
  0.27 \\ \hline
\end{tabular}
\end{table}
}

So far, in all the analyses, we have always defined the time in terms of number of LOB updates (i.e., `tick time'). This means that, for different stocks, there is a different mapping between physical time and tick-time. This aspect constitutes an issue from the point of view of practitioners who are not interested in the forecasts as the result of a mere academic exercise, but are mainly focused on their actual practicability in real-world scenarios. In Table \ref{tab:probability_number_seconds_horizons}, we report the average probability (computed across the $3$-year analysis period) that the number of updates characterising the three horizons $\text{H}{\Delta \tau} \in \{10, 50, 100\}$, happens in a physical time (i) $< 1$ second (s), (ii) $\geq 1$ and $< 10$ seconds, or (iii) $\geq 10$ seconds. For each $\text{H}{\Delta \tau} \in \{10, 50, 100\}$, the probabilities of the three cases sums to $1$. As one can notice, for all the stocks, except MCD (small-tick stock) and ABBV (medium-tick stock), $10$ LOB's updates are more likely to happen in a physical time $< 1$\,s. $50$ LOB's updates, instead, are more likely to happen in a physical time $\geq 1$\,s $\land$ $< 10$\,s except that for CHTR (small-tick stock), MCD (small-tick stock) and AAPL (medium-tick stock). The case of AAPL is particularly notable since it is characterised by a remarkably more frequent trading activity than observed in all the other assets. Lastly, $\text{H}100$ represents the scenario where behavioural clustering is more evident among different classes of stocks. For small-tick stocks, $100$ LOB's updates are more likely to happen in a physical time $\geq 10$\,s. The only two exceptions are GOOG and NVDA, which are characterised by higher trading activity. For medium-tick stocks, ABBV and PM, as per all the other microstructural analyses, show a behaviour which is comparable to the one of small-tick stocks, while AAPL has a behaviour more similar to the one of large-tick stocks. Indeed, for this last class of stocks, $100$ LOB's updates are always more likely to happen in a physical time $\geq 1$\,s $\land$ $< 10$\,s, delineating a trading activity which is higher than the one of small- and medium-tick stocks.

\section{Results}\label{sec:Results}
In this Section, we report the results of our analysis, in particular concerning (i) the assessment of the DeepLOB model performance for mid-price changes direction forecast using traditional machine learning metrics; and (ii) the introduction of a novel, cutting-edge strategy-oriented methodology that computes the probability of correctly predicting a transaction using the model's forecasts. In all the experiments, we assess the behaviour of the three classes of stocks -- small-, medium- and large-tick stocks -- at $3$ predictions horizons $\text{H}{\Delta \tau} \in \{10, 50, 100\}$, at different confidence levels (i.e., adopting various probability thresholds). The goal is to evaluate the forecasting performances in different scenarios and link them to the microstructural properties of the stocks (see Section \ref{sec:Microstructural_Priors}) and the complex underlying LOB dynamics.

\subsection{Assessing model's forecast performances using traditional machine learning metrics}
\begin{figure}[H]
    \centering
    \includegraphics[scale=0.45]{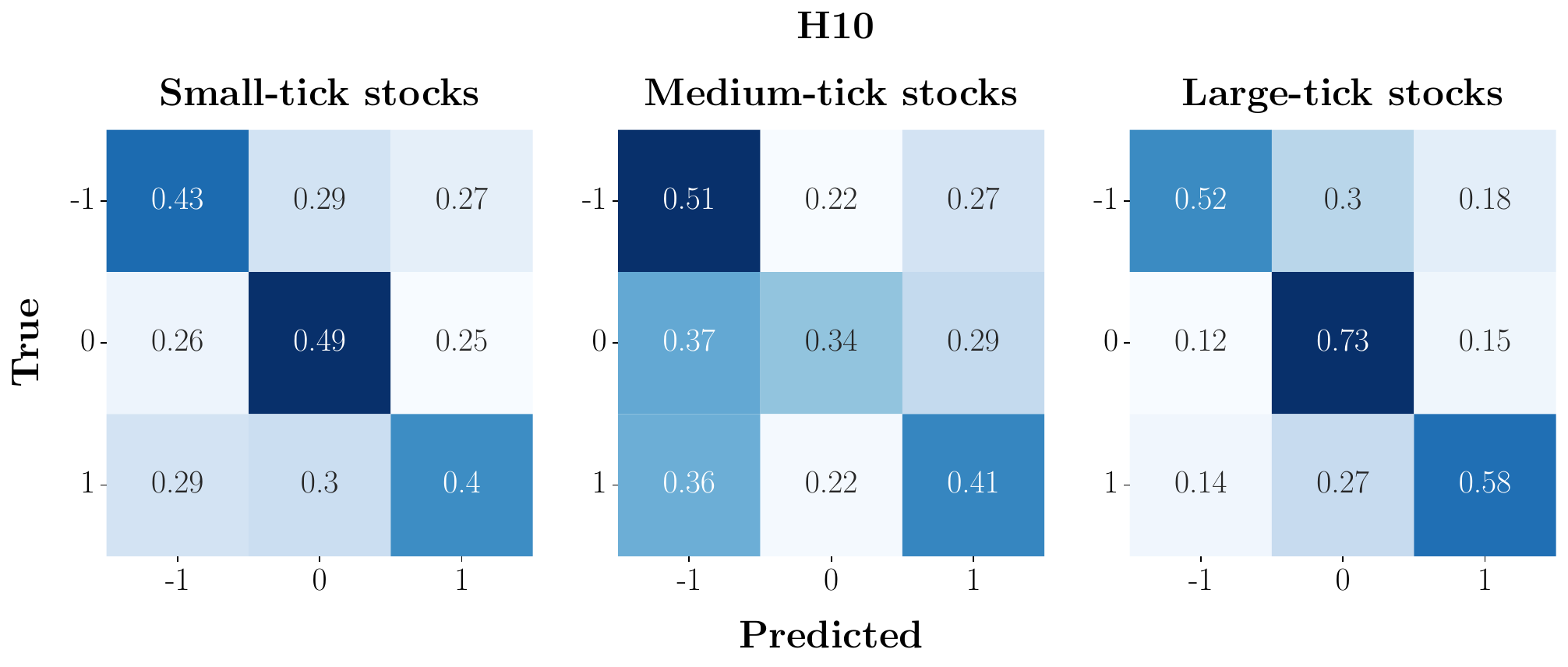}
    \caption{Average confusion matrices at $\text{H}10$. To obtain these compact representations, we firstly compute individual confusion matrices for each stock over the $3$-year analysis period, aggregating them into a list based on the class (i.e., small-, medium- and large-tick stocks). The average matrix is obtained by summing these matrices and dividing by their count, thus reflecting overall performance metrics. This average is finally normalized row-wise, turning counts into proportionate metrics of predictive accuracy and class-specific performance. The final normalized matrix succinctly visualizes the model's average effectiveness in classifying mid-price changes directions, during the period of interest.}
    \label{fig:confusion_matrix_H10}
\end{figure}

To assess the forecasting performances of the DeepLOB model, we start by discussing the confusion matrices computed for each class of stocks  -- small-, medium- and large-tick stocks -- at $3$ predictions horizons $\text{H}{\Delta \tau} \in \{10, 50, 100\}$, across the $3$-year analysis period. In Figure \ref{fig:confusion_matrix_H10}, we report the average confusion matrix for each class of stocks at $\text{H}10$. We observe that models trained on small- and medium-tick stocks demonstrate a non-negligible frequency of reciprocal misclassifications between the extreme classes ($-1$ and $1$), which are the two classes that anticipate a `Down' and `Up' movement, respectively. Specifically, for small-tick stocks, the $29\%$ of true class $1$ is misclassified as class $-1$, and the $27\%$ of true class $-1$ is misclassified as class $1$. Medium-tick stocks exhibit a similar pattern with a slight increase in misclassification for true class $1$ as class $-1$ (i.e., $36\%$). Conversely, for large-tick stocks, the model's predictive performance is markedly distinct with a stronger ability to correctly classify the two extreme classes and most of the errors concentrated towards their misclassification as $0$.

\begin{figure}[H]
    \centering
    \includegraphics[scale=0.45]{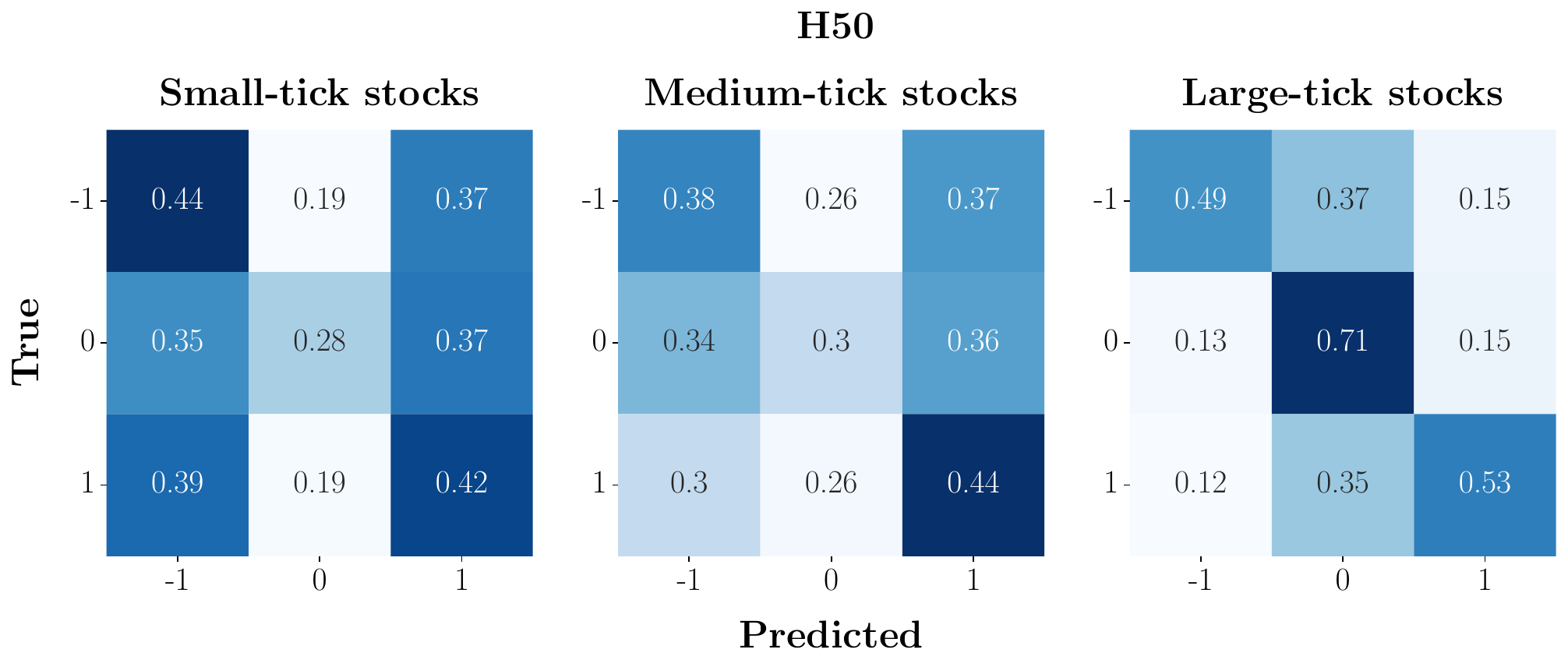}
    \caption{Average confusion matrices at $\text{H}50$. To obtain these compact representations, we firstly compute individual confusion matrices for each stock over the $3$-year analysis period, aggregating them into a list based on the class (i.e., small-, medium- and large-tick stocks). The average matrix is obtained by summing these matrices and dividing by their count, thus reflecting overall performance metrics. This average is finally normalized row-wise, turning counts into proportionate metrics of predictive accuracy and class-specific performance. The final normalized matrix succinctly visualizes the model's average effectiveness in classifying mid-price changes directions, during the period of interest.}
    \label{fig:confusion_matrix_H50}
\end{figure}

In Figure \ref{fig:confusion_matrix_H50}, we report the average confusion matrix for each class of stocks, across the $3$-year analysis period, at $\text{H}50$. In this case, we observe that, compared to what happens at $\text{H}10$, models trained on small- and medium-tick stocks have a higher tendency to mix the extreme classes ($-1$ and $1$), which, we stress again, anticipate a `Down' and `Up' movement. For small-tick stocks, we observe that the $39\%$ of the true class $1$ instances are misclassified as class $-1$, and the $37\%$ of true class $-1$ instances are mistaken for class $1$. 

Medium-tick stocks show a comparable trend, with slightly more misclassification of extreme classes to the central one. On the other hand, for large-tick stocks, the model performance remains similar to the one observed at $\text{H}10$. 

\begin{figure}[H]
    \centering
    \includegraphics[scale=0.45]{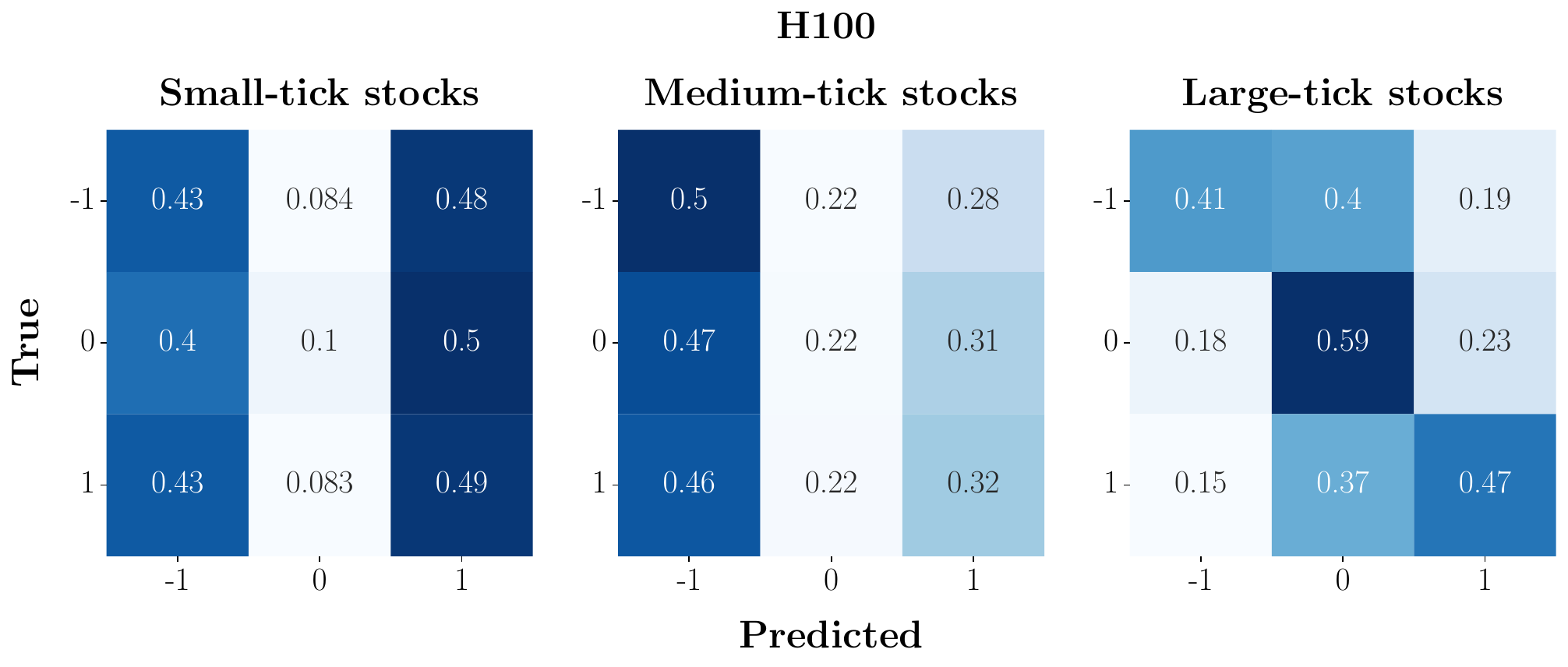}
    \caption{Average confusion matrices at $\text{H}100$. To obtain these compact representations, we firstly compute individual confusion matrices for each stock over the $3$-year analysis period, aggregating them into a list based on the class (i.e., small-, medium- and large-tick stocks). The average matrix is obtained by summing these matrices and dividing by their count, thus reflecting overall performance metrics. This average is finally normalized row-wise, turning counts into proportionate metrics of predictive accuracy and class-specific performance. The final normalized matrix succinctly visualizes the model's average effectiveness in classifying mid-price changes directions, during the period of interest.}
    \label{fig:confusion_matrix_H100}
\end{figure}

In Figure \ref{fig:confusion_matrix_H100}, we report the average confusion matrix for each class of stocks, across the $3$-year analysis period, at $\text{H}100$. In this case, we observe that models trained on small-tick stocks have an equal tendency to correctly classify and misclassify the two extreme classes. In addition to this, in line with what is observed for $\text{H}50$, there is a remarkable tendency to classify class $0$ as $-1$ or $1$, further incrementing the probability of critical errors. A similar scenario is detected for medium-tick stocks, while, for large-tick stocks, one more time, the model's performance remains consistent with the one observed at $\text{H}10$ and $\text{H}50$. 

\begin{figure}[h!]
    \centering
    \includegraphics[scale=0.22]{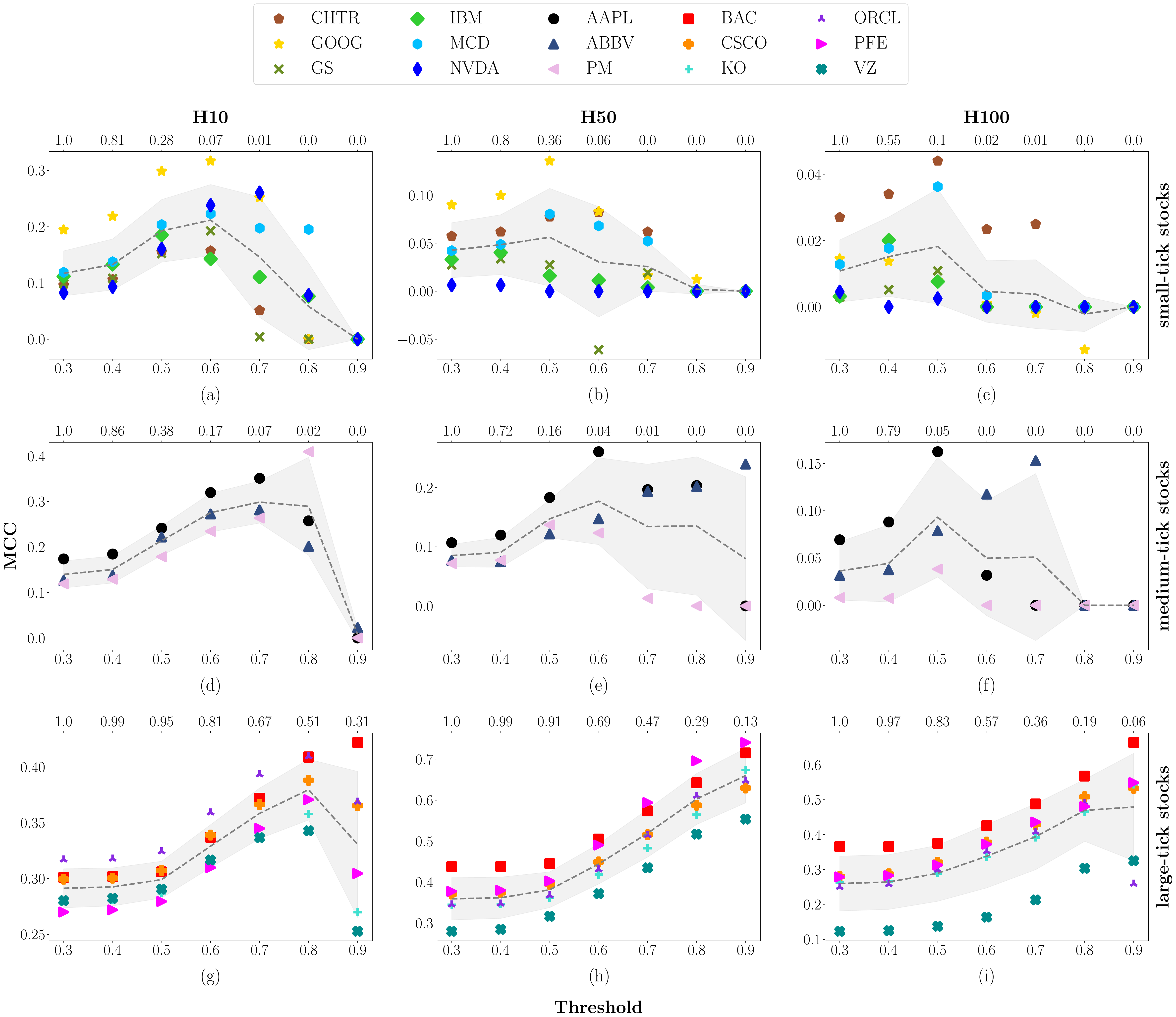}
    \caption{Average Matthews Correlation Coefficient (MCC). Results are organised according to the prediction horizons taken into account (see columns) and stocks’ tick-size (see rows). Each plot contains three main pieces of information: (i) the model’s performance changes applying different thresholds on the probabilities associated with each forecast (shown on the bottom of the x-axis); (ii) the average percentage amount of remaining data after using the threshold (shown on the top of the x-axis); (iii) the performance average pattern and the corresponding standard deviation (shown through the grey line and shadows). All the average values and the standard deviations are computed by considering stocks with the same tick-size, spanning the $3$-year analysis period.}
    \label{fig:cross_years_average_mcc}
\end{figure}

Confusion matrices serve as the foundational instrument for presenting the behavior of predictive models in their broadest context. These matrices provide a detailed breakdown of the model's forecasts which helps in evaluating the performance across different scenarios. However, to gain a deeper insight into a model's capabilities and to make more nuanced assessments of its effectiveness, derived metrics are essential. These metrics offer a more granular view of the model's predictive accuracy and error tendencies, facilitating a comprehensive understanding of its strengths and limitations. By analyzing these derived metrics, researchers and practitioners can better comprehend the potential of each model, enabling them to make informed decisions regarding its application and improvement. To assess the forecasting performances of the DeepLOB model we use the Matthews Correlation Coefficient (MCC) \cite{gorodkin2004comparing}. It is a generalization of Pearson's correlation coefficient between actual and predicted classes; it takes values between $-1$ (in case of inverse prediction) and $+1$ (in case of perfect prediction), while a value of $0$ indicates a random prediction. MCC is generally regarded as a balanced measure which can be used even if the classes are of very different sizes \cite{powers2020evaluation, chicco2017ten}. Figure \ref{fig:cross_years_average_mcc} shows the average MCC computed on the $3$-year analysis period for the different classes of stocks.  Results are organized according to prediction horizons (see columns) and stocks' tick-sizes (see rows). Each plot contains three main pieces of information: (i) the model performance's changes by applying different probability thresholds (shown on the bottom of the $x$-axis); (ii) the average percentage amount of remaining data after using probability thresholds (shown on the top of the $x$-axis); (iii) the performance average trend (computed across stocks belonging to the same class) and the corresponding standard deviation (shown through the grey line and shadow, respectively). Looking at these results, it is worth highlighting the different scales on the $y$-axes for each horizon and each class of stocks.

At $\text{H}10$, without the application of any threshold, the average MCC for \textit{small-tick stocks} is $0.11$, while the standard deviation is $0.04$, with only one stock (GOOG) acting as an outlier with an average MCC value (computed across the $3$-year period of analysis) equal to $0.19$. We observe that, for thresholds $\leq 0.6$, there is an increasing pattern in the average performance. This behaviour is associated with a rapid decrease in the average percentage of data used for the metric's computation. The average performance decreases for thresholds $> 0.6$, and only an average percentage of data $< 1\%$ is used for the metric's computation. Similar findings are detected for \textit{medium-tick stocks}. At $\text{H}10$, without applying any threshold, the average MCC is $0.13$, while the standard deviation equals $0.03$. By increasing the threshold value, we observe an increase in performance and a decrease in the percentage of data used for the metric's computation. Such a decrease is smoother if compared to the one observed in small-tick stocks, but still relevant. Among medium-tick stocks, AAPL performs slightly better than other stocks, remarking an intra-class separation that we already observed from the point of view of  microstructural properties in Section \ref{sec:Microstructural_Priors}. The scenario radically changes for \textit{large-tick stocks}. At $\text{H}10$, without applying any threshold, the average MCC is $0.29$ ($18$ units higher than the one characterizing small-tick stocks and $16$ units higher than the one characterizing medium-tick stocks). At the same time, the standard deviation has a value equal to $0.017$. Also in this case, a threshold-dependent increasing pattern is evident, especially for values $> 0.5$. However, unlike the other two classes of stocks, the average percentage of data used to compute the metric remains remarkably high. In this sense, the case of threshold equal to $0.9$ is meaningful since the metric is still computed using the $31\%$ of available forecasts, hence highlighting an enhanced strength of the signal associated with each forecast. 

Moving to $\text{H}50$, we note that, without applying any threshold, the average MCC for \textit{small-tick stocks} is $0.04$, while the standard deviation is $0.029$. Also in this case, GOOG acts as an outlier with an average MCC value (computed across the $3$-year period of analysis) equal to $0.089$. The same happens for NVDA, but in negative terms: in this case, the average MCC value equals $0.006$. We remark that by varying the threshold, the average performance for small-tick stocks remains almost constant. In contrast, the decrease in the average percentage of data used for the metric's computation is comparable to that observed at $\text{H}10$. For \textit{medium-tick stocks}, at $\text{H}50$, without applying any threshold, the average MCC is $0.085$, while the standard deviation has a value of $0.019$. An average growing pattern is detected for threshold values $\leq 0.6$. In contrast, the average percentage of values used for metric computation decreases with the same velocity as in small-tick stocks. A larger standard deviation is detected for threshold values $> 0.6$. AAPL always performs better than other stocks belonging to the same class. For \textit{large-tick stocks}, at $\text{H}50$, without the application of any threshold, the average MCC is $0.36$ ($32$ units higher than the one of small-tick stocks and $28$ units higher than the one of medium-tick stocks). At the same time, the standard deviation has a value of $0.056$. A clear average growing pattern is detected for all threshold values. In contrast, the average percentage of data used to compute the metric decreases considerably more than what happened at $\text{H}10$, remaining higher than the minimum reached by small- and medium-tick stocks. We remark that the difference between the average performance at threshold $0.9$ and the one without threshold (i.e. threshold equals $0.3$) equals $0.30$. 

Lastly, considering $\text{H}100$, we notice that, without applying any threshold, the average MCC for \textit{small-tick stocks} is $0.01$, while the standard deviation is $0.009$. These results suggest that the model is producing random forecasts. The average performance remains almost constant, varying the threshold, while the decrease in the average percentage of data used for the metric's computation is the steepest if compared to the values at $\text{H}{\Delta \tau} \in {10, 50}$. For \textit{medium-tick stocks}, at $\text{H}100$, without applying any threshold, the average MCC is $0.036$, while the standard deviation has a value of $0.03$. An average growing pattern is detected for threshold values $\leq 0.5$, while the decrease in average percentage is as steeper as in small-tick stocks. A larger standard deviation is detected when the threshold value is $\leq 0.8$, with AAPL stock performing, one more time, better than other class components. For \textit{large-tick stocks}, at $\text{H}100$, without the application of any threshold, the average MCC is $0.26$ ($25$ units higher than the one of small-tick stocks and $23$ units higher than the one of medium-tick stocks). At the same time, the standard deviation has a value of $0.078$. Differently from what happens for the other two classes of stocks, a clear average growing pattern is detected for threshold values $\leq 0.8$, while the average percentage of data used to compute the metric decreases more than what happened at $\text{H}50$, remaining, however, higher than the minima reached for small- and medium-tick stocks.

The analysis reported in this Section is further deepened in \ref{sec:Appendix_B} and \ref{sec: Appendix_C}, where we report (i) the year-wise MCC of the DeepLOB model at $\text{H}\Delta\tau \in \{10, 50, 100\}$, for different confidence levels; (ii) the corresponding statistical significance; and (iii) a replica of the analysis in Figure \ref{fig:cross_years_average_mcc} for the F1 and accuracy score. The coherence of the results bolsters the robustness of the findings discussed earlier in this Section, highlighting that large-tick stocks exhibit a significant predictability rate across all the considered horizons. This is evidenced in Figures \ref{fig:cross_years_average_f1} and \ref{fig:cross_years_average_accuracy}, where we observe that these stocks achieve F1 score realizations greater than $0.45$ without applying any thresholds and surpass $0.7$ when probability thresholds are implemented. Similarly, accuracy scores exceed $0.7$ without thresholds and reach over $0.9$ with the application of probability thresholds. However, as detailed in Section \ref{sec:Practicability_Forecasts}, achieving high scores on these traditional machine learning metrics does not necessarily translate into the generation of actionable trading signals. This distinction remarks the complexity of converting predictive accuracy into practical trading strategies.

\subsection{On the practicability of model's forecasts}\label{sec:Practicability_Forecasts}
The analysis presented above offers significant insights into the DeepLOB model's performance at different horizons for different classes of stocks. However, further discussion is needed to understand the results from the perspective of the microstructural properties of the LOB. To do so, in this Section, we introduce a novel methodology to evaluate the practicability of forecasts.

\begin{figure}[H]
    \centering
    \includegraphics[scale=0.65, trim = 12cm 1cm 12cm 0cm]{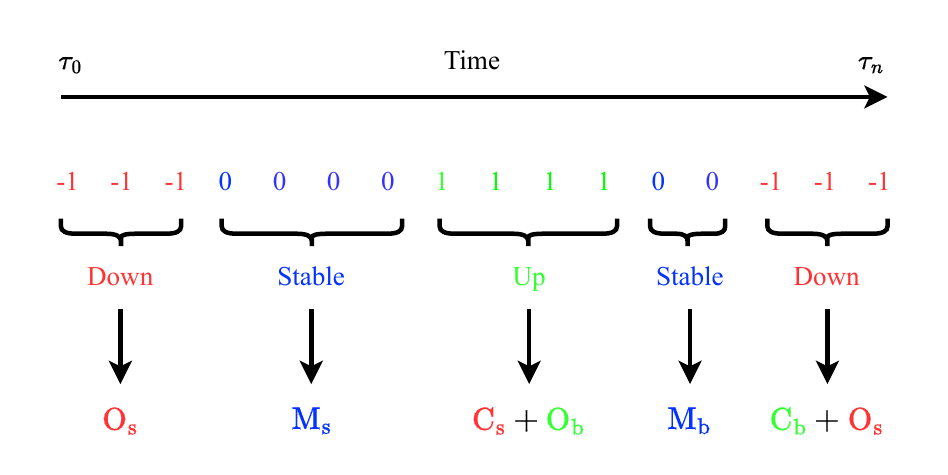}
    \caption{Pictorial representation of a chronologically sorted vector of forecasts. Following the mapping in Equation  \eqref{eq:labelling_step}, we derive a simplified strategy where $\text{O}_{p \, \in \, \{\text{s/b}\}}$ means `opening a new selling/buying position', $\text{M}_{p \, \in \, \{\text{s/b}\}}$ means `maintaining an existing selling/buying position', $\text{C}_{p \, \in \, \{\text{s/b}\}}$ means `closing an existing selling/buying position'.}
    \label{fig:strategy_example_1}
\end{figure}

\begin{figure}[h]
    \centering
    \begin{subfigure}{0.48\textwidth}
        \centering
        \includegraphics[width=\linewidth]{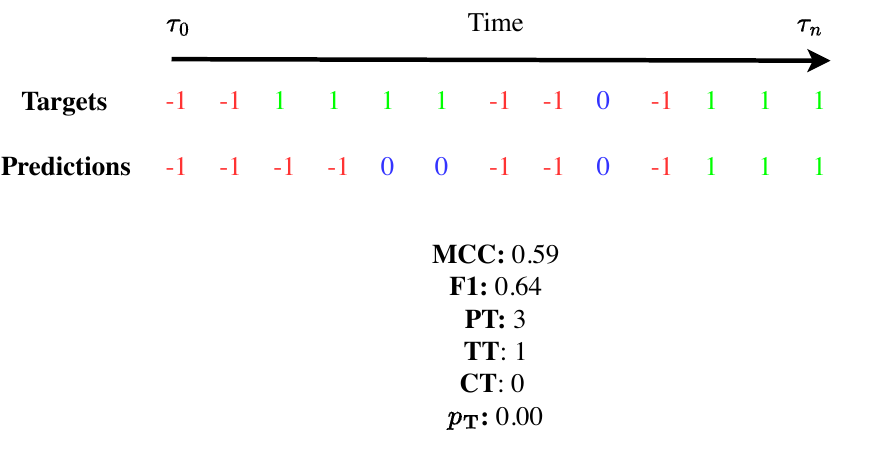}
        \captionsetup{justification=centering, margin=5cm}
        \caption{}
        \label{fig:example_strategy_2a}
    \end{subfigure}
    \hfill
    \begin{subfigure}{0.48\textwidth}
        \centering
        \includegraphics[width=\linewidth, trim = 0cm 0.3cm 0cm 0cm]{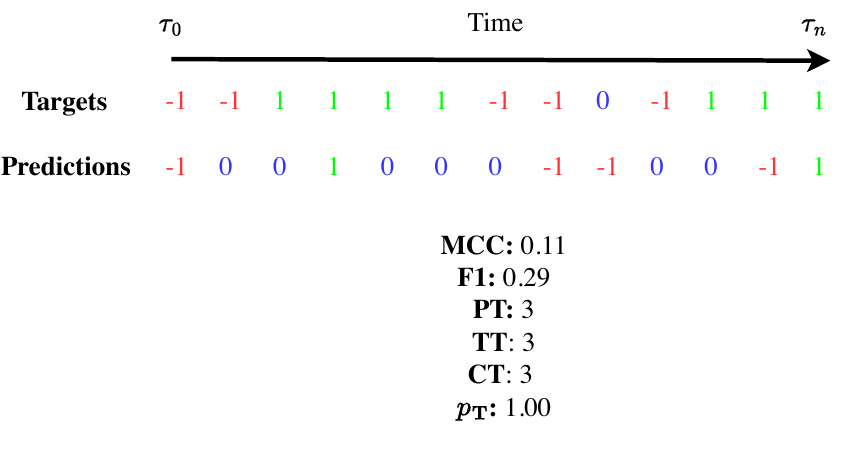}
        \captionsetup{justification=centering, margin=5cm}
        \caption{}
        \label{fig:example_strategy_2b}
    \end{subfigure}
    \caption{Transaction-related (PT,TT, CT, $p_T$) and machine learning metrics (MCC, F1) computed on two chronologically sorted vectors of forecasts and corresponding targets. }
    \label{fig:example_strategy_2}
\end{figure}

Despite the recent attempts made in state-of-the-art research papers \cite{zhang2019deeplob, yin2023deep, wood2021trading}, backtesting a trading strategy based on the outputs of a deep learning model by using historical data-only is not possible. Indeed, several assumptions are needed, including but not limited to (i) having the technical and infrastructural potential to record and process live data, produce forecasts and execute them in due time; (ii) being always executed; (iii) having a zero market-impact; (iv) having zero transaction costs. The combination of all or part of these assumptions ruins any attempt to produce a reliable backtest, and, indeed, it is different from what academics should try to achieve to bridge the gap with the practitioners' community. In this Section, we propose a strategy-oriented analysis of the model's forecasts, which is entirely assumption-free and fully immune to class imbalances. As an introductory example, let us consider a scenario where the mid-price changes' direction forecasts, which, in our case, are always chronologically sorted, are defined as per in Figure \ref{fig:strategy_example_1}. The direct mapping between predictions and trading actions would include (i) opening a selling position in correspondence of the first predicted mid-price `Down' movement (i.e., $\text{O}_{\text{s}}$); (ii) maintaining the selling position in correspondence of the predicted mid-price `Stable' period (i.e., $\text{M}_{\text{s}}$); (iii) closing the existing selling position (i.e., $\text{C}_{\text{s}}$), while opening a new buying position (i.e., $\text{O}_{\text{b}}$) in correspondence of the predicted mid-price `Up' movement; (iv) maintaining the position in correspondence of the predicted mid-price `Stable' period (i.e., $\text{M}_{\text{b}}$); and, (v) closing the existing buying position (i.e., $\text{C}_{\text{b}}$), while opening a selling position (i.e., $\text{O}_{\text{s}}$) in correspondence of the newly predicted mid-price `Down' movement. By performing this simplified strategy, we would have opened $3$ positions and closed $2$ of them, overall completing $2$ transactions (i.e., a transaction is completed when a position is successfully opened and later closed). Using forecasts, however, necessarily implies relying on their `correctness'. To contextualize this concept, let us consider the two examples of chronologically sorted vectors of forecasts presented in Figure \ref{fig:example_strategy_2}. For each of them, we report the MCC, the F1 score, and the following transactions-related metrics:

\newlength{\widestlabel}
\settowidth{\widestlabel}{iv)} 
\addtolength{\widestlabel}{\labelsep} 

\begin{itemize}
    \item The number of potential transactions (PT). Looking at Figure \ref{fig:strategy_example_1}, we remark that a transaction happens when one is able to open a position and then close it (i.e., $\text{O}_\text{s} \rightarrow \text{C}_\text{s} \;\lor\; \text{O}_\text{b} \rightarrow \text{C}_\text{b}$). In this context, we use the term `potential' because transactions are counted on the targets' set.
    \item The total number of executed transactions (TT). This metric is computed in the same manner as PT, but on the predictions' set.
    \item The total number of correctly executed transactions (CT). This metric counts how many times a transaction executed on the predictions' set has a correspondence in the targets' set. In figure \ref{fig:example_strategy_2a}, we show an example where $\text{CT} = 0$, due to discrepancies in the positions' entering/exiting points in the two sets.
    \item The probability $p_{\text{T}}$ to execute a correct transaction.
\end{itemize}

From the definitions provided above, it is evident that the set CT is given by the intersection of PT and TT sets; the probability to execute a correct transaction, $p_{\text{T}}$, is hence computed as follows:

\begin{equation} \label{eq:intersection_of_sets}
    p_{\text{T}} = \frac{\text{CT}}{\text{PT} + \text{TT} - \text{CT}}.
\end{equation}
We remark that, being our approach assumption-free, when we refer to `opening/closing a position' and `executing a transaction', we mean the model's capability to accurately identify an optimal entry point for initiating or concluding a trade, either as separate actions or together, respectively. 

The examples provided in Figure \ref{fig:example_strategy_2} are explicitly designed to highlight the fragility of using traditional machine learning metrics to evaluate the out-of-sample practicability of predictions in the context of LOB forecasting. In particular, they constitute two `extreme' scenarios where traditional machine learning metrics take values far from the ones given by $p_{\text{T}}$, remarking the potential distance of academically acceptable findings and actually practicable ones.
Indeed, in this application domain, we are more interested in the chronological location of prediction errors rather than in the number of their occurrence. To be more specific, we are interested in (i) having at least one correct prediction in correspondence of each `Down' or `Up' movement; and, consequently, (ii) in not having any premature closing signal for an existing open position. The nature and the number of other errors are tolerable when these two conditions are satisfied. In real-world scenarios, also probabilities associated with forecasts should be taken into account. Indeed, we can decide to enter or exit a position based on the probability associated with the forecast (i.e., the strength of the signal). 

\begin{figure}[h]
    \centering
    \begin{subfigure}{0.48\textwidth}
        \centering
        \includegraphics[width=\linewidth]{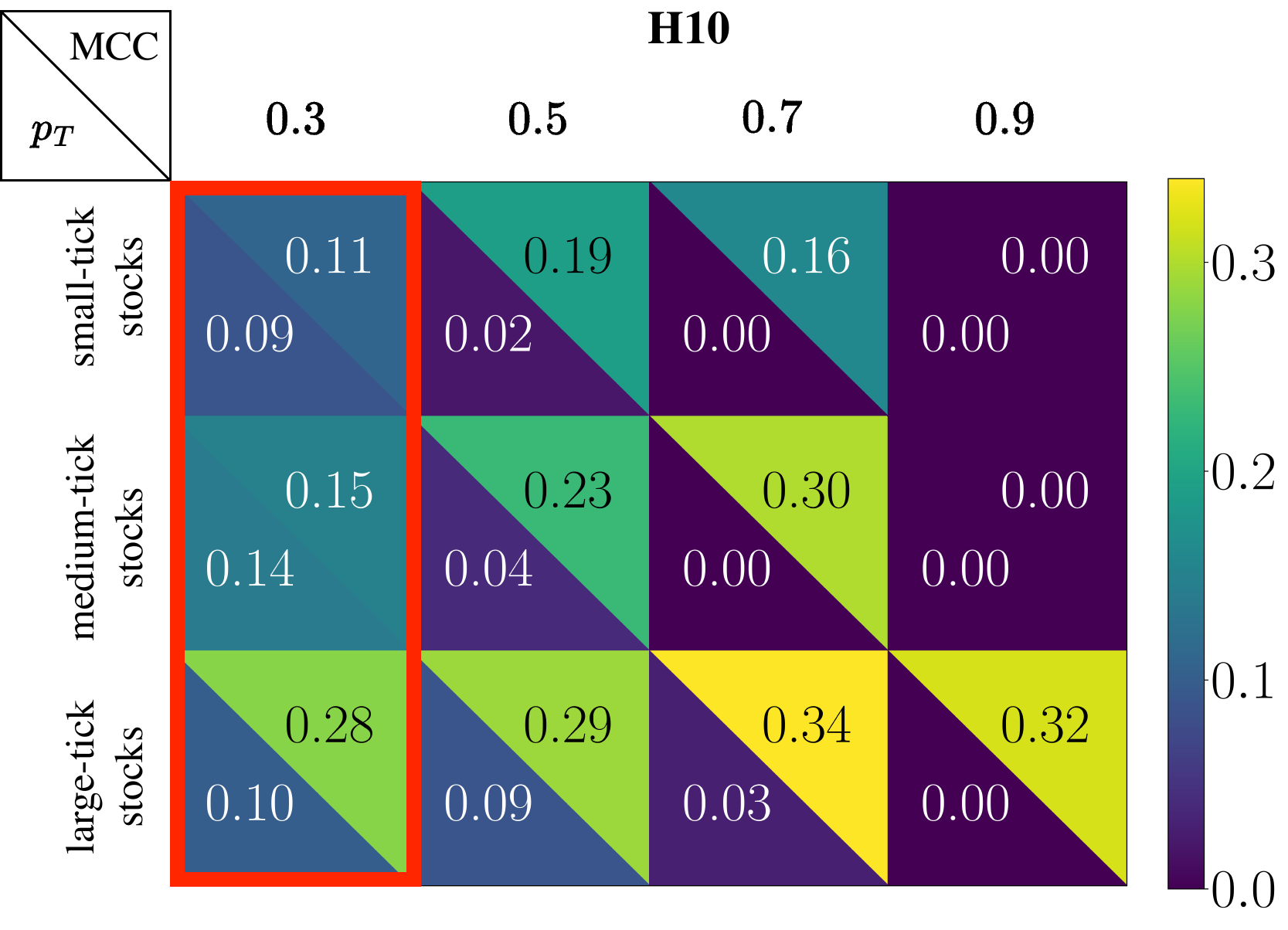}
        \captionsetup{justification=centering, margin=5cm}
        \caption{}
        \label{fig:pt_a}
    \end{subfigure}
    \hfill
    \begin{subfigure}{0.48\textwidth}
        \centering
        \includegraphics[width=\linewidth, trim = 0cm 0.0cm 0cm 0cm]{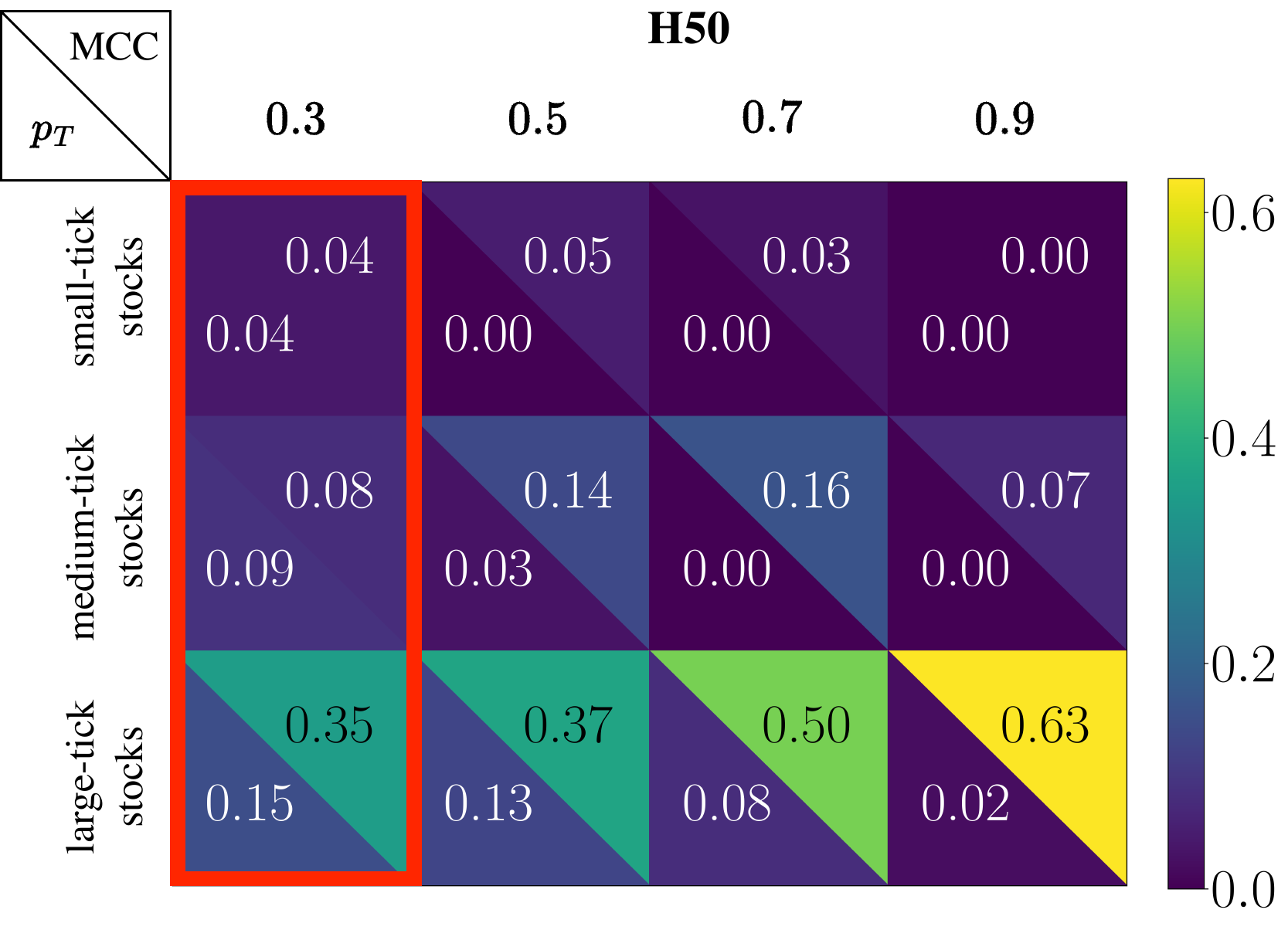}
        \captionsetup{justification=centering, margin=5cm}
        \caption{}
        \label{fig:pt_b}
    \end{subfigure}
    \hfill
    \begin{subfigure}{0.48\textwidth}
        \centering
        \includegraphics[width=\linewidth, trim = 0cm 0.0cm 0cm 0cm]{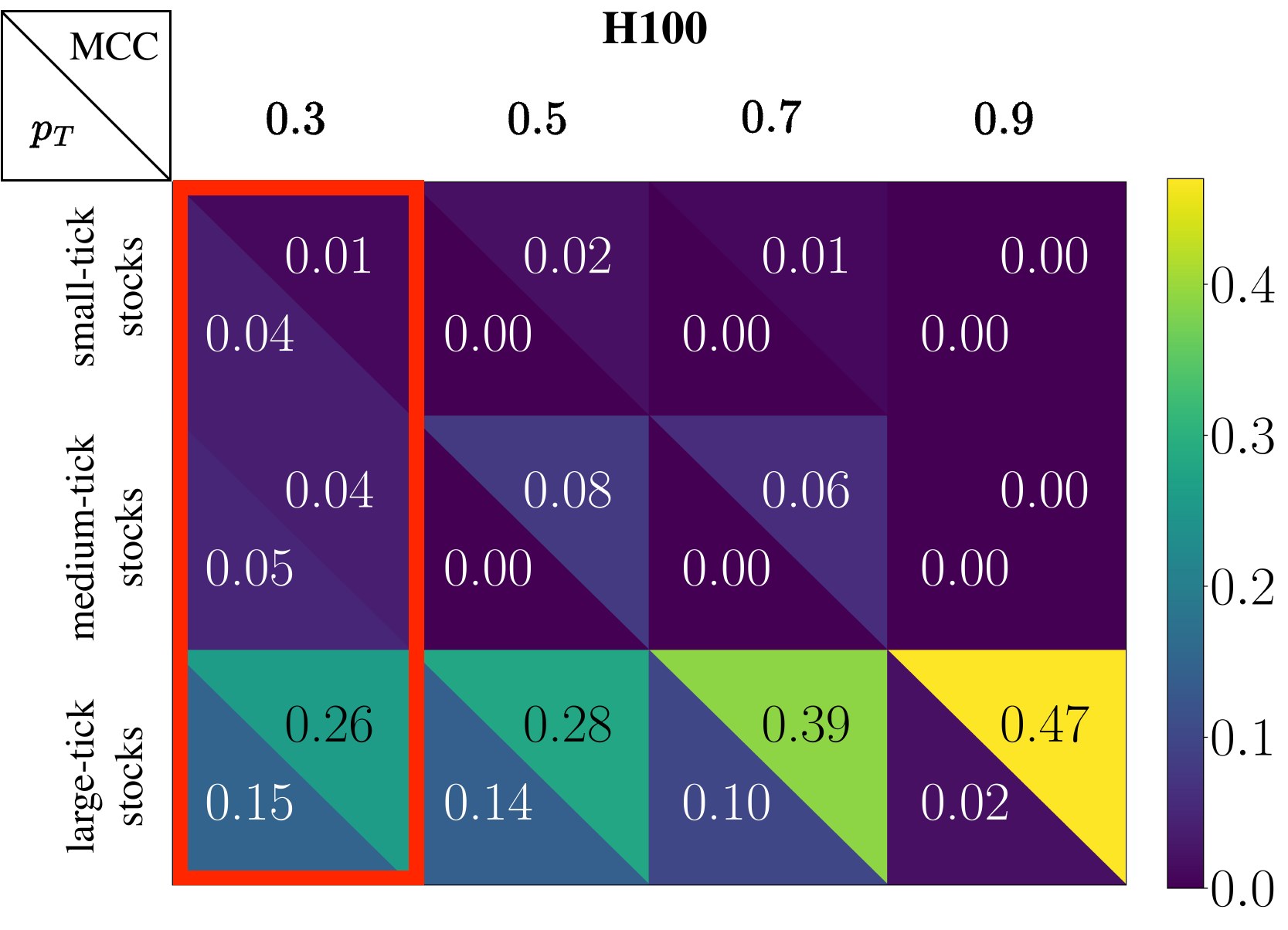}
        \captionsetup{justification=centering, margin=5cm}
        \caption{}
        \label{fig:pt_c}
    \end{subfigure}
    \caption{Coarse-grained representation of the behaviour of the average $p_{\text{T}}$ and MCC at $\text{H}{\Delta \tau} \in \{10, 50, 100\}$. For each class of stocks, we compute the average value for both metrics applying different probability thresholds (i.e., $0.3$, $0.5$, $0.7$, $0.9$). We notice two different behaviours that remain consistent across different scenarios: (i) $p_{\text{T}}$ decreases for increasing probability thresholds and increases moving from small-tick stocks to large-tick stocks; (ii) the MCC increases for increasing probability thresholds (this is more evident moving to longer prediction horizons) and increases also moving from small-tick stocks to large-tick stocks. We highlight in red the scenarios where no probability threshold is applied (i.e., the signal's sequence is untouched).}
    \label{fig:coincise_pt}
\end{figure}

These metrics are studied at different granularity levels in Figure \ref{fig:coincise_pt} (i.e., coarse-grained representation) and in Table \ref{tab:main_table_strategies} (i.e., fine-grained representation). Specifically, in Figure \ref{fig:coincise_pt}, for each class of stocks, we compute the average value for $p_{\text{T}}$ and MCC applying different probability thresholds ($0.3$, $0.5$, $0.7$, $0.9$) and we notice two different behaviours that remain consistent across different scenarios: (i) $p_{\text{T}}$ decreases for increasing probability thresholds and increases moving from small-tick stocks to large-tick stocks; (ii) the MCC increases for increasing probability thresholds (this is more evident moving to longer prediction horizons) and increases also moving from small-tick stocks to large-tick stocks. On one side these findings highlight the relevance of the positioning of the signal. By applying different probability thresholds we may break the signal's sequence, and even if the performance of classical machine learning metrics increases because of the increase of the strength of the signal, we are not able to correctly manage positions. On the other hand they highlight, one more time, the impact of the microstructural properties of the stocks on the signal's usability: overall large-tick stocks demonstrate to offer higher probabilities to actually operate trading in a fully automated way compared to small-tick stocks).

{
\renewcommand{\arraystretch}{1.7}
\begin{table}[h!]
    \centering
    \caption{Strategy-oriented, assumption-free study on the practicability of deep learning forecasts. For each $\text{H}{\Delta \tau} \in \{10, 50, 100\}$, we report the stock's $\text{PT}$, $p_{\text{T}}$, MCC and F1 score with the application of a probability threshold equal to $0.3$ (i.e. no threshold), $0.5$, $0.7$ and $0.9$. Being $p_{\text{T}}$ a computational expensive metric, results reported in this Table are refers to the first $50\%$ of available data only.}
    \label{tab:main_table_strategies}

    \begin{subtable}{.5\linewidth}
        \centering
        \tiny
        \setlength{\tabcolsep}{1pt}
        \begin{tabular}{cccccccccccccc}
            \hline
            \multicolumn{14}{c}{\textbf{H10}} \\ \hline
            \multicolumn{1}{c|}{\textbf{Ticker}} &
              \multicolumn{1}{c|}{\textbf{PT}} &
              \multicolumn{3}{c|}{\textbf{0.3}} &
              \multicolumn{3}{c|}{\textbf{0.5}} &
              \multicolumn{3}{c|}{\textbf{0.7}} &
              \multicolumn{3}{c}{\textbf{0.9}} \\ \cline{3-14} 
            \multicolumn{1}{c|}{} &
              \multicolumn{1}{c|}{} &
              \textbf{$p_{\mathbf{\text{T}}}$} &
              \textbf{MCC} &
              \multicolumn{1}{c|}{\textbf{F1}} &
              \textbf{$p_{\mathbf{\text{T}}}$} &
              \textbf{MCC} &
              \multicolumn{1}{c|}{\textbf{F1}} &
              \textbf{$p_{\mathbf{\text{T}}}$} &
              \textbf{MCC} &
              \multicolumn{1}{c|}{\textbf{F1}} &
              \textbf{$p_{\mathbf{\text{T}}}$} &
              \textbf{MCC} &
              \textbf{F1} \\ \hline
            \multicolumn{1}{c|}{CHTR} &
              \multicolumn{1}{c|}{84751} &
              0.06 &
              0.09 &
              \multicolumn{1}{c|}{0.37} &
              0.00 &
              0.13 &
              \multicolumn{1}{c|}{0.37} &
              0.00 &
              0.04 &
              \multicolumn{1}{c|}{0.29} &
              0.00 &
              0.00 &
              0.00 \\
            \multicolumn{1}{c|}{GOOG} &
              \multicolumn{1}{c|}{297533} &
              0.04 &
              0.20 &
              \multicolumn{1}{c|}{0.46} &
              0.00 &
              0.33 &
              \multicolumn{1}{c|}{0.53} &
              0.00 &
              0.28 &
              \multicolumn{1}{c|}{0.39} &
              0.00 &
              0.00 &
              0.00 \\
            \multicolumn{1}{c|}{GS} &
              \multicolumn{1}{c|}{99858} &
              0.08 &
              0.08 &
              \multicolumn{1}{c|}{0.30} &
              0.01 &
              0.13 &
              \multicolumn{1}{c|}{0.33} &
              0.00 &
              0.00 &
              \multicolumn{1}{c|}{0.22} &
              0.00 &
              0.00 &
              0.00 \\
            \multicolumn{1}{c|}{IBM} &
              \multicolumn{1}{c|}{131162} &
              0.14 &
              0.10 &
              \multicolumn{1}{c|}{0.36} &
              0.04 &
              0.17 &
              \multicolumn{1}{c|}{0.35} &
              0.00 &
              0.12 &
              \multicolumn{1}{c|}{0.30} &
              0.00 &
              0.00 &
              0.00 \\
            \multicolumn{1}{c|}{MCD} &
              \multicolumn{1}{c|}{132481} &
              0.10 &
              0.12 &
              \multicolumn{1}{c|}{0.41} &
              0.02 &
              0.20 &
              \multicolumn{1}{c|}{0.44} &
              0.00 &
              0.29 &
              \multicolumn{1}{c|}{0.51} &
              0.00 &
              0.00 &
              0.00 \\
            \multicolumn{1}{c|}{NVDA} &
              \multicolumn{1}{c|}{233588} &
              0.12 &
              0.08 &
              \multicolumn{1}{c|}{0.35} &
              0.02 &
              0.16 &
              \multicolumn{1}{c|}{0.37} &
              0.00 &
              0.23 &
              \multicolumn{1}{c|}{0.42} &
              0.00 &
              0.00 &
              0.13 \\ \hline
            \multicolumn{1}{c|}{AAPL} &
              \multicolumn{1}{c|}{483853} &
              0.15 &
              0.17 &
              \multicolumn{1}{c|}{0.42} &
              0.05 &
              0.23 &
              \multicolumn{1}{c|}{0.46} &
              0.00 &
              0.33 &
              \multicolumn{1}{c|}{0.50} &
              0.00 &
              0.00 &
              0.29 \\
            \multicolumn{1}{c|}{ABBV} &
              \multicolumn{1}{c|}{79782} &
              0.14 &
              0.15 &
              \multicolumn{1}{c|}{0.41} &
              0.04 &
              0.25 &
              \multicolumn{1}{c|}{0.47} &
              0.00 &
              0.23 &
              \multicolumn{1}{c|}{0.41} &
              0.00 &
              0.02 &
              0.33 \\
            \multicolumn{1}{c|}{PM} &
              \multicolumn{1}{c|}{80043} &
              0.13 &
              0.13 &
              \multicolumn{1}{c|}{0.38} &
              0.04 &
              0.22 &
              \multicolumn{1}{c|}{0.43} &
              0.00 &
              0.35 &
              \multicolumn{1}{c|}{0.47} &
              0.00 &
              0.00 &
              0.00 \\ \hline
            \multicolumn{1}{c|}{BAC} &
              \multicolumn{1}{c|}{27155} &
              0.07 &
              0.30 &
              \multicolumn{1}{c|}{0.45} &
              0.06 &
              0.30 &
              \multicolumn{1}{c|}{0.46} &
              0.04 &
              0.36 &
              \multicolumn{1}{c|}{0.50} &
              0.01 &
              0.40 &
              0.54 \\
            \multicolumn{1}{c|}{CSCO} &
              \multicolumn{1}{c|}{69914} &
              0.10 &
              0.31 &
              \multicolumn{1}{c|}{0.50} &
              0.08 &
              0.32 &
              \multicolumn{1}{c|}{0.51} &
              0.03 &
              0.37 &
              \multicolumn{1}{c|}{0.55} &
              0.00 &
              0.33 &
              0.54 \\
            \multicolumn{1}{c|}{KO} &
              \multicolumn{1}{c|}{37589} &
              0.11 &
              0.24 &
              \multicolumn{1}{c|}{0.46} &
              0.09 &
              0.25 &
              \multicolumn{1}{c|}{0.47} &
              0.04 &
              0.28 &
              \multicolumn{1}{c|}{0.50} &
              0.00 &
              0.20 &
              0.45 \\
            \multicolumn{1}{c|}{ORCL} &
              \multicolumn{1}{c|}{62514} &
              0.12 &
              0.32 &
              \multicolumn{1}{c|}{0.49} &
              0.11 &
              0.33 &
              \multicolumn{1}{c|}{0.50} &
              0.04 &
              0.40 &
              \multicolumn{1}{c|}{0.56} &
              0.00 &
              0.39 &
              0.50 \\
            \multicolumn{1}{c|}{PFE} &
              \multicolumn{1}{c|}{38853} &
              0.09 &
              0.28 &
              \multicolumn{1}{c|}{0.46} &
              0.08 &
              0.29 &
              \multicolumn{1}{c|}{0.47} &
              0.03 &
              0.36 &
              \multicolumn{1}{c|}{0.54} &
              0.00 &
              0.36 &
              0.52 \\
            \multicolumn{1}{c|}{VZ} &
              \multicolumn{1}{c|}{87383} &
              0.11 &
              0.27 &
              \multicolumn{1}{c|}{0.48} &
              0.09 &
              0.27 &
              \multicolumn{1}{c|}{0.49} &
              0.02 &
              0.31 &
              \multicolumn{1}{c|}{0.52} &
              0.00 &
              0.21 &
              0.44 \\ \hline
            \end{tabular}
        \vspace{5pt}
        \captionsetup{justification=centering}
        \caption{}
        \label{tab:sub_table_a}
    \end{subtable}%
    \hfill
    \begin{subtable}{.5\linewidth}
        \centering
        \tiny
        \setlength{\tabcolsep}{1pt}
        \begin{tabular}{cccccccccccccc}
            \hline
            \multicolumn{14}{c}{\textbf{H50}} \\ \hline
            \multicolumn{1}{c|}{\textbf{Ticker}} &
              \multicolumn{1}{c|}{\textbf{PT}} &
              \multicolumn{3}{c|}{\textbf{0.3}} &
              \multicolumn{3}{c|}{\textbf{0.5}} &
              \multicolumn{3}{c|}{\textbf{0.7}} &
              \multicolumn{3}{c}{\textbf{0.9}} \\ \cline{3-14} 
            \multicolumn{1}{c|}{} &
              \multicolumn{1}{c|}{} &
              \textbf{$p_{\mathbf{\text{T}}}$} &
              \textbf{MCC} &
              \multicolumn{1}{c|}{\textbf{F1}} &
              \textbf{$p_{\mathbf{\text{T}}}$} &
              \textbf{MCC} &
              \multicolumn{1}{c|}{\textbf{F1}} &
              \textbf{$p_{\mathbf{\text{T}}}$} &
              \textbf{MCC} &
              \multicolumn{1}{c|}{\textbf{F1}} &
              \textbf{$p_{\mathbf{\text{T}}}$} &
              \textbf{MCC} &
              \textbf{F1} \\ \hline
            \multicolumn{1}{c|}{CHTR} &
              \multicolumn{1}{c|}{54303} &
              0.03 &
              0.05 &
              \multicolumn{1}{c|}{0.35} &
              0.01 &
              0.06 &
              \multicolumn{1}{c|}{0.34} &
              0.00 &
              0.09 &
              \multicolumn{1}{c|}{0.21} &
              0.00 &
              0.00 &
              0.06 \\
            \multicolumn{1}{c|}{GOOG} &
              \multicolumn{1}{c|}{196149} &
              0.06 &
              0.09 &
              \multicolumn{1}{c|}{0.36} &
              0.01 &
              0.13 &
              \multicolumn{1}{c|}{0.36} &
              0.00 &
              -0.03 &
              \multicolumn{1}{c|}{0.26} &
              0.00 &
              0.00 &
              0.00 \\
            \multicolumn{1}{c|}{GS} &
              \multicolumn{1}{c|}{55020} &
              0.05 &
              0.02 &
              \multicolumn{1}{c|}{0.29} &
              0.00 &
              0.02 &
              \multicolumn{1}{c|}{0.22} &
              0.00 &
              0.02 &
              \multicolumn{1}{c|}{0.13} &
              0.00 &
              0.00 &
              0.02 \\
            \multicolumn{1}{c|}{IBM} &
              \multicolumn{1}{c|}{64859} &
              0.06 &
              0.03 &
              \multicolumn{1}{c|}{0.28} &
              0.01 &
              0.02 &
              \multicolumn{1}{c|}{0.20} &
              0.00 &
              0.00 &
              \multicolumn{1}{c|}{0.22} &
              0.00 &
              0.00 &
              0.00 \\
            \multicolumn{1}{c|}{MCD} &
              \multicolumn{1}{c|}{73427} &
              0.04 &
              0.04 &
              \multicolumn{1}{c|}{0.31} &
              0.01 &
              0.07 &
              \multicolumn{1}{c|}{0.32} &
              0.00 &
              0.07 &
              \multicolumn{1}{c|}{0.36} &
              0.00 &
              0.00 &
              0.00 \\
            \multicolumn{1}{c|}{NVDA} &
              \multicolumn{1}{c|}{104414} &
              0.01 &
              0.01 &
              \multicolumn{1}{c|}{0.19} &
              0.00 &
              0.00 &
              \multicolumn{1}{c|}{0.12} &
              0.00 &
              0.00 &
              \multicolumn{1}{c|}{0.14} &
              0.00 &
              0.00 &
              0.00 \\ \hline
            \multicolumn{1}{c|}{AAPL} &
              \multicolumn{1}{c|}{218318} &
              0.08 &
              0.10 &
              \multicolumn{1}{c|}{0.34} &
              0.04 &
              0.18 &
              \multicolumn{1}{c|}{0.45} &
              0.00 &
              0.16 &
              \multicolumn{1}{c|}{0.47} &
              0.00 &
              0.00 &
              0.00 \\
            \multicolumn{1}{c|}{ABBV} &
              \multicolumn{1}{c|}{38321} &
              0.08 &
              0.08 &
              \multicolumn{1}{c|}{0.28} &
              0.03 &
              0.12 &
              \multicolumn{1}{c|}{0.27} &
              0.00 &
              0.23 &
              \multicolumn{1}{c|}{0.48} &
              0.00 &
              0.21 &
              0.54 \\
            \multicolumn{1}{c|}{PM} &
              \multicolumn{1}{c|}{42523} &
              0.10 &
              0.07 &
              \multicolumn{1}{c|}{0.37} &
              0.01 &
              0.12 &
              \multicolumn{1}{c|}{0.34} &
              0.00 &
              0.10 &
              \multicolumn{1}{c|}{0.31} &
              0.00 &
              0.00 &
              0.00 \\ \hline
            \multicolumn{1}{c|}{BAC} &
              \multicolumn{1}{c|}{20711} &
              0.11 &
              0.43 &
              \multicolumn{1}{c|}{0.61} &
              0.10 &
              0.44 &
              \multicolumn{1}{c|}{0.62} &
              0.07 &
              0.57 &
              \multicolumn{1}{c|}{0.71} &
              0.02 &
              0.69 &
              0.76 \\
            \multicolumn{1}{c|}{CSCO} &
              \multicolumn{1}{c|}{48673} &
              0.17 &
              0.36 &
              \multicolumn{1}{c|}{0.57} &
              0.15 &
              0.38 &
              \multicolumn{1}{c|}{0.58} &
              0.12 &
              0.49 &
              \multicolumn{1}{c|}{0.65} &
              0.06 &
              0.60 &
              0.68 \\
            \multicolumn{1}{c|}{KO} &
              \multicolumn{1}{c|}{25581} &
              0.16 &
              0.33 &
              \multicolumn{1}{c|}{0.54} &
              0.15 &
              0.34 &
              \multicolumn{1}{c|}{0.55} &
              0.10 &
              0.45 &
              \multicolumn{1}{c|}{0.63} &
              0.03 &
              0.64 &
              0.75 \\
            \multicolumn{1}{c|}{ORCL} &
              \multicolumn{1}{c|}{41516} &
              0.16 &
              0.34 &
              \multicolumn{1}{c|}{0.55} &
              0.15 &
              0.37 &
              \multicolumn{1}{c|}{0.57} &
              0.08 &
              0.51 &
              \multicolumn{1}{c|}{0.67} &
              0.00 &
              0.65 &
              0.71 \\
            \multicolumn{1}{c|}{PFE} &
              \multicolumn{1}{c|}{25519} &
              0.15 &
              0.38 &
              \multicolumn{1}{c|}{0.57} &
              0.14 &
              0.40 &
              \multicolumn{1}{c|}{0.59} &
              0.08 &
              0.59 &
              \multicolumn{1}{c|}{0.72} &
              0.00 &
              0.74 &
              0.79 \\
            \multicolumn{1}{c|}{VZ} &
              \multicolumn{1}{c|}{51748} &
              0.12 &
              0.25 &
              \multicolumn{1}{c|}{0.48} &
              0.11 &
              0.29 &
              \multicolumn{1}{c|}{0.50} &
              0.05 &
              0.38 &
              \multicolumn{1}{c|}{0.56} &
              0.00 &
              0.47 &
              0.61 \\ \hline
            \end{tabular}
        \vspace{5pt}
        \captionsetup{justification=centering}
        \caption{}
        \label{tab:sub_table_b}
    \end{subtable}

    \vspace{0.2cm}

    \begin{subtable}{\linewidth}
        \centering
        \tiny
        \setlength{\tabcolsep}{1pt}
        \begin{tabular}{cccccccccccccc}
            \hline
            \multicolumn{14}{c}{\textbf{H100}} \\ \hline
            \multicolumn{1}{c|}{\textbf{Ticker}} &
              \multicolumn{1}{c|}{\textbf{PT}} &
              \multicolumn{3}{c|}{\textbf{0.3}} &
              \multicolumn{3}{c|}{\textbf{0.5}} &
              \multicolumn{3}{c|}{\textbf{0.7}} &
              \multicolumn{3}{c}{\textbf{0.9}} \\ \cline{3-14} 
            \multicolumn{1}{c|}{} &
              \multicolumn{1}{c|}{} &
              \textbf{$p_{\mathbf{\text{T}}}$} &
              \textbf{MCC} &
              \multicolumn{1}{c|}{\textbf{F1}} &
              \textbf{$p_{\mathbf{\text{T}}}$} &
              \textbf{MCC} &
              \multicolumn{1}{c|}{\textbf{F1}} &
              \textbf{$p_{\mathbf{\text{T}}}$} &
              \textbf{MCC} &
              \multicolumn{1}{c|}{\textbf{F1}} &
              \textbf{$p_{\mathbf{\text{T}}}$} &
              \textbf{MCC} &
              \textbf{F1} \\ \hline
            \multicolumn{1}{c|}{CHTR} &
              \multicolumn{1}{c|}{42794} &
              0.04 &
              0.03 &
              \multicolumn{1}{c|}{0.34} &
              0.00 &
              0.03 &
              \multicolumn{1}{c|}{0.23} &
              0.00 &
              0.03 &
              \multicolumn{1}{c|}{0.08} &
              0.00 &
              0.00 &
              0.16 \\
            \multicolumn{1}{c|}{GOOG} &
              \multicolumn{1}{c|}{150428} &
              0.06 &
              0.02 &
              \multicolumn{1}{c|}{0.32} &
              0.00 &
              0.01 &
              \multicolumn{1}{c|}{0.21} &
              0.00 &
              -0.00 &
              \multicolumn{1}{c|}{0.19} &
              0.00 &
              0.00 &
              0.01 \\
            \multicolumn{1}{c|}{GS} &
              \multicolumn{1}{c|}{39687} &
              0.06 &
              0.01 &
              \multicolumn{1}{c|}{0.29} &
              0.00 &
              0.00 &
              \multicolumn{1}{c|}{0.04} &
              0.00 &
              0.00 &
              \multicolumn{1}{c|}{0.03} &
              0.00 &
              0.00 &
              0.00 \\
            \multicolumn{1}{c|}{IBM} &
              \multicolumn{1}{c|}{45758} &
              0.03 &
              -0.01 &
              \multicolumn{1}{c|}{0.27} &
              0.00 &
              0.01 &
              \multicolumn{1}{c|}{0.21} &
              0.00 &
              0.00 &
              \multicolumn{1}{c|}{0.06} &
              0.00 &
              0.00 &
              0.00 \\
            \multicolumn{1}{c|}{MCD} &
              \multicolumn{1}{c|}{52440} &
              0.04 &
              0.01 &
              \multicolumn{1}{c|}{0.28} &
              0.00 &
              0.04 &
              \multicolumn{1}{c|}{0.21} &
              0.00 &
              0.00 &
              \multicolumn{1}{c|}{0.02} &
              0.00 &
              0.00 &
              0.00 \\
            \multicolumn{1}{c|}{NVDA} &
              \multicolumn{1}{c|}{70535} &
              0.03 &
              0.00 &
              \multicolumn{1}{c|}{0.31} &
              0.00 &
              0.01 &
              \multicolumn{1}{c|}{0.07} &
              0.00 &
              0.00 &
              \multicolumn{1}{c|}{0.00} &
              0.00 &
              0.00 &
              0.00 \\ \hline
            \multicolumn{1}{c|}{AAPL} &
              \multicolumn{1}{c|}{153620} &
              0.06 &
              0.06 &
              \multicolumn{1}{c|}{0.32} &
              0.00 &
              0.17 &
              \multicolumn{1}{c|}{0.46} &
              0.00 &
              0.00 &
              \multicolumn{1}{c|}{0.30} &
              0.00 &
              0.00 &
              0.00 \\
            \multicolumn{1}{c|}{ABBV} &
              \multicolumn{1}{c|}{26504} &
              0.05 &
              0.04 &
              \multicolumn{1}{c|}{0.30} &
              0.01 &
              0.06 &
              \multicolumn{1}{c|}{0.20} &
              0.00 &
              0.18 &
              \multicolumn{1}{c|}{0.29} &
              0.00 &
              0.00 &
              0.00 \\
            \multicolumn{1}{c|}{PM} &
              \multicolumn{1}{c|}{30333} &
              0.05 &
              0.01 &
              \multicolumn{1}{c|}{0.30} &
              0.00 &
              0.01 &
              \multicolumn{1}{c|}{0.16} &
              0.00 &
              0.00 &
              \multicolumn{1}{c|}{0.05} &
              0.00 &
              0.00 &
              0.00 \\ \hline
            \multicolumn{1}{c|}{BAC} &
              \multicolumn{1}{c|}{15693} &
              0.19 &
              0.35 &
              \multicolumn{1}{c|}{0.55} &
              0.19 &
              0.36 &
              \multicolumn{1}{c|}{0.55} &
              0.16 &
              0.46 &
              \multicolumn{1}{c|}{0.60} &
              0.09 &
              0.62 &
              0.70 \\
            \multicolumn{1}{c|}{CSCO} &
              \multicolumn{1}{c|}{39307} &
              0.16 &
              0.30 &
              \multicolumn{1}{c|}{0.53} &
              0.15 &
              0.33 &
              \multicolumn{1}{c|}{0.56} &
              0.13 &
              0.44 &
              \multicolumn{1}{c|}{0.61} &
              0.02 &
              0.56 &
              0.65 \\
            \multicolumn{1}{c|}{KO} &
              \multicolumn{1}{c|}{19119} &
              0.13 &
              0.25 &
              \multicolumn{1}{c|}{0.50} &
              0.12 &
              0.27 &
              \multicolumn{1}{c|}{0.51} &
              0.08 &
              0.37 &
              \multicolumn{1}{c|}{0.57} &
              0.02 &
              0.52 &
              0.62 \\
            \multicolumn{1}{c|}{ORCL} &
              \multicolumn{1}{c|}{31524} &
              0.17 &
              0.25 &
              \multicolumn{1}{c|}{0.48} &
              0.16 &
              0.30 &
              \multicolumn{1}{c|}{0.51} &
              0.11 &
              0.43 &
              \multicolumn{1}{c|}{0.55} &
              0.00 &
              0.25 &
              0.49 \\
            \multicolumn{1}{c|}{PFE} &
              \multicolumn{1}{c|}{19621} &
              0.13 &
              0.28 &
              \multicolumn{1}{c|}{0.50} &
              0.12 &
              0.31 &
              \multicolumn{1}{c|}{0.52} &
              0.08 &
              0.42 &
              \multicolumn{1}{c|}{0.59} &
              0.01 &
              0.55 &
              0.58 \\
            \multicolumn{1}{c|}{VZ} &
              \multicolumn{1}{c|}{37859} &
              0.09 &
              0.11 &
              \multicolumn{1}{c|}{0.31} &
              0.08 &
              0.13 &
              \multicolumn{1}{c|}{0.31} &
              0.03 &
              0.20 &
              \multicolumn{1}{c|}{0.35} &
              0.00 &
              0.31 &
              0.40 \\ \hline
        \end{tabular}
        \vspace{5pt}
        \captionsetup{justification=centering}
        \caption{}
        \label{tab:sub_table_c}
    \end{subtable}
\end{table}
}

Deepening the analysis at the the level of specific stocks, looking at Table \ref{tab:main_table_strategies}, we notice that, for the class of {\em small-tick stocks}, at $\text{H}10$, the average $\text{PT}$ is $1.63 \times 10^5$. Without the application of any threshold, looking at $p_{\text{T}}$, we observe a separation between stocks: the first set is made of CHTR, GOOG and GS and is characterized by an average $p_{\text{T}}$ equal to $0.06$, while the second one is made of IBM, MCD and NVDA and is characterized by an average $p_{\text{T}}$ equal to $0.12$. Such a separation, which was not evident in Figure \ref{fig:cross_years_average_mcc}, is directly linked to the microstructural properties of the considered stocks. Indeed, as observed in Figures \ref{fig:pdf_spreads_in_ticks_partial} and \ref{fig:pdf_relative_number_levels_partial}, assets belonging to the second family present less extreme realizations of the spread and of the actual LOB's depth, making them structurally more similar to large-tick stocks and more suitable to be treated as input for a deep learning model\,\footnote{These findings are also coherent with the ones observed in Table \ref{tab:information_richness}, where for stocks belonging to the second family, we observe an average (computed across years) IR value equal to $1.85$, which is higher than the one observed for the stocks belonging to the first family of stocks (i.e., $1.71$).}. We remark that the above-mentioned statistical properties of the LOB can be effectively exploited by the deep learning model thanks to the rough balancing in class distribution observed at $\text{H}10$ (see Table \ref{tab:average_test_classes}). As we point out later in this Section, moving to $\text{H}{\Delta \tau} \in \{50, 100\}$, this effect will vanish due to a stronger class imbalance.  For all the small-tick stocks, the decrease of $p_{\text{T}}$ is fast when applying probability thresholds. Specifically, net of minor oscillations, the probability of correctly executing a trade at a threshold larger than $0.5$ is zero. For \textit{medium-tick stocks}, the average $\text{PT}$ (i.e., $2.1 \times 10^5$) is strongly biased by the behaviour of AAPL. In contrast, $p_{\text{T}}$ (which has an average value equal to $0.14$) has similar realizations for all the stocks. Also in this case, the decrease in $p_{\text{T}}$ is fast when thresholds are applied, and the probability of correctly executing a transaction at a threshold larger than $0.5$ is $0$. The behaviour is different when we analyze \textit{large-tick stocks}. In this case, the average $\text{PT}$ equals $5.3 \times 10^4$, which is almost $1/3$ of the one detected for small-tick stocks. Even if the number of LOB updates is much higher for large-tick stocks than for small-tick stocks, the number of mid-price changes, and, consequently, the number of potentially exploitable transactions, follows an inverse pattern, being, on average, one order of magnitude higher for small-tick stocks than for large-tick stocks. Also the application of probability thresholds in large-tick stocks leads to different results. Indeed, without applying any threshold, the average $p_{\text{T}}$ value for this class of stocks equals $0.10$, with a smoother decrease for higher threshold values. In this case, the probability of correctly executing a transaction is $\approx 0$ only for a threshold equal to $0.9$. To deepen the $p_{\text{T}}$-related results discussed for $\text{H}10$, it is useful to exploit the average confusion matrices in Figure \ref{fig:confusion_matrix_H10} as an instrument to understand the distribution of forecasting errors. In this context, indeed, we observe that the non-negligible frequency of reciprocal misclassifications between the extreme classes ($-1$ and $1$) for models trained on small- and medium-tick stocks, directly determines a sub-optimal management of the opening/closing of existing or new positions. Conversely, for large-tick stocks, errors' concentration towards the misclassification of the two extreme classes as $0$ guarantees a reduced impact on the management of the opening/closing of existing or new positions. Moving to $\text{H}50$, for small-tick stocks, we observe a decrease in the average $\text{PT}$, which is equal to $9.1 \times 10^4$. Also the $p_{\text{T}}$, for all the probability thresholds, is conststenly lower than the one observed at $\text{H}10$. Indeed, without the application of any threshold (i.e. $0.3$), the average $p_{\text{T}}$ is equal to $0.04$, while the probability of correctly executing a transaction at a threshold larger than $ 0.5$ is always $0$. Differently from what is observed at $\text{H}10$, stocks have no intra-class separation. These findings are also true for medium-tick stocks. In this case, the average $\text{PT}$ is equal to $9.9 \times 10^4$, while the average $p_{\text{T}}$ is equal to $0.09$. One more time, a symmetrically different trend is observed for large-tick stocks. In this case, it remains true that the average number of potentially executable transactions (i.e. $3.6 \times 10^4$) decreases when compared to the one observed at $\text{H}10$, however, curiously, the average $p_{\text{T}}$ increases reaching a value of $0.15$. The analysis of confusion matrices (see Figure \ref{fig:confusion_matrix_H50}) reveals that, for small- and medium-tick stocks, the lowest realizations of $p_{\text{T}}$ compared to $\text{H}10$, are directly linked to a more evident tendency to mix the extreme classes ($-1$ and $1$) which directly determine the opening/closing of existing or new positions. Conversely, for large-tick stocks, a decrease of these types of errors in favour of a misclassification of extreme classes toward the central one (i.e., class $0$), determines an increase in realizations of $p_{\text{T}}$ compared to $\text{H}10$. Similar findings can be detected moving to $\text{H}100$. In this case, for small-tick stocks, we observe a further decrease in the average $\text{PT}$, which is equal to $6.7 \times 10^4$. Without applying any threshold, the average $p_{\text{T}}$ is in line with the one observed at $\text{H}50$, while the probability of correctly executing a transaction with the application of a threshold larger than $0.3$ is always $0$. Also in this case, there is no intra-class separation among stocks. For medium-tick stocks, the average $\text{PT}$ is equal to $7.0 \times 10^4$, while the average $p_{\text{T}}$ is lower than the one observed at $\text{H}50$, with a value equal to $0.05$. One more time, for large-tick stocks, we observe that even if it remains true that the average number of potentially executable transactions (i.e., $2.7 \times 10^4$) decreases when compared to the one observed at $\text{H}50$, the average $p_{\text{T}}$ remains unchanged with a value equal to $0.15$. In this case, the probability of correctly executing a transaction is remarkably higher than $0$ for probability thresholds $\leq 0.7$. In this case, results of the analysis of confusion matrices (see Figure \ref{fig:confusion_matrix_H100}) are identical to the ones performed at $\text{H}50$.

To conclude the analysis of the results, we draw the attention to the uniqueness of the pattern of $p_{\text{T}}$ observed for large-tick stocks across the different horizons. Microstructural properties alone cannot fully explain this behaviour. Instead, we must also consider the primary role of class distributions at $\text{H}{\Delta \tau} \in \{10, 50, 100\}$ in determining this trend. As we have previously noted in Tables \ref{tab:average_training_classes}, \ref{tab:average_validation_classes} and \ref{tab:average_test_classes}, class imbalances follow two symmetrically different patterns for small- and large-tick stocks. The first class of assets has a more balanced distribution at $\text{H}{10}$, while the second class of assets achieves a stable balance at $\text{H}{\Delta \tau} \in \{50, 100\}$. Overall, this result, combined with the balanced sampling technique used during the training stage, as well as the aggregate statistical properties of the LOB for different classes of stocks, sheds lights on the practicability of forecasts and issues related to the use of deep learning forecasting techniques on LOB data. 

\section{Conclusion and Future Work}\label{sec:Conclusion}
This paper addresses the issue of bridging Limit Order Book (LOB) microstructural analysis and forecasting. To achieve this goal, we first collect high-quality LOB data for a heterogeneous set of $15$ liquid stocks traded on the NASDAQ exchange over a period of $3$ years ($2017-2019$) and classify them based on the tick size. By doing so, we establish quantitative bounds that allow us to separate between small-tick, medium-tick, and large-tick stocks. On the microstructural side, we analyse several properties of the selected stocks, remarking that -- although new and increasingly refined measures have been introduced (e.g., the `information richness' ratio \cite{kolm2023deep}) -- the observed behavioural clusters can ultimately be entirely traced back to tick-size-driven effects. On the forecasting side, we introduce `LOBFrame', a novel, open-source software framework designed to offer a standardized protocol for processing large-scale LOB data. This framework embodies the latest scientific advancements, offering a unified system that encompasses an extensive pipeline for data transformation and processing, rapid training, validation, and testing phases, alongside with the assessment of model output quality via trading simulations. Its design not only encapsulates a comprehensive solution for this paper's research purposes, but also ensures easy integration of future models, highlighting its adaptability and readiness for forthcoming technological progress in the field. More specifically, in this work, we build upon a state-of-the-art model (i.e., DeepLOB) for multivariate time series forecasting explicitly designed to handle LOB dynamics, and we propose a labelling procedure that improves the model's usability in the development of high-frequency trading strategies. We also create a data-parsimonious pipeline that efficiently handles inherent data imbalances. The forecasting performances are assessed measuring the  Matthews Correlation Coefficient (MCC) across three different prediction horizons (expressed in the number of LOB updates) at different confidence levels (i.e., probability thresholds). Our results demonstrate that the chosen deep learning model performs significantly differently depending on the stocks' tick size. Specifically, we find that large-tick stocks are easier to predict, they are overall characterized by a stronger prediction signal, and their forecast performance is robust across prediction horizons. However, by examining the likelihood of observing a mid-price update over progressively shorter physical time intervals, we highlight that the utility of this signal is significantly dependent on the presence of low-latency hardware infrastructures.

Finally, going deeper with the study the practicability of obtained forecasts in real-world scenarios, we develop a strategy-oriented, assumptions-free and class imbalances-immune methodology to compute the probability of executing a correct transaction using the forecasts of the chosen model. We argue that this approach is more general than the one based on estimating the PnL of a single strategy on historical data, which is often based on unrealistic assumptions. We show that assessing the probability of executing a correct transaction is a more robust procedure compared to those used in traditional deep learning, as it correctly takes into account the impact of the chronological location of errors on the performance.

Our paper provides a robust methodology and a data pipeline that bridges the analysis and modelling of microstructural properties of LOB data with the forecast of LOB dynamics, providing specific indications to practitioners on the stocks characteristics and factors driving the forecast performance. Indeed, there are a number of research avenues yet to be explored. Specifically, a cross-exchange validation of our results is needed. In addition, we remark the need for structured testing of different deep learning models on heterogeneous classes of stocks. This analysis would aim to unveil how the models' architectural peculiarities can be exploited to handle specific challenges coming from, for example, the sparser LOB structure characterising small- to medium-tick stocks. This includes further studies on the potentialities of transformer models \cite{vaswani2017attention, zeng2023transformers, zhou2021informer, wen2022transformers}, diffusion models \cite{sohl2015deep, ho2020denoising, song2019generative, nichol2021improved} and graph-based models \cite{briola2023homological, wang2023homological, wang2022sparsification} in the application domain considered in this paper.

\section*{Acknowledgments}
The author, T.A., acknowledges the financial support from ESRC (ES/K002309/1), EPSRC (EP/P031730/1) and EC (H2020-ICT-2018-2 825215). The authors declare no conflict of interest. The funders had no role in the design of the study; in the collection, analyses, or interpretation of data; in the writing of the manuscript; or in the decision to publish the results. The author, A.B., acknowledges Dr Riccardo Marcaccioli and Giulio Bellini for the valuable feedback on the paper's content. 

\newpage

\bibliographystyle{unsrt}
\bibliography{references}

\pagebreak

\appendix

\section{Information Richness Ratio}\label{sec:Appendix_A}

{
\renewcommand{\arraystretch}{1.35}
\begin{table}[H]
\centering
\caption{In this Table, for each stock, we report the total number of LOB updates, the total number of price changes and the associated `information-richness' (IR) ratio.}
\label{tab:information_richness}
\scriptsize
\begin{tabular}{c|ccc|ccc|ccc}
\hline
\textbf{Ticker} & \multicolumn{3}{c|}{\textbf{2017}} & \multicolumn{3}{c|}{\textbf{2018}} & \multicolumn{3}{c}{\textbf{2019}} \\ \cline{2-10} 
\textbf{} &
  \textbf{\begin{tabular}[c]{@{}c@{}}LOB\\ Updates\end{tabular}} &
  \textbf{\begin{tabular}[c]{@{}c@{}}Price \\ Changes\end{tabular}} &
  \textbf{IR} &
  \textbf{\begin{tabular}[c]{@{}c@{}}LOB\\ Updates\end{tabular}} &
  \textbf{\begin{tabular}[c]{@{}c@{}}Price \\ Changes\end{tabular}} &
  \textbf{IR} &
  \textbf{\begin{tabular}[c]{@{}c@{}}LOB\\ Updates\end{tabular}} &
  \textbf{\begin{tabular}[c]{@{}c@{}}Price \\ Changes\end{tabular}} &
  \textbf{IR} \\ \hline
CHTR            & 1.25e+07    & 2.89e+06    & 1.46   & 1.84e+07    & 4.12e+06    & 1.49   & 1.52e+07    & 3.84e+06   & 1.38   \\
GOOG            & 3.39e+07    & 5.23e+06    & 1.87   & 1.37e+08    & 1.47e+07    & 2.24   & 1.10e+08    & 1.27e+07   & 2.16   \\
GS              & 1.84e+07    & 3.97e+06    & 1.54   & 3.55e+07    & 6.69e+06    & 1.67   & 2.79e+07    & 5.50e+06   & 1.62   \\
IBM             & 2.35e+07    & 2.84e+06    & 2.11   & 4.10e+07    & 6.17e+06    & 1.89   & 4.18e+07    & 5.13e+06   & 2.10   \\
MCD             & 2.44e+07    & 2.70e+06    & 2.20   & 3.48e+07    & 6.09e+06    & 1.74   & 2.69e+07    & 4.19e+06   & 1.86   \\
NVDA            & 7.31e+07    & 1.26e+07    & 1.76   & 9.05e+07    & 2.16e+07    & 1.43   & 8.37e+07    & 1.62e+07   & 1.64   \\ \hline
AAPL            & 1.91e+08    & 1.25e+07    & 2.72   & 2.22e+08    & 2.88e+07    & 2.04   & 2.30e+08    & 2.47e+07   & 2.23   \\
ABBV            & 3.53e+07    & 3.38e+06    & 2.35   & 3.40e+07    & 5.85e+06    & 1.76   & 5.03e+07    & 4.54e+06   & 2.40   \\
PM              & 3.40e+07    & 3.66e+06    & 2.23   & 3.88e+07    & 4.48e+06    & 2.16   & 3.96e+07    & 3.48e+06   & 2.43   \\ \hline
BAC             & 9.58e+07    & 5.91e+05    & 5.09   & 1.62e+08    & 1.26e+06    & 4.85   & 1.23e+08    & 7.06e+05   & 5.16   \\
CSCO            & 6.76e+07    & 4.29e+05    & 5.06   & 1.69e+08    & 2.59e+06    & 4.18   & 1.42e+08    & 1.80e+06   & 4.37   \\
KO              & 3.71e+07    & 4.59e+05    & 4.39   & 6.93e+07    & 1.53e+06    & 3.82   & 7.27e+07    & 1.13e+06   & 4.17   \\
ORCL            & 4.80e+07    & 9.23e+05    & 3.95   & 1.11e+08    & 2.86e+06    & 3.66   & 1.03e+08    & 1.96e+06   & 3.96   \\
PFE             & 4.40e+07    & 5.13e+05    & 4.45   & 9.68e+07    & 1.84e+06    & 3.96   & 9.70e+07    & 1.18e+06   & 4.41   \\
VZ              & 4.84e+07    & 1.18e+06    & 3.71   & 9.16e+07    & 3.19e+06    & 3.36   & 9.24e+07    & 1.81e+06   & 3.93   \\ \hline
\end{tabular}
\end{table}
}

An example of a derived microstructural property is the stocks' `information richness' (IR) ratio. This measure was introduced by Kolm et al. \cite{kolm2023deep}, and is defined as the logarithm of the ratio of the total number of LOB's updates to price changes (i.e., events occurring at the best levels of the LOB).

As one can notice from  Table \ref{tab:information_richness}, where we report for  for each stock (belonging to one of the three tick size classes) the total number of LOB updates, the total number of price changes and the associated `information-richness' (IR) ratio, an evident clustering behaviour is detectable based on the stocks' tick-size. 

As expected from results of analyses in Section \ref{sec:Microstructural_Priors}, small-tick stocks, inherently characterised by a lower number of updates, are exposed to a less granular and information-poor price discovery. In this case, the maximum IR value (i.e., $2.24$) is reached by GOOG during $2018$, while the minimum IR value (i.e., $1.38$) is reached by CHTR in $2019$. Medium-tick stocks are instead characterized by higher IR values with a maximum value of $2.72$ achieved by AAPL in $2017$ and a minimum value of $1.76$ for ABBV in $2018$. It is worth noting that, differently from previous analyses, a clear separation between AAPL and the other two stocks is not evident. Lastly, large-tick stocks are characterised by higher values of IR and, consequently, by a more granular and information-rich price discovery. The maximum realization (i.e., $5.16$) is detected for BAC during $2017$, while the minimum realization (i.e., $3.36$) is detected for VZ during $2018$.  

The analysis of the stocks' `information richness' needs further discussion. Indeed, in the original paper \cite{kolm2023deep}, the authors claim it is a measure for stocks' predictability; this is only partially true. As we empirically show here, there is a direct mapping between the `information-richness' of a stock and its tick-size; consequently, the tick-size itself could be used as a proxy measure of a stock's predictability. 

\pagebreak

\section{Statistical Significance of Traditional Machine Learning Metrics}\label{sec:Appendix_B}

{
\renewcommand{\arraystretch}{1.1}
\begin{table}[H]
\caption{Year-wise Matthews Correlation Coefficient (MCC) of the DeepLOB model at $\text{H}10$ for different confidence levels (i.e., probability thresholds). Statistical significance obtained through a parametric $t$-test is represented through asterisks. $p$-values $> 0.05$ are not marked. $p$-values $< 0.001$ are marked as $^{***}$. $0.001 \leq$ $p$-values $< 0.01$ are marked as $^{**}$. $0.01 \leq$ $p$-values $< 0.05$ are marked as $^{*}$. The `$/$' symbol is used when applying a probability thresholds leads to the absence of any remaining forecast.}
\label{tab:p_values_h10}
\setlength{\tabcolsep}{7pt}
\begin{tabular}{c|c|ccccccc}
\hline
\textbf{Ticker} &
  \textbf{Year} &
  \multicolumn{7}{c}{\textbf{H10}} \\ \cline{3-9} 
\textbf{} &
  \textbf{} &
  \multicolumn{1}{c|}{\textbf{0.3}} &
  \multicolumn{1}{c|}{\textbf{0.4}} &
  \multicolumn{1}{c|}{\textbf{0.5}} &
  \multicolumn{1}{c|}{\textbf{0.6}} &
  \multicolumn{1}{c|}{\textbf{0.7}} &
  \multicolumn{1}{c|}{\textbf{0.8}} &
  \textbf{0.9} \\ \hline
\multirow{3}{*}{CHTR} &
  2017 &
  \multicolumn{1}{c|}{$7.60\text{e-}02^{***}$} &
  \multicolumn{1}{c|}{$7.97\text{e-}02^{***}$} &
  \multicolumn{1}{c|}{$8.75\text{e-}02^{***}$} &
  \multicolumn{1}{c|}{$9.32\text{e-}02^{***}$} &
  \multicolumn{1}{c|}{$-1.88\text{e-}02   $} &
  \multicolumn{1}{c|}{/} &
  / \\
 &
  2018 &
  \multicolumn{1}{c|}{$1.07\text{e-}01^{***}$} &
  \multicolumn{1}{c|}{$1.13\text{e-}01^{***}$} &
  \multicolumn{1}{c|}{$2.13\text{e-}01^{***}$} &
  \multicolumn{1}{c|}{$1.48\text{e-}01^{***}$} &
  \multicolumn{1}{c|}{$0.00\text{e+}00   $} &
  \multicolumn{1}{c|}{$0.00\text{e+}00   $} &
  / \\
 &
  2019 &
  \multicolumn{1}{c|}{$1.05\text{e-}01^{***}$} &
  \multicolumn{1}{c|}{$1.26\text{e-}01^{***}$} &
  \multicolumn{1}{c|}{$1.65\text{e-}01^{***}$} &
  \multicolumn{1}{c|}{$2.30\text{e-}01^{***}$} &
  \multicolumn{1}{c|}{$1.73\text{e-}01^{***}$} &
  \multicolumn{1}{c|}{/} &
  / \\ \cline{2-9} 
\multirow{3}{*}{GOOG} &
  2017 &
  \multicolumn{1}{c|}{$1.40\text{e-}01^{***}$} &
  \multicolumn{1}{c|}{$1.62\text{e-}01^{***}$} &
  \multicolumn{1}{c|}{$2.46\text{e-}01^{***}$} &
  \multicolumn{1}{c|}{$3.30\text{e-}01^{***}$} &
  \multicolumn{1}{c|}{$5.53\text{e-}01^{***}$} &
  \multicolumn{1}{c|}{/} &
  / \\
 &
  2018 &
  \multicolumn{1}{c|}{$2.42\text{e-}01^{***}$} &
  \multicolumn{1}{c|}{$2.66\text{e-}01^{***}$} &
  \multicolumn{1}{c|}{$3.43\text{e-}01^{***}$} &
  \multicolumn{1}{c|}{$2.52\text{e-}01^{***}$} &
  \multicolumn{1}{c|}{$9.73\text{e-}02^{***}$} &
  \multicolumn{1}{c|}{$0.00\text{e+}00   $} &
  / \\
 &
  2019 &
  \multicolumn{1}{c|}{$2.01\text{e-}01^{***}$} &
  \multicolumn{1}{c|}{$2.28\text{e-}01^{***}$} &
  \multicolumn{1}{c|}{$3.06\text{e-}01^{***}$} &
  \multicolumn{1}{c|}{$3.69\text{e-}01^{***}$} &
  \multicolumn{1}{c|}{$1.05\text{e-}01^{***}$} &
  \multicolumn{1}{c|}{$0.00\text{e+}00   $} &
  / \\ \cline{2-9} 
\multirow{3}{*}{GS} &
  2017 &
  \multicolumn{1}{c|}{$1.87\text{e-}01^{***}$} &
  \multicolumn{1}{c|}{$2.13\text{e-}01^{***}$} &
  \multicolumn{1}{c|}{$3.17\text{e-}01^{***}$} &
  \multicolumn{1}{c|}{$3.52\text{e-}01^{***}$} &
  \multicolumn{1}{c|}{$1.24\text{e-}02   $} &
  \multicolumn{1}{c|}{$0.00\text{e+}00   $} &
  / \\
 &
  2018 &
  \multicolumn{1}{c|}{$5.80\text{e-}02^{***}$} &
  \multicolumn{1}{c|}{$6.25\text{e-}02^{***}$} &
  \multicolumn{1}{c|}{$1.31\text{e-}01^{***}$} &
  \multicolumn{1}{c|}{$2.22\text{e-}01^{***}$} &
  \multicolumn{1}{c|}{/} &
  \multicolumn{1}{c|}{/} &
  / \\
 &
  2019 &
  \multicolumn{1}{c|}{$5.13\text{e-}02^{***}$} &
  \multicolumn{1}{c|}{$4.89\text{e-}02^{***}$} &
  \multicolumn{1}{c|}{$8.72\text{e-}03^{***}$} &
  \multicolumn{1}{c|}{$4.34\text{e-}03   $} &
  \multicolumn{1}{c|}{$0.00\text{e+}00   $} &
  \multicolumn{1}{c|}{/} &
  / \\ \cline{2-9} 
\multirow{3}{*}{IBM} &
  2017 &
  \multicolumn{1}{c|}{$1.17\text{e-}01^{***}$} &
  \multicolumn{1}{c|}{$1.38\text{e-}01^{***}$} &
  \multicolumn{1}{c|}{$1.81\text{e-}01^{***}$} &
  \multicolumn{1}{c|}{$3.51\text{e-}02^{**} $} &
  \multicolumn{1}{c|}{$0.00\text{e+}00   $} &
  \multicolumn{1}{c|}{/} &
  / \\
 &
  2018 &
  \multicolumn{1}{c|}{$9.31\text{e-}02^{***}$} &
  \multicolumn{1}{c|}{$9.72\text{e-}02^{***}$} &
  \multicolumn{1}{c|}{$1.36\text{e-}01^{***}$} &
  \multicolumn{1}{c|}{$1.97\text{e-}01^{***}$} &
  \multicolumn{1}{c|}{$2.53\text{e-}01^{***}$} &
  \multicolumn{1}{c|}{$2.61\text{e-}01^{***}$} &
  / \\
 &
  2019 &
  \multicolumn{1}{c|}{$1.25\text{e-}01^{***}$} &
  \multicolumn{1}{c|}{$1.64\text{e-}01^{***}$} &
  \multicolumn{1}{c|}{$2.39\text{e-}01^{***}$} &
  \multicolumn{1}{c|}{$1.96\text{e-}01^{***}$} &
  \multicolumn{1}{c|}{$7.88\text{e-}02^{***}$} &
  \multicolumn{1}{c|}{$-3.32\text{e-}02   $} &
  / \\ \cline{2-9} 
\multirow{3}{*}{MCD} &
  2017 &
  \multicolumn{1}{c|}{$1.98\text{e-}01^{***}$} &
  \multicolumn{1}{c|}{$2.16\text{e-}01^{***}$} &
  \multicolumn{1}{c|}{$2.97\text{e-}01^{***}$} &
  \multicolumn{1}{c|}{$4.12\text{e-}01^{***}$} &
  \multicolumn{1}{c|}{$4.74\text{e-}01^{***}$} &
  \multicolumn{1}{c|}{$5.85\text{e-}01^{***}$} &
  / \\
 &
  2018 &
  \multicolumn{1}{c|}{$7.05\text{e-}02^{***}$} &
  \multicolumn{1}{c|}{$9.29\text{e-}02^{***}$} &
  \multicolumn{1}{c|}{$1.60\text{e-}01^{***}$} &
  \multicolumn{1}{c|}{$1.30\text{e-}01^{***}$} &
  \multicolumn{1}{c|}{$1.18\text{e-}01^{**} $} &
  \multicolumn{1}{c|}{/} &
  / \\
 &
  2019 &
  \multicolumn{1}{c|}{$8.84\text{e-}02^{***}$} &
  \multicolumn{1}{c|}{$1.04\text{e-}01^{***}$} &
  \multicolumn{1}{c|}{$1.54\text{e-}01^{***}$} &
  \multicolumn{1}{c|}{$1.27\text{e-}01^{**} $} &
  \multicolumn{1}{c|}{/} &
  \multicolumn{1}{c|}{/} &
  / \\ \cline{2-9} 
\multirow{3}{*}{NVDA} &
  2017 &
  \multicolumn{1}{c|}{$2.40\text{e-}02^{***}$} &
  \multicolumn{1}{c|}{$2.34\text{e-}02^{***}$} &
  \multicolumn{1}{c|}{$2.95\text{e-}02^{***}$} &
  \multicolumn{1}{c|}{$3.05\text{e-}02^{***}$} &
  \multicolumn{1}{c|}{$1.54\text{e-}01^{***}$} &
  \multicolumn{1}{c|}{$6.14\text{e-}02   $} &
  / \\
 &
  2018 &
  \multicolumn{1}{c|}{$1.16\text{e-}01^{***}$} &
  \multicolumn{1}{c|}{$1.20\text{e-}01^{***}$} &
  \multicolumn{1}{c|}{$2.33\text{e-}01^{***}$} &
  \multicolumn{1}{c|}{$3.98\text{e-}01^{***}$} &
  \multicolumn{1}{c|}{$3.83\text{e-}01^{***}$} &
  \multicolumn{1}{c|}{/} &
  / \\
 &
  2019 &
  \multicolumn{1}{c|}{$1.08\text{e-}01^{***}$} &
  \multicolumn{1}{c|}{$1.36\text{e-}01^{***}$} &
  \multicolumn{1}{c|}{$2.18\text{e-}01^{***}$} &
  \multicolumn{1}{c|}{$2.86\text{e-}01^{***}$} &
  \multicolumn{1}{c|}{$2.44\text{e-}01^{***}$} &
  \multicolumn{1}{c|}{$1.73\text{e-}01^{***}$} &
  $0.00\text{e+}00   $ \\ \hline
\multirow{3}{*}{AAPL} &
  2017 &
  \multicolumn{1}{c|}{$2.80\text{e-}01^{***}$} &
  \multicolumn{1}{c|}{$2.88\text{e-}01^{***}$} &
  \multicolumn{1}{c|}{$3.30\text{e-}01^{***}$} &
  \multicolumn{1}{c|}{$4.10\text{e-}01^{***}$} &
  \multicolumn{1}{c|}{$5.07\text{e-}01^{***}$} &
  \multicolumn{1}{c|}{$6.14\text{e-}01^{***}$} &
  $0.00\text{e+}00   $ \\
 &
  2018 &
  \multicolumn{1}{c|}{$5.92\text{e-}02^{***}$} &
  \multicolumn{1}{c|}{$7.34\text{e-}02^{***}$} &
  \multicolumn{1}{c|}{$1.35\text{e-}01^{***}$} &
  \multicolumn{1}{c|}{$2.20\text{e-}01^{***}$} &
  \multicolumn{1}{c|}{$1.59\text{e-}01^{***}$} &
  \multicolumn{1}{c|}{$1.75\text{e-}02   $} &
  $0.00\text{e+}00   $ \\
 &
  2019 &
  \multicolumn{1}{c|}{$1.82\text{e-}01^{***}$} &
  \multicolumn{1}{c|}{$1.93\text{e-}01^{***}$} &
  \multicolumn{1}{c|}{$2.59\text{e-}01^{***}$} &
  \multicolumn{1}{c|}{$3.28\text{e-}01^{***}$} &
  \multicolumn{1}{c|}{$3.87\text{e-}01^{***}$} &
  \multicolumn{1}{c|}{$1.41\text{e-}01^{***}$} &
  $0.00\text{e+}00   $ \\ \cline{2-9} 
\multirow{3}{*}{ABBV} &
  2017 &
  \multicolumn{1}{c|}{$1.66\text{e-}01^{***}$} &
  \multicolumn{1}{c|}{$1.74\text{e-}01^{***}$} &
  \multicolumn{1}{c|}{$2.66\text{e-}01^{***}$} &
  \multicolumn{1}{c|}{$3.11\text{e-}01^{***}$} &
  \multicolumn{1}{c|}{$3.40\text{e-}01^{***}$} &
  \multicolumn{1}{c|}{$3.64\text{e-}01^{***}$} &
  $6.90\text{e-}02^{***}$ \\
 &
  2018 &
  \multicolumn{1}{c|}{$8.02\text{e-}02^{***}$} &
  \multicolumn{1}{c|}{$9.28\text{e-}02^{***}$} &
  \multicolumn{1}{c|}{$1.84\text{e-}01^{***}$} &
  \multicolumn{1}{c|}{$2.12\text{e-}01^{***}$} &
  \multicolumn{1}{c|}{$1.56\text{e-}01^{**} $} &
  \multicolumn{1}{c|}{/} &
  / \\
 &
  2019 &
  \multicolumn{1}{c|}{$1.34\text{e-}01^{***}$} &
  \multicolumn{1}{c|}{$1.47\text{e-}01^{***}$} &
  \multicolumn{1}{c|}{$2.16\text{e-}01^{***}$} &
  \multicolumn{1}{c|}{$2.94\text{e-}01^{***}$} &
  \multicolumn{1}{c|}{$3.48\text{e-}01^{***}$} &
  \multicolumn{1}{c|}{$2.40\text{e-}01^{***}$} &
  / \\ \cline{2-9} 
\multirow{3}{*}{PM} &
  2017 &
  \multicolumn{1}{c|}{$1.60\text{e-}01^{***}$} &
  \multicolumn{1}{c|}{$1.78\text{e-}01^{***}$} &
  \multicolumn{1}{c|}{$2.61\text{e-}01^{***}$} &
  \multicolumn{1}{c|}{$3.71\text{e-}01^{***}$} &
  \multicolumn{1}{c|}{$4.34\text{e-}01^{***}$} &
  \multicolumn{1}{c|}{$4.80\text{e-}01^{***}$} &
  / \\
 &
  2018 &
  \multicolumn{1}{c|}{$7.73\text{e-}02^{***}$} &
  \multicolumn{1}{c|}{$8.37\text{e-}02^{***}$} &
  \multicolumn{1}{c|}{$1.10\text{e-}01^{***}$} &
  \multicolumn{1}{c|}{$1.14\text{e-}01^{***}$} &
  \multicolumn{1}{c|}{$9.59\text{e-}02^{***}$} &
  \multicolumn{1}{c|}{$3.90\text{e-}01^{**} $} &
  / \\
 &
  2019 &
  \multicolumn{1}{c|}{$1.19\text{e-}01^{***}$} &
  \multicolumn{1}{c|}{$1.27\text{e-}01^{***}$} &
  \multicolumn{1}{c|}{$1.65\text{e-}01^{***}$} &
  \multicolumn{1}{c|}{$2.18\text{e-}01^{***}$} &
  \multicolumn{1}{c|}{$2.61\text{e-}01^{***}$} &
  \multicolumn{1}{c|}{$3.57\text{e-}01^{***}$} &
  / \\ \hline
\multirow{3}{*}{BAC} &
  2017 &
  \multicolumn{1}{c|}{$2.85\text{e-}01^{***}$} &
  \multicolumn{1}{c|}{$2.86\text{e-}01^{***}$} &
  \multicolumn{1}{c|}{$2.93\text{e-}01^{***}$} &
  \multicolumn{1}{c|}{$3.30\text{e-}01^{***}$} &
  \multicolumn{1}{c|}{$3.78\text{e-}01^{***}$} &
  \multicolumn{1}{c|}{$4.38\text{e-}01^{***}$} &
  $4.53\text{e-}01^{***}$ \\
 &
  2018 &
  \multicolumn{1}{c|}{$3.33\text{e-}01^{***}$} &
  \multicolumn{1}{c|}{$3.33\text{e-}01^{***}$} &
  \multicolumn{1}{c|}{$3.37\text{e-}01^{***}$} &
  \multicolumn{1}{c|}{$3.65\text{e-}01^{***}$} &
  \multicolumn{1}{c|}{$3.93\text{e-}01^{***}$} &
  \multicolumn{1}{c|}{$4.26\text{e-}01^{***}$} &
  $4.65\text{e-}01^{***}$ \\
 &
  2019 &
  \multicolumn{1}{c|}{$2.86\text{e-}01^{***}$} &
  \multicolumn{1}{c|}{$2.86\text{e-}01^{***}$} &
  \multicolumn{1}{c|}{$2.88\text{e-}01^{***}$} &
  \multicolumn{1}{c|}{$3.17\text{e-}01^{***}$} &
  \multicolumn{1}{c|}{$3.46\text{e-}01^{***}$} &
  \multicolumn{1}{c|}{$3.64\text{e-}01^{***}$} &
  $3.48\text{e-}01^{***}$ \\ \cline{2-9} 
\multirow{3}{*}{CSCO} &
  2017 &
  \multicolumn{1}{c|}{$2.51\text{e-}01^{***}$} &
  \multicolumn{1}{c|}{$2.51\text{e-}01^{***}$} &
  \multicolumn{1}{c|}{$2.52\text{e-}01^{***}$} &
  \multicolumn{1}{c|}{$2.78\text{e-}01^{***}$} &
  \multicolumn{1}{c|}{$2.92\text{e-}01^{***}$} &
  \multicolumn{1}{c|}{$2.90\text{e-}01^{***}$} &
  $2.83\text{e-}01^{***}$ \\
 &
  2018 &
  \multicolumn{1}{c|}{$2.96\text{e-}01^{***}$} &
  \multicolumn{1}{c|}{$2.97\text{e-}01^{***}$} &
  \multicolumn{1}{c|}{$3.04\text{e-}01^{***}$} &
  \multicolumn{1}{c|}{$3.32\text{e-}01^{***}$} &
  \multicolumn{1}{c|}{$3.61\text{e-}01^{***}$} &
  \multicolumn{1}{c|}{$3.96\text{e-}01^{***}$} &
  $4.54\text{e-}01^{***}$ \\
 &
  2019 &
  \multicolumn{1}{c|}{$3.52\text{e-}01^{***}$} &
  \multicolumn{1}{c|}{$3.54\text{e-}01^{***}$} &
  \multicolumn{1}{c|}{$3.66\text{e-}01^{***}$} &
  \multicolumn{1}{c|}{$4.07\text{e-}01^{***}$} &
  \multicolumn{1}{c|}{$4.47\text{e-}01^{***}$} &
  \multicolumn{1}{c|}{$4.79\text{e-}01^{***}$} &
  $3.60\text{e-}01^{***}$ \\ \cline{2-9} 
\multirow{3}{*}{KO} &
  2017 &
  \multicolumn{1}{c|}{$2.37\text{e-}01^{***}$} &
  \multicolumn{1}{c|}{$2.37\text{e-}01^{***}$} &
  \multicolumn{1}{c|}{$2.42\text{e-}01^{***}$} &
  \multicolumn{1}{c|}{$2.68\text{e-}01^{***}$} &
  \multicolumn{1}{c|}{$2.98\text{e-}01^{***}$} &
  \multicolumn{1}{c|}{$3.42\text{e-}01^{***}$} &
  $3.83\text{e-}01^{***}$ \\
 &
  2018 &
  \multicolumn{1}{c|}{$2.93\text{e-}01^{***}$} &
  \multicolumn{1}{c|}{$2.94\text{e-}01^{***}$} &
  \multicolumn{1}{c|}{$2.99\text{e-}01^{***}$} &
  \multicolumn{1}{c|}{$3.16\text{e-}01^{***}$} &
  \multicolumn{1}{c|}{$3.29\text{e-}01^{***}$} &
  \multicolumn{1}{c|}{$3.28\text{e-}01^{***}$} &
  $2.55\text{e-}01^{***}$ \\
 &
  2019 &
  \multicolumn{1}{c|}{$3.10\text{e-}01^{***}$} &
  \multicolumn{1}{c|}{$3.11\text{e-}01^{***}$} &
  \multicolumn{1}{c|}{$3.20\text{e-}01^{***}$} &
  \multicolumn{1}{c|}{$3.55\text{e-}01^{***}$} &
  \multicolumn{1}{c|}{$3.86\text{e-}01^{***}$} &
  \multicolumn{1}{c|}{$4.04\text{e-}01^{***}$} &
  $1.73\text{e-}01^{***}$ \\ \cline{2-9} 
\multirow{3}{*}{ORCL} &
  2017 &
  \multicolumn{1}{c|}{$3.18\text{e-}01^{***}$} &
  \multicolumn{1}{c|}{$3.19\text{e-}01^{***}$} &
  \multicolumn{1}{c|}{$3.23\text{e-}01^{***}$} &
  \multicolumn{1}{c|}{$3.72\text{e-}01^{***}$} &
  \multicolumn{1}{c|}{$4.29\text{e-}01^{***}$} &
  \multicolumn{1}{c|}{$4.74\text{e-}01^{***}$} &
  $4.73\text{e-}01^{***}$ \\
 &
  2018 &
  \multicolumn{1}{c|}{$3.23\text{e-}01^{***}$} &
  \multicolumn{1}{c|}{$3.24\text{e-}01^{***}$} &
  \multicolumn{1}{c|}{$3.30\text{e-}01^{***}$} &
  \multicolumn{1}{c|}{$3.58\text{e-}01^{***}$} &
  \multicolumn{1}{c|}{$3.72\text{e-}01^{***}$} &
  \multicolumn{1}{c|}{$3.32\text{e-}01^{***}$} &
  $1.18\text{e-}01^{***}$ \\
 &
  2019 &
  \multicolumn{1}{c|}{$3.10\text{e-}01^{***}$} &
  \multicolumn{1}{c|}{$3.11\text{e-}01^{***}$} &
  \multicolumn{1}{c|}{$3.20\text{e-}01^{***}$} &
  \multicolumn{1}{c|}{$3.49\text{e-}01^{***}$} &
  \multicolumn{1}{c|}{$3.79\text{e-}01^{***}$} &
  \multicolumn{1}{c|}{$4.23\text{e-}01^{***}$} &
  $5.15\text{e-}01^{***}$ \\ \cline{2-9} 
\multirow{3}{*}{PF\text{E}} &
  2017 &
  \multicolumn{1}{c|}{$2.52\text{e-}01^{***}$} &
  \multicolumn{1}{c|}{$2.55\text{e-}01^{***}$} &
  \multicolumn{1}{c|}{$2.67\text{e-}01^{***}$} &
  \multicolumn{1}{c|}{$3.10\text{e-}01^{***}$} &
  \multicolumn{1}{c|}{$3.69\text{e-}01^{***}$} &
  \multicolumn{1}{c|}{$4.09\text{e-}01^{***}$} &
  $3.29\text{e-}01^{***}$ \\
 &
  2018 &
  \multicolumn{1}{c|}{$2.72\text{e-}01^{***}$} &
  \multicolumn{1}{c|}{$2.73\text{e-}01^{***}$} &
  \multicolumn{1}{c|}{$2.80\text{e-}01^{***}$} &
  \multicolumn{1}{c|}{$2.99\text{e-}01^{***}$} &
  \multicolumn{1}{c|}{$3.13\text{e-}01^{***}$} &
  \multicolumn{1}{c|}{$3.07\text{e-}01^{***}$} &
  $1.53\text{e-}01^{***}$ \\
 &
  2019 &
  \multicolumn{1}{c|}{$2.86\text{e-}01^{***}$} &
  \multicolumn{1}{c|}{$2.87\text{e-}01^{***}$} &
  \multicolumn{1}{c|}{$2.92\text{e-}01^{***}$} &
  \multicolumn{1}{c|}{$3.21\text{e-}01^{***}$} &
  \multicolumn{1}{c|}{$3.53\text{e-}01^{***}$} &
  \multicolumn{1}{c|}{$3.96\text{e-}01^{***}$} &
  $4.32\text{e-}01^{***}$ \\ \cline{2-9} 
\multirow{3}{*}{VZ} &
  2017 &
  \multicolumn{1}{c|}{$3.13\text{e-}01^{***}$} &
  \multicolumn{1}{c|}{$3.16\text{e-}01^{***}$} &
  \multicolumn{1}{c|}{$3.23\text{e-}01^{***}$} &
  \multicolumn{1}{c|}{$3.63\text{e-}01^{***}$} &
  \multicolumn{1}{c|}{$4.08\text{e-}01^{***}$} &
  \multicolumn{1}{c|}{$4.52\text{e-}01^{***}$} &
  $3.85\text{e-}01^{***}$ \\
 &
  2018 &
  \multicolumn{1}{c|}{$2.42\text{e-}01^{***}$} &
  \multicolumn{1}{c|}{$2.45\text{e-}01^{***}$} &
  \multicolumn{1}{c|}{$2.61\text{e-}01^{***}$} &
  \multicolumn{1}{c|}{$2.88\text{e-}01^{***}$} &
  \multicolumn{1}{c|}{$3.06\text{e-}01^{***}$} &
  \multicolumn{1}{c|}{$2.97\text{e-}01^{***}$} &
  $1.81\text{e-}01^{***}$ \\
 &
  2019 &
  \multicolumn{1}{c|}{$2.86\text{e-}01^{***}$} &
  \multicolumn{1}{c|}{$2.87\text{e-}01^{***}$} &
  \multicolumn{1}{c|}{$2.88\text{e-}01^{***}$} &
  \multicolumn{1}{c|}{$2.99\text{e-}01^{***}$} &
  \multicolumn{1}{c|}{$2.97\text{e-}01^{***}$} &
  \multicolumn{1}{c|}{$2.80\text{e-}01^{***}$} &
  $1.92\text{e-}01^{***}$ \\ \hline
\end{tabular}
\end{table}
}

{
\renewcommand{\arraystretch}{1.1}
\begin{table}[H]
\caption{Year-wise Matthews Correlation Coefficient (MCC) of the DeepLOB model at $\text{H}50$ for different confidence levels (i.e., probability thresholds). Statistical significance obtained through a parametric t-test is represented through asterisks. $p$-values $> 0.05$ are not marked. $p$-values $< 0.001$ are marked as $^{***}$. $0.001 \leq$ $p$-values $< 0.01$ are marked as $^{**}$. $0.01 \leq$ $p$-values $< 0.05$ are marked as $^{*}$. The `$/$' symbol is used when the application of a probability thresholds implies the absence of any remaining forecast.}
\label{tab:p_values_H50}
\setlength{\tabcolsep}{7pt}
\begin{tabular}{c|c|ccccccc}
\hline
\textbf{Ticker} &
  \textbf{Year} &
  \multicolumn{7}{c}{\textbf{H50}} \\ \cline{3-9} 
\textbf{} &
  \textbf{} &
  \multicolumn{1}{c|}{\textbf{0.3}} &
  \multicolumn{1}{c|}{\textbf{0.4}} &
  \multicolumn{1}{c|}{\textbf{0.5}} &
  \multicolumn{1}{c|}{\textbf{0.6}} &
  \multicolumn{1}{c|}{\textbf{0.7}} &
  \multicolumn{1}{c|}{\textbf{0.8}} &
  \textbf{0.9} \\ \hline
\multirow{3}{*}{CHTR} &
  2017 &
  \multicolumn{1}{c|}{$1.15\text{e-}01^{***}$} &
  \multicolumn{1}{c|}{$1.22\text{e-}01^{***}$} &
  \multicolumn{1}{c|}{$1.40\text{e-}01^{***}$} &
  \multicolumn{1}{c|}{$1.56\text{e-}01^{***}$} &
  \multicolumn{1}{c|}{$1.32\text{e-}01^{***}$} &
  \multicolumn{1}{c|}{$0.00\text{e+}00$} &
  $0.00\text{e+}00$ \\
 &
  2018 &
  \multicolumn{1}{c|}{$4.04\text{e-}02^{***}$} &
  \multicolumn{1}{c|}{$4.42\text{e-}02^{***}$} &
  \multicolumn{1}{c|}{$6.39\text{e-}02^{***}$} &
  \multicolumn{1}{c|}{$9.08\text{e-}02^{***}$} &
  \multicolumn{1}{c|}{$5.37\text{e-}02^{***}$} &
  \multicolumn{1}{c|}{$0.00\text{e+}00$} &
  $0.00\text{e+}00$ \\
 &
  2019 &
  \multicolumn{1}{c|}{$1.69\text{e-}02^{***}$} &
  \multicolumn{1}{c|}{$1.90\text{e-}02^{***}$} &
  \multicolumn{1}{c|}{$2.87\text{e-}02^{***}$} &
  \multicolumn{1}{c|}{$0.00\text{e+}00$} &
  \multicolumn{1}{c|}{$0.00\text{e+}00$} &
  \multicolumn{1}{c|}{$0.00\text{e+}00$} &
  $0.00\text{e+}00$ \\ \cline{2-9} 
\multirow{3}{*}{GOOG} &
  2017 &
  \multicolumn{1}{c|}{$2.02\text{e-}02^{***}$} &
  \multicolumn{1}{c|}{$2.36\text{e-}02^{***}$} &
  \multicolumn{1}{c|}{$4.45\text{e-}02^{***}$} &
  \multicolumn{1}{c|}{$6.33\text{e-}02^{***}$} &
  \multicolumn{1}{c|}{$4.95\text{e-}02^{***}$} &
  \multicolumn{1}{c|}{$3.74\text{e-}02$} &
  $0.00\text{e+}00$ \\
 &
  2018 &
  \multicolumn{1}{c|}{$1.34\text{e-}01^{***}$} &
  \multicolumn{1}{c|}{$1.44\text{e-}01^{***}$} &
  \multicolumn{1}{c|}{$1.48\text{e-}01^{***}$} &
  \multicolumn{1}{c|}{$1.86\text{e-}01^{***}$} &
  \multicolumn{1}{c|}{$0.00\text{e+}00$} &
  \multicolumn{1}{c|}{/} &
  / \\
 &
  2019 &
  \multicolumn{1}{c|}{$1.16\text{e-}01^{***}$} &
  \multicolumn{1}{c|}{$1.32\text{e-}01^{***}$} &
  \multicolumn{1}{c|}{$2.15\text{e-}01^{***}$} &
  \multicolumn{1}{c|}{$0.00\text{e+}00$} &
  \multicolumn{1}{c|}{/} &
  \multicolumn{1}{c|}{/} &
  / \\ \cline{2-9} 
\multirow{3}{*}{GS} &
  2017 &
  \multicolumn{1}{c|}{$6.44\text{e-}02^{***}$} &
  \multicolumn{1}{c|}{$7.38\text{e-}02^{***}$} &
  \multicolumn{1}{c|}{$8.36\text{e-}02^{***}$} &
  \multicolumn{1}{c|}{$3.82\text{e-}02^{***}$} &
  \multicolumn{1}{c|}{$5.81\text{e-}02^{***}$} &
  \multicolumn{1}{c|}{$0.00\text{e+}00$} &
  $0.00\text{e+}00$ \\
 &
  2018 &
  \multicolumn{1}{c|}{$1.80\text{e-}02^{***}$} &
  \multicolumn{1}{c|}{$2.72\text{e-}02^{***}$} &
  \multicolumn{1}{c|}{$0.00\text{e+}00$} &
  \multicolumn{1}{c|}{$0.00\text{e+}00$} &
  \multicolumn{1}{c|}{$0.00\text{e+}00$} &
  \multicolumn{1}{c|}{/} &
  / \\
 &
  2019 &
  \multicolumn{1}{c|}{$-2.80\text{e-}04$} &
  \multicolumn{1}{c|}{$4.74\text{e-}04$} &
  \multicolumn{1}{c|}{$-8.48\text{e-}04$} &
  \multicolumn{1}{c|}{$-2.21\text{e-}01$} &
  \multicolumn{1}{c|}{/} &
  \multicolumn{1}{c|}{/} &
  / \\ \cline{2-9} 
\multirow{3}{*}{IBM} &
  2017 &
  \multicolumn{1}{c|}{$3.47\text{e-}02^{***}$} &
  \multicolumn{1}{c|}{$3.50\text{e-}02^{***}$} &
  \multicolumn{1}{c|}{$3.54\text{e-}02^{***}$} &
  \multicolumn{1}{c|}{$3.16\text{e-}02^{***}$} &
  \multicolumn{1}{c|}{$0.00\text{e+}00$} &
  \multicolumn{1}{c|}{$0.00\text{e+}00$} &
  / \\
 &
  2018 &
  \multicolumn{1}{c|}{$4.73\text{e-}02^{***}$} &
  \multicolumn{1}{c|}{$4.71\text{e-}02^{***}$} &
  \multicolumn{1}{c|}{$1.32\text{e-}02^{***}$} &
  \multicolumn{1}{c|}{$2.50\text{e-}03^{**}$} &
  \multicolumn{1}{c|}{$1.17\text{e-}02$} &
  \multicolumn{1}{c|}{/} &
  / \\
 &
  2019 &
  \multicolumn{1}{c|}{$1.72\text{e-}02^{***}$} &
  \multicolumn{1}{c|}{$3.89\text{e-}02^{***}$} &
  \multicolumn{1}{c|}{$0.00\text{e+}00$} &
  \multicolumn{1}{c|}{/} &
  \multicolumn{1}{c|}{/} &
  \multicolumn{1}{c|}{/} &
  / \\ \cline{2-9} 
\multirow{3}{*}{MCD} &
  2017 &
  \multicolumn{1}{c|}{$9.19\text{e-}02^{***}$} &
  \multicolumn{1}{c|}{$1.03\text{e-}01^{***}$} &
  \multicolumn{1}{c|}{$1.47\text{e-}01^{***}$} &
  \multicolumn{1}{c|}{$1.57\text{e-}01^{***}$} &
  \multicolumn{1}{c|}{$1.57\text{e-}01^{***}$} &
  \multicolumn{1}{c|}{/} &
  / \\
 &
  2018 &
  \multicolumn{1}{c|}{$2.98\text{e-}02^{***}$} &
  \multicolumn{1}{c|}{$3.95\text{e-}02^{***}$} &
  \multicolumn{1}{c|}{$9.07\text{e-}02^{***}$} &
  \multicolumn{1}{c|}{$0.00\text{e+}00$} &
  \multicolumn{1}{c|}{/} &
  \multicolumn{1}{c|}{/} &
  / \\
 &
  2019 &
  \multicolumn{1}{c|}{$4.90\text{e-}03^{***}$} &
  \multicolumn{1}{c|}{$3.75\text{e-}03^{***}$} &
  \multicolumn{1}{c|}{$3.24\text{e-}03^{**}$} &
  \multicolumn{1}{c|}{$4.72\text{e-}02$} &
  \multicolumn{1}{c|}{$0.00\text{e+}00$} &
  \multicolumn{1}{c|}{$0.00\text{e+}00$} &
  / \\ \cline{2-9} 
\multirow{3}{*}{NVDA} &
  2017 &
  \multicolumn{1}{c|}{$-1.64\text{e-}03^{**}$} &
  \multicolumn{1}{c|}{$-9.20\text{e-}04$} &
  \multicolumn{1}{c|}{$0.00\text{e+}00$} &
  \multicolumn{1}{c|}{$0.00\text{e+}00$} &
  \multicolumn{1}{c|}{/} &
  \multicolumn{1}{c|}{/} &
  / \\
 &
  2018 &
  \multicolumn{1}{c|}{$1.16\text{e-}02^{***}$} &
  \multicolumn{1}{c|}{$1.87\text{e-}02^{***}$} &
  \multicolumn{1}{c|}{$0.00\text{e+}00$} &
  \multicolumn{1}{c|}{$0.00\text{e+}00$} &
  \multicolumn{1}{c|}{$0.00\text{e+}00$} &
  \multicolumn{1}{c|}{$0.00\text{e+}00$} &
  / \\
 &
  2019 &
  \multicolumn{1}{c|}{$9.21\text{e-}03^{***}$} &
  \multicolumn{1}{c|}{$1.39\text{e-}03$} &
  \multicolumn{1}{c|}{$0.00\text{e+}00$} &
  \multicolumn{1}{c|}{/} &
  \multicolumn{1}{c|}{/} &
  \multicolumn{1}{c|}{/} &
  / \\ \hline
\multirow{3}{*}{AAPL} &
  2017 &
  \multicolumn{1}{c|}{$1.97\text{e-}01^{***}$} &
  \multicolumn{1}{c|}{$2.28\text{e-}01^{***}$} &
  \multicolumn{1}{c|}{$3.15\text{e-}01^{***}$} &
  \multicolumn{1}{c|}{$4.16\text{e-}01^{***}$} &
  \multicolumn{1}{c|}{$5.09\text{e-}01^{***}$} &
  \multicolumn{1}{c|}{$6.09\text{e-}01^{***}$} &
  / \\
 &
  2018 &
  \multicolumn{1}{c|}{$1.05\text{e-}02^{***}$} &
  \multicolumn{1}{c|}{$1.31\text{e-}02^{***}$} &
  \multicolumn{1}{c|}{/} &
  \multicolumn{1}{c|}{/} &
  \multicolumn{1}{c|}{/} &
  \multicolumn{1}{c|}{/} &
  / \\
 &
  2019 &
  \multicolumn{1}{c|}{$1.12\text{e-}01^{***}$} &
  \multicolumn{1}{c|}{$1.18\text{e-}01^{***}$} &
  \multicolumn{1}{c|}{$2.34\text{e-}01^{***}$} &
  \multicolumn{1}{c|}{$3.65\text{e-}01^{***}$} &
  \multicolumn{1}{c|}{$7.96\text{e-}02^{***}$} &
  \multicolumn{1}{c|}{$0.00\text{e+}00$} &
  / \\ \cline{2-9} 
\multirow{3}{*}{ABBV} &
  2017 &
  \multicolumn{1}{c|}{$1.22\text{e-}01^{***}$} &
  \multicolumn{1}{c|}{$1.24\text{e-}01^{***}$} &
  \multicolumn{1}{c|}{$1.75\text{e-}01^{***}$} &
  \multicolumn{1}{c|}{$3.01\text{e-}01^{***}$} &
  \multicolumn{1}{c|}{$4.69\text{e-}01^{***}$} &
  \multicolumn{1}{c|}{$6.05\text{e-}01^{***}$} &
  $7.18\text{e-}01^{***}$ \\
 &
  2018 &
  \multicolumn{1}{c|}{$1.06\text{e-}02^{***}$} &
  \multicolumn{1}{c|}{$-2.10\text{e-}03$} &
  \multicolumn{1}{c|}{$0.00\text{e+}00$} &
  \multicolumn{1}{c|}{$0.00\text{e+}00$} &
  \multicolumn{1}{c|}{/} &
  \multicolumn{1}{c|}{/} &
  / \\
 &
  2019 &
  \multicolumn{1}{c|}{$9.77\text{e-}02^{***}$} &
  \multicolumn{1}{c|}{$1.02\text{e-}01^{***}$} &
  \multicolumn{1}{c|}{$1.89\text{e-}01^{***}$} &
  \multicolumn{1}{c|}{$1.40\text{e-}01^{***}$} &
  \multicolumn{1}{c|}{$1.10\text{e-}01^{***}$} &
  \multicolumn{1}{c|}{$0.00\text{e+}00$} &
  / \\ \cline{2-9} 
\multirow{3}{*}{PM} &
  2017 &
  \multicolumn{1}{c|}{$8.66\text{e-}02^{***}$} &
  \multicolumn{1}{c|}{$8.90\text{e-}02^{***}$} &
  \multicolumn{1}{c|}{$1.94\text{e-}01^{***}$} &
  \multicolumn{1}{c|}{$2.57\text{e-}01^{***}$} &
  \multicolumn{1}{c|}{$0.00\text{e+}00$} &
  \multicolumn{1}{c|}{/} &
  / \\
 &
  2018 &
  \multicolumn{1}{c|}{$5.71\text{e-}02^{***}$} &
  \multicolumn{1}{c|}{$6.06\text{e-}02^{***}$} &
  \multicolumn{1}{c|}{$9.23\text{e-}02^{***}$} &
  \multicolumn{1}{c|}{$3.85\text{e-}02^{***}$} &
  \multicolumn{1}{c|}{$3.81\text{e-}02^{**}$} &
  \multicolumn{1}{c|}{/} &
  / \\
 &
  2019 &
  \multicolumn{1}{c|}{$7.08\text{e-}02^{***}$} &
  \multicolumn{1}{c|}{$8.18\text{e-}02^{***}$} &
  \multicolumn{1}{c|}{$1.24\text{e-}01^{***}$} &
  \multicolumn{1}{c|}{$7.35\text{e-}02^{***}$} &
  \multicolumn{1}{c|}{/} &
  \multicolumn{1}{c|}{/} &
  / \\ \hline
\multirow{3}{*}{BAC} &
  2017 &
  \multicolumn{1}{c|}{$4.53\text{e-}01^{***}$} &
  \multicolumn{1}{c|}{$4.53\text{e-}01^{***}$} &
  \multicolumn{1}{c|}{$4.59\text{e-}01^{***}$} &
  \multicolumn{1}{c|}{$5.28\text{e-}01^{***}$} &
  \multicolumn{1}{c|}{$6.16\text{e-}01^{***}$} &
  \multicolumn{1}{c|}{$6.98\text{e-}01^{***}$} &
  $7.80\text{e-}01^{***}$ \\
 &
  2018 &
  \multicolumn{1}{c|}{$4.57\text{e-}01^{***}$} &
  \multicolumn{1}{c|}{$4.57\text{e-}01^{***}$} &
  \multicolumn{1}{c|}{$4.59\text{e-}01^{***}$} &
  \multicolumn{1}{c|}{$4.94\text{e-}01^{***}$} &
  \multicolumn{1}{c|}{$5.25\text{e-}01^{***}$} &
  \multicolumn{1}{c|}{$5.54\text{e-}01^{***}$} &
  $5.90\text{e-}01^{***}$ \\
 &
  2019 &
  \multicolumn{1}{c|}{$4.03\text{e-}01^{***}$} &
  \multicolumn{1}{c|}{$4.05\text{e-}01^{***}$} &
  \multicolumn{1}{c|}{$4.18\text{e-}01^{***}$} &
  \multicolumn{1}{c|}{$4.94\text{e-}01^{***}$} &
  \multicolumn{1}{c|}{$5.81\text{e-}01^{***}$} &
  \multicolumn{1}{c|}{$6.75\text{e-}01^{***}$} &
  $7.77\text{e-}01^{***}$ \\ \cline{2-9} 
\multirow{3}{*}{CSCO} &
  2017 &
  \multicolumn{1}{c|}{$3.86\text{e-}01^{***}$} &
  \multicolumn{1}{c|}{$3.86\text{e-}01^{***}$} &
  \multicolumn{1}{c|}{$3.87\text{e-}01^{***}$} &
  \multicolumn{1}{c|}{$4.16\text{e-}01^{***}$} &
  \multicolumn{1}{c|}{$4.53\text{e-}01^{***}$} &
  \multicolumn{1}{c|}{$5.22\text{e-}01^{***}$} &
  $6.57\text{e-}01^{***}$ \\
 &
  2018 &
  \multicolumn{1}{c|}{$3.58\text{e-}01^{***}$} &
  \multicolumn{1}{c|}{$3.61\text{e-}01^{***}$} &
  \multicolumn{1}{c|}{$3.85\text{e-}01^{***}$} &
  \multicolumn{1}{c|}{$4.32\text{e-}01^{***}$} &
  \multicolumn{1}{c|}{$4.86\text{e-}01^{***}$} &
  \multicolumn{1}{c|}{$5.52\text{e-}01^{***}$} &
  $6.49\text{e-}01^{***}$ \\
 &
  2019 &
  \multicolumn{1}{c|}{$3.69\text{e-}01^{***}$} &
  \multicolumn{1}{c|}{$3.75\text{e-}01^{***}$} &
  \multicolumn{1}{c|}{$4.12\text{e-}01^{***}$} &
  \multicolumn{1}{c|}{$5.00\text{e-}01^{***}$} &
  \multicolumn{1}{c|}{$6.07\text{e-}01^{***}$} &
  \multicolumn{1}{c|}{$6.89\text{e-}01^{***}$} &
  $5.83\text{e-}01^{***}$ \\ \cline{2-9} 
\multirow{3}{*}{KO} &
  2017 &
  \multicolumn{1}{c|}{$4.21\text{e-}01^{***}$} &
  \multicolumn{1}{c|}{$4.22\text{e-}01^{***}$} &
  \multicolumn{1}{c|}{$4.34\text{e-}01^{***}$} &
  \multicolumn{1}{c|}{$5.11\text{e-}01^{***}$} &
  \multicolumn{1}{c|}{$5.98\text{e-}01^{***}$} &
  \multicolumn{1}{c|}{$6.94\text{e-}01^{***}$} &
  $7.68\text{e-}01^{***}$ \\
 &
  2018 &
  \multicolumn{1}{c|}{$2.94\text{e-}01^{***}$} &
  \multicolumn{1}{c|}{$2.97\text{e-}01^{***}$} &
  \multicolumn{1}{c|}{$3.13\text{e-}01^{***}$} &
  \multicolumn{1}{c|}{$3.52\text{e-}01^{***}$} &
  \multicolumn{1}{c|}{$3.93\text{e-}01^{***}$} &
  \multicolumn{1}{c|}{$4.49\text{e-}01^{***}$} &
  $5.60\text{e-}01^{***}$ \\
 &
  2019 &
  \multicolumn{1}{c|}{$3.18\text{e-}01^{***}$} &
  \multicolumn{1}{c|}{$3.21\text{e-}01^{***}$} &
  \multicolumn{1}{c|}{$3.37\text{e-}01^{***}$} &
  \multicolumn{1}{c|}{$3.91\text{e-}01^{***}$} &
  \multicolumn{1}{c|}{$4.58\text{e-}01^{***}$} &
  \multicolumn{1}{c|}{$5.50\text{e-}01^{***}$} &
  $6.93\text{e-}01^{***}$ \\ \cline{2-9} 
\multirow{3}{*}{ORCL} &
  2017 &
  \multicolumn{1}{c|}{$4.04\text{e-}01^{***}$} &
  \multicolumn{1}{c|}{$4.07\text{e-}01^{***}$} &
  \multicolumn{1}{c|}{$4.34\text{e-}01^{***}$} &
  \multicolumn{1}{c|}{$5.10\text{e-}01^{***}$} &
  \multicolumn{1}{c|}{$6.00\text{e-}01^{***}$} &
  \multicolumn{1}{c|}{$6.99\text{e-}01^{***}$} &
  $7.92\text{e-}01^{***}$ \\
 &
  2018 &
  \multicolumn{1}{c|}{$3.15\text{e-}01^{***}$} &
  \multicolumn{1}{c|}{$3.16\text{e-}01^{***}$} &
  \multicolumn{1}{c|}{$3.29\text{e-}01^{***}$} &
  \multicolumn{1}{c|}{$3.80\text{e-}01^{***}$} &
  \multicolumn{1}{c|}{$4.67\text{e-}01^{***}$} &
  \multicolumn{1}{c|}{$6.04\text{e-}01^{***}$} &
  $6.40\text{e-}01^{***}$ \\
 &
  2019 &
  \multicolumn{1}{c|}{$3.17\text{e-}01^{***}$} &
  \multicolumn{1}{c|}{$3.19\text{e-}01^{***}$} &
  \multicolumn{1}{c|}{$3.39\text{e-}01^{***}$} &
  \multicolumn{1}{c|}{$4.02\text{e-}01^{***}$} &
  \multicolumn{1}{c|}{$4.77\text{e-}01^{***}$} &
  \multicolumn{1}{c|}{$5.28\text{e-}01^{***}$} &
  $5.10\text{e-}01^{***}$ \\ \cline{2-9} 
\multirow{3}{*}{PF\text{E}} &
  2017 &
  \multicolumn{1}{c|}{$4.12\text{e-}01^{***}$} &
  \multicolumn{1}{c|}{$4.13\text{e-}01^{***}$} &
  \multicolumn{1}{c|}{$4.26\text{e-}01^{***}$} &
  \multicolumn{1}{c|}{$5.09\text{e-}01^{***}$} &
  \multicolumn{1}{c|}{$6.21\text{e-}01^{***}$} &
  \multicolumn{1}{c|}{$7.50\text{e-}01^{***}$} &
  $8.44\text{e-}01^{***}$ \\
 &
  2018 &
  \multicolumn{1}{c|}{$3.07\text{e-}01^{***}$} &
  \multicolumn{1}{c|}{$3.10\text{e-}01^{***}$} &
  \multicolumn{1}{c|}{$3.40\text{e-}01^{***}$} &
  \multicolumn{1}{c|}{$4.24\text{e-}01^{***}$} &
  \multicolumn{1}{c|}{$5.11\text{e-}01^{***}$} &
  \multicolumn{1}{c|}{$5.82\text{e-}01^{***}$} &
  $6.46\text{e-}01^{***}$ \\
 &
  2019 &
  \multicolumn{1}{c|}{$4.13\text{e-}01^{***}$} &
  \multicolumn{1}{c|}{$4.17\text{e-}01^{***}$} &
  \multicolumn{1}{c|}{$4.42\text{e-}01^{***}$} &
  \multicolumn{1}{c|}{$5.38\text{e-}01^{***}$} &
  \multicolumn{1}{c|}{$6.51\text{e-}01^{***}$} &
  \multicolumn{1}{c|}{$7.56\text{e-}01^{***}$} &
  $7.33\text{e-}01^{***}$ \\ \cline{2-9} 
\multirow{3}{*}{VZ} &
  2017 &
  \multicolumn{1}{c|}{$3.27\text{e-}01^{***}$} &
  \multicolumn{1}{c|}{$3.33\text{e-}01^{***}$} &
  \multicolumn{1}{c|}{$3.71\text{e-}01^{***}$} &
  \multicolumn{1}{c|}{$4.26\text{e-}01^{***}$} &
  \multicolumn{1}{c|}{$4.78\text{e-}01^{***}$} &
  \multicolumn{1}{c|}{$5.36\text{e-}01^{***}$} &
  $6.30\text{e-}01^{***}$ \\
 &
  2018 &
  \multicolumn{1}{c|}{$2.06\text{e-}01^{***}$} &
  \multicolumn{1}{c|}{$2.11\text{e-}01^{***}$} &
  \multicolumn{1}{c|}{$2.46\text{e-}01^{***}$} &
  \multicolumn{1}{c|}{$3.00\text{e-}01^{***}$} &
  \multicolumn{1}{c|}{$3.68\text{e-}01^{***}$} &
  \multicolumn{1}{c|}{$4.76\text{e-}01^{***}$} &
  $4.48\text{e-}01^{***}$ \\
 &
  2019 &
  \multicolumn{1}{c|}{$3.07\text{e-}01^{***}$} &
  \multicolumn{1}{c|}{$3.09\text{e-}01^{***}$} &
  \multicolumn{1}{c|}{$3.32\text{e-}01^{***}$} &
  \multicolumn{1}{c|}{$3.90\text{e-}01^{***}$} &
  \multicolumn{1}{c|}{$4.59\text{e-}01^{***}$} &
  \multicolumn{1}{c|}{$5.39\text{e-}01^{***}$} &
  $5.83\text{e-}01^{***}$ \\ \hline
\end{tabular}
\end{table}
}

{
\renewcommand{\arraystretch}{1.1}
\begin{table}[H]
\caption{Year-wise Matthews Correlation Coefficient (MCC) of the DeepLOB model at $\text{H}100$ for different confidence levels (i.e., probability thresholds). Statistical significance obtained through a parametric t-test is represented through asterisks. $p$-values $> 0.05$ are not marked. $p$-values $< 0.001$ are marked as $^{***}$. $0.001 \leq$ $p$-values $< 0.01$ are marked as $^{**}$. $0.01 \leq$ $p$-values $< 0.05$ are marked as $^{*}$. The `$/$' symbol is used when the application of a probability thresholds implies the absence of any remaining forecast.}
\label{tab:p_values_H100}
\setlength{\tabcolsep}{5pt}
\begin{tabular}{c|c|ccccccc}
\hline
\textbf{Ticker} &
  \textbf{Year} &
  \multicolumn{7}{c}{\textbf{H100}} \\ \cline{3-9} 
\textbf{} &
  \textbf{} &
  \multicolumn{1}{c|}{\textbf{0.3}} &
  \multicolumn{1}{c|}{\textbf{0.4}} &
  \multicolumn{1}{c|}{\textbf{0.5}} &
  \multicolumn{1}{c|}{\textbf{0.6}} &
  \multicolumn{1}{c|}{\textbf{0.7}} &
  \multicolumn{1}{c|}{\textbf{0.8}} &
  \textbf{0.9} \\ \hline
\multirow{3}{*}{CHTR} &
  2017 &
  \multicolumn{1}{c|}{$4.66\text{e-}02^{***}$} &
  \multicolumn{1}{c|}{$7.13\text{e-}02^{***}$} &
  \multicolumn{1}{c|}{$7.76\text{e-}02^{***}$} &
  \multicolumn{1}{c|}{$0.00\text{e+}00$} &
  \multicolumn{1}{c|}{$0.00\text{e+}00$} &
  \multicolumn{1}{c|}{$0.00\text{e+}00$} &
  / \\
 &
  2018 &
  \multicolumn{1}{c|}{$3.08\text{e-}02^{***}$} &
  \multicolumn{1}{c|}{$3.25\text{e-}02^{***}$} &
  \multicolumn{1}{c|}{$4.73\text{e-}02^{***}$} &
  \multicolumn{1}{c|}{$7.03\text{e-}02^{***}$} &
  \multicolumn{1}{c|}{$7.49\text{e-}02^{***}$} &
  \multicolumn{1}{c|}{$0.00\text{e+}00$} &
  $0.00\text{e+}00$ \\
 &
  2019 &
  \multicolumn{1}{c|}{$3.58\text{e-}03^{***}$} &
  \multicolumn{1}{c|}{$-1.57\text{e-}03$} &
  \multicolumn{1}{c|}{$7.23\text{e-}03$} &
  \multicolumn{1}{c|}{$0.00\text{e+}00$} &
  \multicolumn{1}{c|}{$0.00\text{e+}00$} &
  \multicolumn{1}{c|}{$0.00\text{e+}00$} &
  / \\ \cline{2-9} 
\multirow{3}{*}{GOOG} &
  2017 &
  \multicolumn{1}{c|}{$1.25\text{e-}02^{***}$} &
  \multicolumn{1}{c|}{$1.28\text{e-}02^{***}$} &
  \multicolumn{1}{c|}{$8.49\text{e-}03^{***}$} &
  \multicolumn{1}{c|}{$3.16\text{e-}03$} &
  \multicolumn{1}{c|}{$-5.84\text{e-}03^{**}$} &
  \multicolumn{1}{c|}{$-3.88\text{e-}02^{***}$} &
  $0.00\text{e+}00$ \\
 &
  2018 &
  \multicolumn{1}{c|}{$7.10\text{e-}03^{***}$} &
  \multicolumn{1}{c|}{$6.52\text{e-}03^{***}$} &
  \multicolumn{1}{c|}{$0.00\text{e+}00$} &
  \multicolumn{1}{c|}{$0.00\text{e+}00$} &
  \multicolumn{1}{c|}{/} &
  \multicolumn{1}{c|}{/} &
  / \\
 &
  2019 &
  \multicolumn{1}{c|}{$2.39\text{e-}02^{***}$} &
  \multicolumn{1}{c|}{$2.20\text{e-}02^{***}$} &
  \multicolumn{1}{c|}{$1.56\text{e-}02^{***}$} &
  \multicolumn{1}{c|}{$0.00\text{e+}00$} &
  \multicolumn{1}{c|}{$0.00\text{e+}00$} &
  \multicolumn{1}{c|}{$0.00\text{e+}00$} &
  / \\ \cline{2-9} 
\multirow{3}{*}{GS} &
  2017 &
  \multicolumn{1}{c|}{$1.43\text{e-}02^{***}$} &
  \multicolumn{1}{c|}{$1.42\text{e-}02^{***}$} &
  \multicolumn{1}{c|}{$3.26\text{e-}02^{***}$} &
  \multicolumn{1}{c|}{$0.00\text{e+}00$} &
  \multicolumn{1}{c|}{$0.00\text{e+}00$} &
  \multicolumn{1}{c|}{$0.00\text{e+}00$} &
  $0.00\text{e+}00$ \\
 &
  2018 &
  \multicolumn{1}{c|}{$1.48\text{e-}03^{**}$} &
  \multicolumn{1}{c|}{$8.64\text{e-}03^{***}$} &
  \multicolumn{1}{c|}{$0.00\text{e+}00$} &
  \multicolumn{1}{c|}{$0.00\text{e+}00$} &
  \multicolumn{1}{c|}{$0.00\text{e+}00$} &
  \multicolumn{1}{c|}{$0.00\text{e+}00$} &
  / \\
 &
  2019 &
  \multicolumn{1}{c|}{$-6.92\text{e-}03^{***}$} &
  \multicolumn{1}{c|}{$-7.32\text{e-}03^{***}$} &
  \multicolumn{1}{c|}{$0.00\text{e+}00$} &
  \multicolumn{1}{c|}{$0.00\text{e+}00$} &
  \multicolumn{1}{c|}{$0.00\text{e+}00$} &
  \multicolumn{1}{c|}{$0.00\text{e+}00$} &
  / \\ \cline{2-9} 
\multirow{3}{*}{IBM} &
  2017 &
  \multicolumn{1}{c|}{$3.78\text{e-}03^{***}$} &
  \multicolumn{1}{c|}{$4.26\text{e-}03^{***}$} &
  \multicolumn{1}{c|}{$1.81\text{e-}02^{***}$} &
  \multicolumn{1}{c|}{$0.00\text{e+}00$} &
  \multicolumn{1}{c|}{$0.00\text{e+}00$} &
  \multicolumn{1}{c|}{$0.00\text{e+}00$} &
  $0.00\text{e+}00$ \\
 &
  2018 &
  \multicolumn{1}{c|}{$4.20\text{e-}03^{***}$} &
  \multicolumn{1}{c|}{$4.35\text{e-}02^{***}$} &
  \multicolumn{1}{c|}{$-7.08\text{e-}03$} &
  \multicolumn{1}{c|}{$0.00\text{e+}00$} &
  \multicolumn{1}{c|}{/} &
  \multicolumn{1}{c|}{/} &
  / \\
 &
  2019 &
  \multicolumn{1}{c|}{$1.46\text{e-}03$} &
  \multicolumn{1}{c|}{$1.25\text{e-}02^{***}$} &
  \multicolumn{1}{c|}{$1.19\text{e-}02^{***}$} &
  \multicolumn{1}{c|}{$0.00\text{e+}00$} &
  \multicolumn{1}{c|}{$0.00\text{e+}00$} &
  \multicolumn{1}{c|}{/} &
  / \\ \cline{2-9} 
\multirow{3}{*}{MCD} &
  2017 &
  \multicolumn{1}{c|}{$2.85\text{e-}02^{***}$} &
  \multicolumn{1}{c|}{$3.70\text{e-}02^{***}$} &
  \multicolumn{1}{c|}{$9.68\text{e-}02^{***}$} &
  \multicolumn{1}{c|}{$1.03\text{e-}02$} &
  \multicolumn{1}{c|}{$0.00\text{e+}00$} &
  \multicolumn{1}{c|}{/} &
  / \\
 &
  2018 &
  \multicolumn{1}{c|}{$8.15\text{e-}03^{***}$} &
  \multicolumn{1}{c|}{$9.12\text{e-}03^{***}$} &
  \multicolumn{1}{c|}{$6.63\text{e-}03$} &
  \multicolumn{1}{c|}{$0.00\text{e+}00$} &
  \multicolumn{1}{c|}{$0.00\text{e+}00$} &
  \multicolumn{1}{c|}{$0.00\text{e+}00$} &
  / \\
 &
  2019 &
  \multicolumn{1}{c|}{$1.81\text{e-}03^{**}$} &
  \multicolumn{1}{c|}{$6.88\text{e-}03^{***}$} &
  \multicolumn{1}{c|}{$5.37\text{e-}03$} &
  \multicolumn{1}{c|}{$0.00\text{e+}00$} &
  \multicolumn{1}{c|}{$0.00\text{e+}00$} &
  \multicolumn{1}{c|}{/} &
  / \\ \cline{2-9} 
\multirow{3}{*}{NVDA} &
  2017 &
  \multicolumn{1}{c|}{$4.80\text{e-}03^{***}$} &
  \multicolumn{1}{c|}{$2.14\text{e-}03^{**}$} &
  \multicolumn{1}{c|}{$0.00\text{e+}00$} &
  \multicolumn{1}{c|}{/} &
  \multicolumn{1}{c|}{/} &
  \multicolumn{1}{c|}{/} &
  / \\
 &
  2018 &
  \multicolumn{1}{c|}{$4.84\text{e-}03^{***}$} &
  \multicolumn{1}{c|}{$1.42\text{e-}03$} &
  \multicolumn{1}{c|}{$7.53\text{e-}03$} &
  \multicolumn{1}{c|}{$0.00\text{e+}00$} &
  \multicolumn{1}{c|}{/} &
  \multicolumn{1}{c|}{/} &
  / \\
 &
  2019 &
  \multicolumn{1}{c|}{$3.98\text{e-}03^{***}$} &
  \multicolumn{1}{c|}{$-3.50\text{e-}03^{***}$} &
  \multicolumn{1}{c|}{$0.00\text{e+}00$} &
  \multicolumn{1}{c|}{/} &
  \multicolumn{1}{c|}{/} &
  \multicolumn{1}{c|}{/} &
  / \\ \hline
\multirow{3}{*}{AAPL} &
  2017 &
  \multicolumn{1}{c|}{$1.36\text{e-}01^{***}$} &
  \multicolumn{1}{c|}{$1.84\text{e-}01^{***}$} &
  \multicolumn{1}{c|}{$3.21\text{e-}01^{***}$} &
  \multicolumn{1}{c|}{$9.54\text{e-}02^{***}$} &
  \multicolumn{1}{c|}{$0.00\text{e+}00$} &
  \multicolumn{1}{c|}{/} &
  / \\
 &
  2018 &
  \multicolumn{1}{c|}{$7.56\text{e-}03^{***}$} &
  \multicolumn{1}{c|}{$1.17\text{e-}02^{***}$} &
  \multicolumn{1}{c|}{/} &
  \multicolumn{1}{c|}{/} &
  \multicolumn{1}{c|}{/} &
  \multicolumn{1}{c|}{/} &
  / \\
 &
  2019 &
  \multicolumn{1}{c|}{$6.38\text{e-}02^{***}$} &
  \multicolumn{1}{c|}{$6.84\text{e-}02^{***}$} &
  \multicolumn{1}{c|}{$1.67\text{e-}01^{***}$} &
  \multicolumn{1}{c|}{$0.00\text{e+}00$} &
  \multicolumn{1}{c|}{/} &
  \multicolumn{1}{c|}{/} &
  / \\ \cline{2-9} 
\multirow{3}{*}{ABBV} &
  2017 &
  \multicolumn{1}{c|}{$7.07\text{e-}02^{***}$} &
  \multicolumn{1}{c|}{$7.59\text{e-}02^{***}$} &
  \multicolumn{1}{c|}{$1.81\text{e-}01^{***}$} &
  \multicolumn{1}{c|}{$3.53\text{e-}01^{***}$} &
  \multicolumn{1}{c|}{$4.59\text{e-}01^{***}$} &
  \multicolumn{1}{c|}{$0.00\text{e+}00$} &
  / \\
 &
  2018 &
  \multicolumn{1}{c|}{$9.28\text{e-}03^{***}$} &
  \multicolumn{1}{c|}{$1.38\text{e-}02^{***}$} &
  \multicolumn{1}{c|}{$7.35\text{e-}03$} &
  \multicolumn{1}{c|}{$0.00\text{e+}00$} &
  \multicolumn{1}{c|}{$0.00\text{e+}00$} &
  \multicolumn{1}{c|}{/} &
  / \\
 &
  2019 &
  \multicolumn{1}{c|}{$1.50\text{e-}02^{***}$} &
  \multicolumn{1}{c|}{$2.29\text{e-}02^{***}$} &
  \multicolumn{1}{c|}{$4.74\text{e-}02^{***}$} &
  \multicolumn{1}{c|}{/} &
  \multicolumn{1}{c|}{/} &
  \multicolumn{1}{c|}{/} &
  / \\ \cline{2-9} 
\multirow{3}{*}{PM} &
  2017 &
  \multicolumn{1}{c|}{$1.94\text{e-}02^{***}$} &
  \multicolumn{1}{c|}{$1.88\text{e-}02^{***}$} &
  \multicolumn{1}{c|}{$3.29\text{e-}02^{***}$} &
  \multicolumn{1}{c|}{$0.00\text{e+}00$} &
  \multicolumn{1}{c|}{$0.00\text{e+}00$} &
  \multicolumn{1}{c|}{/} &
  / \\
 &
  2018 &
  \multicolumn{1}{c|}{$3.15\text{e-}03^{***}$} &
  \multicolumn{1}{c|}{$2.49\text{e-}03^{***}$} &
  \multicolumn{1}{c|}{$3.40\text{e-}03$} &
  \multicolumn{1}{c|}{$0.00\text{e+}00$} &
  \multicolumn{1}{c|}{$0.00\text{e+}00$} &
  \multicolumn{1}{c|}{$0.00\text{e+}00$} &
  / \\
 &
  2019 &
  \multicolumn{1}{c|}{$1.32\text{e-}03$} &
  \multicolumn{1}{c|}{$8.93\text{e-}04$} &
  \multicolumn{1}{c|}{$7.86\text{e-}02^{***}$} &
  \multicolumn{1}{c|}{$0.00\text{e+}00$} &
  \multicolumn{1}{c|}{/} &
  \multicolumn{1}{c|}{/} &
  / \\ \hline
\multirow{3}{*}{BAC} &
  2017 &
  \multicolumn{1}{c|}{$4.48\text{e-}01^{***}$} &
  \multicolumn{1}{c|}{$4.48\text{e-}01^{***}$} &
  \multicolumn{1}{c|}{$4.66\text{e-}01^{***}$} &
  \multicolumn{1}{c|}{$5.60\text{e-}01^{***}$} &
  \multicolumn{1}{c|}{$6.74\text{e-}01^{***}$} &
  \multicolumn{1}{c|}{$8.04\text{e-}01^{***}$} &
  $8.83\text{e-}01^{***}$ \\
 &
  2018 &
  \multicolumn{1}{c|}{$3.87\text{e-}01^{***}$} &
  \multicolumn{1}{c|}{$3.87\text{e-}01^{***}$} &
  \multicolumn{1}{c|}{$3.91\text{e-}01^{***}$} &
  \multicolumn{1}{c|}{$4.27\text{e-}01^{***}$} &
  \multicolumn{1}{c|}{$4.64\text{e-}01^{***}$} &
  \multicolumn{1}{c|}{$5.10\text{e-}01^{***}$} &
  $5.99\text{e-}01^{***}$ \\
 &
  2019 &
  \multicolumn{1}{c|}{$2.62\text{e-}01^{***}$} &
  \multicolumn{1}{c|}{$2.63\text{e-}01^{***}$} &
  \multicolumn{1}{c|}{$2.69\text{e-}01^{***}$} &
  \multicolumn{1}{c|}{$2.90\text{e-}01^{***}$} &
  \multicolumn{1}{c|}{$3.27\text{e-}01^{***}$} &
  \multicolumn{1}{c|}{$3.90\text{e-}01^{***}$} &
  $5.10\text{e-}01^{***}$ \\ \cline{2-9} 
\multirow{3}{*}{CSCO} &
  2017 &
  \multicolumn{1}{c|}{$3.30\text{e-}01^{***}$} &
  \multicolumn{1}{c|}{$3.30\text{e-}01^{***}$} &
  \multicolumn{1}{c|}{$3.33\text{e-}01^{***}$} &
  \multicolumn{1}{c|}{$3.94\text{e-}01^{***}$} &
  \multicolumn{1}{c|}{$4.66\text{e-}01^{***}$} &
  \multicolumn{1}{c|}{$5.89\text{e-}01^{***}$} &
  $8.06\text{e-}01^{***}$ \\
 &
  2018 &
  \multicolumn{1}{c|}{$2.30\text{e-}01^{***}$} &
  \multicolumn{1}{c|}{$2.36\text{e-}01^{***}$} &
  \multicolumn{1}{c|}{$2.66\text{e-}01^{***}$} &
  \multicolumn{1}{c|}{$3.05\text{e-}01^{***}$} &
  \multicolumn{1}{c|}{$3.52\text{e-}01^{***}$} &
  \multicolumn{1}{c|}{$4.29\text{e-}01^{***}$} &
  $5.43\text{e-}01^{***}$ \\
 &
  2019 &
  \multicolumn{1}{c|}{$2.78\text{e-}01^{***}$} &
  \multicolumn{1}{c|}{$2.95\text{e-}01^{***}$} &
  \multicolumn{1}{c|}{$3.65\text{e-}01^{***}$} &
  \multicolumn{1}{c|}{$4.36\text{e-}01^{***}$} &
  \multicolumn{1}{c|}{$4.68\text{e-}01^{***}$} &
  \multicolumn{1}{c|}{$5.07\text{e-}01^{***}$} &
  $2.49\text{e-}01^{***}$ \\ \cline{2-9} 
\multirow{3}{*}{KO} &
  2017 &
  \multicolumn{1}{c|}{$3.36\text{e-}01^{***}$} &
  \multicolumn{1}{c|}{$3.38\text{e-}01^{***}$} &
  \multicolumn{1}{c|}{$3.53\text{e-}01^{***}$} &
  \multicolumn{1}{c|}{$4.00\text{e-}01^{***}$} &
  \multicolumn{1}{c|}{$4.58\text{e-}01^{***}$} &
  \multicolumn{1}{c|}{$5.38\text{e-}01^{***}$} &
  $6.44\text{e-}01^{***}$ \\
 &
  2018 &
  \multicolumn{1}{c|}{$2.25\text{e-}01^{***}$} &
  \multicolumn{1}{c|}{$2.31\text{e-}01^{***}$} &
  \multicolumn{1}{c|}{$2.60\text{e-}01^{***}$} &
  \multicolumn{1}{c|}{$3.06\text{e-}01^{***}$} &
  \multicolumn{1}{c|}{$3.56\text{e-}01^{***}$} &
  \multicolumn{1}{c|}{$4.13\text{e-}01^{***}$} &
  $4.97\text{e-}01^{***}$ \\
 &
  2019 &
  \multicolumn{1}{c|}{$2.24\text{e-}01^{***}$} &
  \multicolumn{1}{c|}{$2.29\text{e-}01^{***}$} &
  \multicolumn{1}{c|}{$2.55\text{e-}01^{***}$} &
  \multicolumn{1}{c|}{$3.02\text{e-}01^{***}$} &
  \multicolumn{1}{c|}{$3.61\text{e-}01^{***}$} &
  \multicolumn{1}{c|}{$4.46\text{e-}01^{***}$} &
  $4.91\text{e-}01^{***}$ \\ \cline{2-9} 
\multirow{3}{*}{ORCL} &
  2017 &
  \multicolumn{1}{c|}{$3.39\text{e-}01^{***}$} &
  \multicolumn{1}{c|}{$3.49\text{e-}01^{***}$} &
  \multicolumn{1}{c|}{$4.03\text{e-}01^{***}$} &
  \multicolumn{1}{c|}{$4.75\text{e-}01^{***}$} &
  \multicolumn{1}{c|}{$5.44\text{e-}01^{***}$} &
  \multicolumn{1}{c|}{$6.21\text{e-}01^{***}$} &
  $6.97\text{e-}01^{***}$ \\
 &
  2018 &
  \multicolumn{1}{c|}{$1.73\text{e-}01^{***}$} &
  \multicolumn{1}{c|}{$1.76\text{e-}01^{***}$} &
  \multicolumn{1}{c|}{$1.92\text{e-}01^{***}$} &
  \multicolumn{1}{c|}{$2.27\text{e-}01^{***}$} &
  \multicolumn{1}{c|}{$2.85\text{e-}01^{***}$} &
  \multicolumn{1}{c|}{$3.66\text{e-}01^{***}$} &
  $8.05\text{e-}02^{***}$ \\
 &
  2019 &
  \multicolumn{1}{c|}{$2.38\text{e-}01^{***}$} &
  \multicolumn{1}{c|}{$2.48\text{e-}01^{***}$} &
  \multicolumn{1}{c|}{$3.01\text{e-}01^{***}$} &
  \multicolumn{1}{c|}{$3.55\text{e-}01^{***}$} &
  \multicolumn{1}{c|}{$3.94\text{e-}01^{***}$} &
  \multicolumn{1}{c|}{$4.86\text{e-}01^{***}$} &
  / \\ \cline{2-9} 
\multirow{3}{*}{PF\text{E}} &
  2017 &
  \multicolumn{1}{c|}{$3.24\text{e-}01^{***}$} &
  \multicolumn{1}{c|}{$3.27\text{e-}01^{***}$} &
  \multicolumn{1}{c|}{$3.55\text{e-}01^{***}$} &
  \multicolumn{1}{c|}{$4.19\text{e-}01^{***}$} &
  \multicolumn{1}{c|}{$4.97\text{e-}01^{***}$} &
  \multicolumn{1}{c|}{$6.17\text{e-}01^{***}$} &
  $7.76\text{e-}01^{***}$ \\
 &
  2018 &
  \multicolumn{1}{c|}{$1.79\text{e-}01^{***}$} &
  \multicolumn{1}{c|}{$1.82\text{e-}01^{***}$} &
  \multicolumn{1}{c|}{$2.07\text{e-}01^{***}$} &
  \multicolumn{1}{c|}{$2.51\text{e-}01^{***}$} &
  \multicolumn{1}{c|}{$2.90\text{e-}01^{***}$} &
  \multicolumn{1}{c|}{$2.69\text{e-}01^{***}$} &
  $2.09\text{e-}01^{***}$ \\
 &
  2019 &
  \multicolumn{1}{c|}{$3.32\text{e-}01^{***}$} &
  \multicolumn{1}{c|}{$3.39\text{e-}01^{***}$} &
  \multicolumn{1}{c|}{$3.77\text{e-}01^{***}$} &
  \multicolumn{1}{c|}{$4.47\text{e-}01^{***}$} &
  \multicolumn{1}{c|}{$5.21\text{e-}01^{***}$} &
  \multicolumn{1}{c|}{$5.55\text{e-}01^{***}$} &
  $6.62\text{e-}01^{***}$ \\ \cline{2-9} 
\multirow{3}{*}{VZ} &
  2017 &
  \multicolumn{1}{c|}{$7.78\text{e-}02^{***}$} &
  \multicolumn{1}{c|}{$7.80\text{e-}02^{***}$} &
  \multicolumn{1}{c|}{$8.10\text{e-}02^{***}$} &
  \multicolumn{1}{c|}{$1.06\text{e-}01^{***}$} &
  \multicolumn{1}{c|}{$1.47\text{e-}01^{***}$} &
  \multicolumn{1}{c|}{$1.82\text{e-}01^{***}$} &
  $9.13\text{e-}02^{***}$ \\
 &
  2018 &
  \multicolumn{1}{c|}{$1.03\text{e-}01^{***}$} &
  \multicolumn{1}{c|}{$1.05\text{e-}01^{***}$} &
  \multicolumn{1}{c|}{$1.14\text{e-}01^{***}$} &
  \multicolumn{1}{c|}{$1.23\text{e-}01^{***}$} &
  \multicolumn{1}{c|}{$1.79\text{e-}01^{***}$} &
  \multicolumn{1}{c|}{$3.51\text{e-}01^{***}$} &
  $3.54\text{e-}01^{***}$ \\
 &
  2019 &
  \multicolumn{1}{c|}{$1.87\text{e-}01^{***}$} &
  \multicolumn{1}{c|}{$1.91\text{e-}01^{***}$} &
  \multicolumn{1}{c|}{$2.17\text{e-}01^{***}$} &
  \multicolumn{1}{c|}{$2.61\text{e-}01^{***}$} &
  \multicolumn{1}{c|}{$3.13\text{e-}01^{***}$} &
  \multicolumn{1}{c|}{$3.77\text{e-}01^{***}$} &
  $5.30\text{e-}01^{***}$ \\ \hline
\end{tabular}
\end{table}
}

\section{Additional Analyses using Traditional Machine Learning Metrics}\label{sec: Appendix_C}

\begin{figure}[h!]
    \centering
    \includegraphics[width=\textwidth, height=20cm]{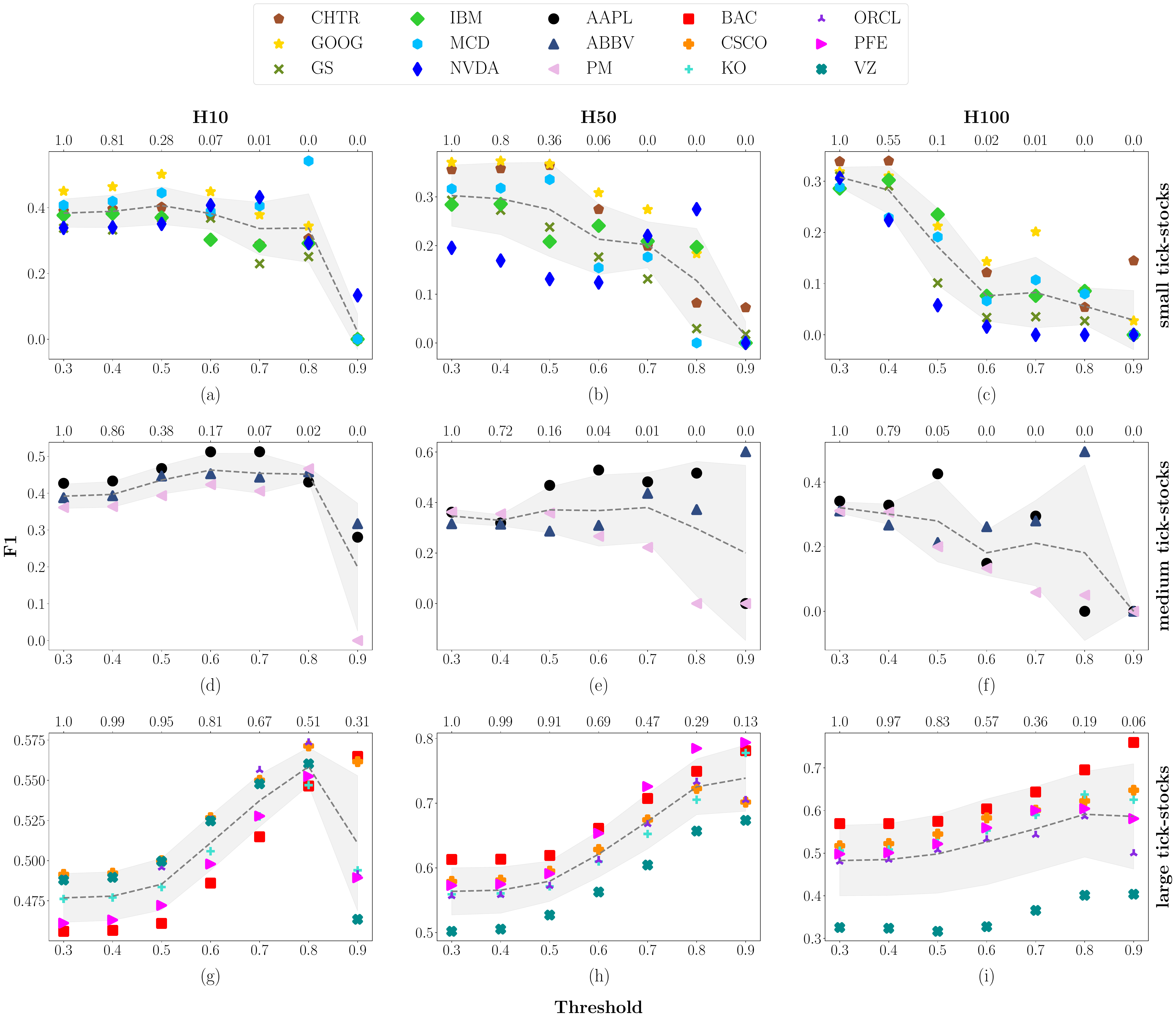}
    \caption{Average F1 score. Results are organised according to the prediction horizons taken into account (see columns) and stocks’ tick-size (see rows). Each plot contains three main pieces of information: (i) the model’s performance changes applying different thresholds on the probabilities associated with each forecast (shown on the bottom of the x-axis); (ii) the average percentage amount of remaining data after using the threshold (shown on the top of the x-axis); (iii) the performance average pattern and the corresponding standard deviation (shown through the grey line and shadows). All the average values and the standard deviations are computed by considering stocks with the same tick-size, spanning the $3$-year analysis period.}
    \label{fig:cross_years_average_f1}
\end{figure}

\begin{figure}[h!]
    \centering
    \includegraphics[width=\textwidth, height=20cm]{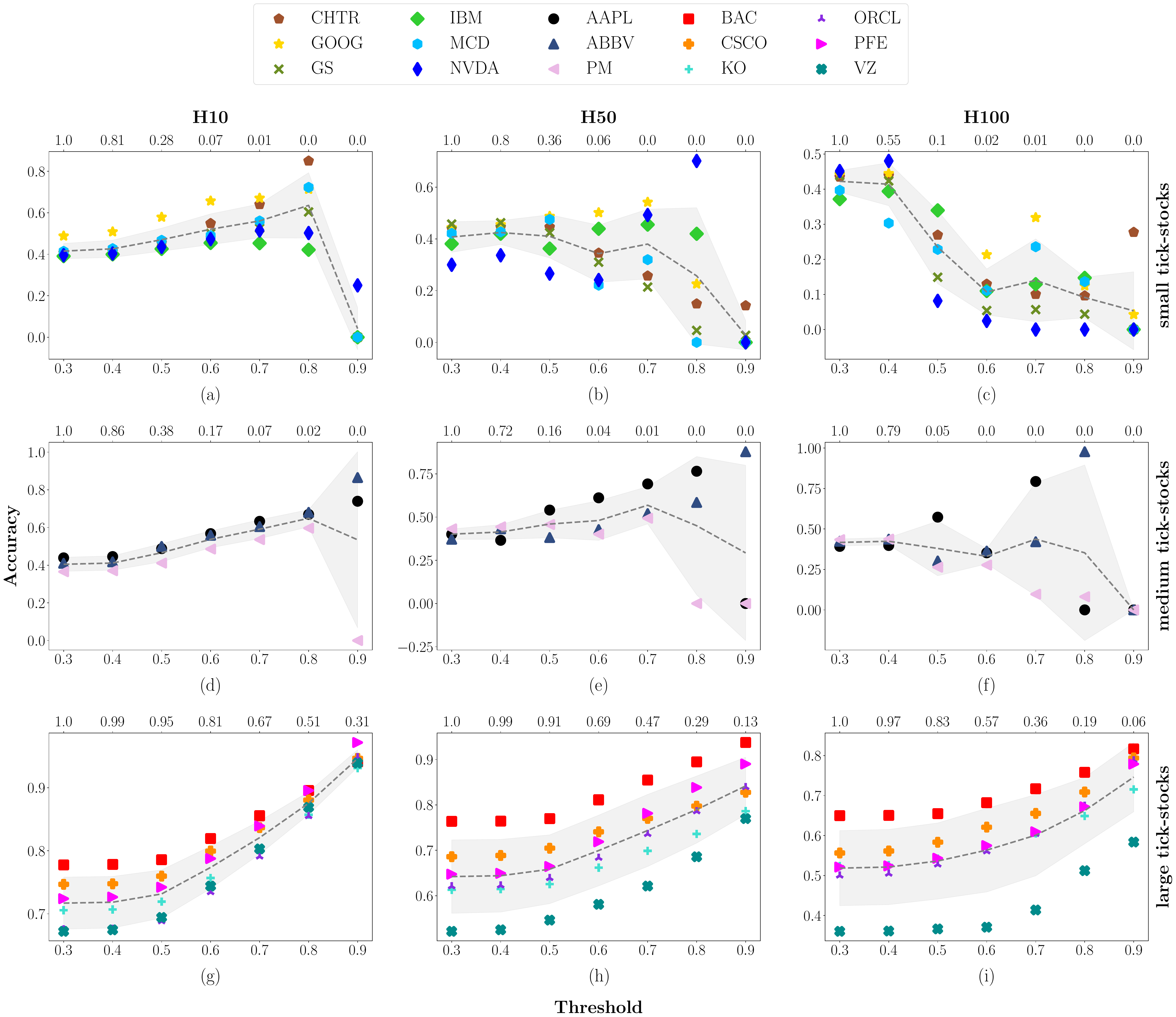}
    \caption{Average accuracy score. Results are organised according to the prediction horizons taken into account (see columns) and stocks’ tick-size (see rows). Each plot contains three main pieces of information: (i) the model’s performance changes applying different thresholds on the probabilities associated with each forecast (shown on the bottom of the x-axis); (ii) the average percentage amount of remaining data after using the threshold (shown on the top of the x-axis); (iii) the performance average pattern and the corresponding standard deviation (shown through the grey line and shadows). All the average values and the standard deviations are computed by considering stocks with the same tick-size, spanning the $3$-year analysis period.}
    \label{fig:cross_years_average_accuracy}
\end{figure}

\end{document}